\documentclass[twocolumn, switch]{article} 

\usepackage{preprint}

\usepackage{amsmath, amsthm, amssymb, amsfonts}

\usepackage[numbers,square]{natbib}
\usepackage[utf8]{inputenc}	
\usepackage[T1]{fontenc}	
\usepackage{xcolor}		
\usepackage[colorlinks = true,
            linkcolor = purple,
            urlcolor  = blue,
            citecolor = cyan,
            anchorcolor = black]{hyperref}	
\usepackage{booktabs} 		
\usepackage{nicefrac}		
\usepackage{microtype}		
\usepackage{lineno}		
\usepackage{float}			

\usepackage[utf8]{inputenc} 
\usepackage{tikz}
\usepackage{pgfplots}
\usepackage[noend]{algpseudocode}
\usepackage{mathtools} 
\usepackage{csquotes}
\usepackage[Algorithm]{algorithm} 
\usepackage{booktabs}
\usepackage{tabularx}
\usepackage{multicol}
\usepackage{array}
\usepackage{multirow}
\usepackage[T1]{fontenc}
\usepackage{subcaption}
\usepackage{graphicx}
\usepackage{url}
\usepackage[official]{eurosym}
\usepackage{cleveref}
\usetikzlibrary{arrows, arrows.meta, shapes}
\usepackage{placeins}
\usepackage{nameref}

\usepackage{newfloat}
\DeclareFloatingEnvironment[name={Supplementary Figure}]{suppfigure}
\usepackage{sidecap}
\sidecaptionvpos{figure}{c}

\usepackage{titlesec}
\titlespacing\section{0pt}{12pt plus 3pt minus 3pt}{1pt plus 1pt minus 1pt}
\titlespacing\subsection{0pt}{10pt plus 3pt minus 3pt}{1pt plus 1pt minus 1pt}
\titlespacing\subsubsection{0pt}{8pt plus 3pt minus 3pt}{1pt plus 1pt minus 1pt}

\usepackage{tikz,xcolor,hyperref}

{ \vspace{-0.15cm}%
    \small\noindent{\bfseries Availability of Data and Material:}\par%
    \noindent\ignorespaces}%
{ \par\noindent%
\ignorespacesafterend }%

\newcommand{\gSet}[1]{\{\,#1\,\}}
\newcommand{\norm}[1]{\left\lVert\,#1\,\right\rVert}

\DeclarePairedDelimiter\abs{\lvert}{\rvert}
\makeatletter
\let\oldabs\abs
\def\abs{\@ifstar{\oldabs}{\oldabs*}}
\newcolumntype{x}[1]{>{\centering\let\newline\\\arraybackslash\hspace{0pt}}p{#1}}

\newcommand{\algparam}[9]{
\setlength{\tabcolsep}{8pt}
\begin{table*}[htpb]\caption{Algorithm parameters}\label{tab:alg_param}\centering
\begin{tabular}{x{4cm}ccc}
	\toprule \\
	objective & $\populationsize$ & $\numberofparents$ & $\numberofgenerations$ \\
	\midrule
	arbitrage & #1 & #2 & #3 \\
	peak shaving & #4 & #5 & #6 \\
	local SDM & #7 & #8 & #9 \\
	\bottomrule
\end{tabular}

\end{table*}
\setlength{\tabcolsep}{6pt}
}
\newcommand{\schedulevar}{U}
\newcommand{\choosevar}{x}
\newcommand{\targetvar}{c^{\rm target}}
\newcommand{\powersell}{P^{\rm sell}}
\newcommand{\powerbuy}{P^{\rm buy}}
\newcommand{\powerlocal}{P^{\rm lu}}
\newcommand{\marketchargevar}{c}
\newcommand{\marketdischargevar}{d}
\newcommand{\householdprofile}{H}
\newcommand{\dischargestorage}{D_e}
\newcommand{\current}{I}

\newcommand{\populationsize}{\mu}
\newcommand{\numberofgenerations}{\kappa}
\newcommand{\generationsize}{\lambda}
\newcommand{\numberofparents}{\rho}
\newcommand{\stepsize}{\sigma}
\newcommand{\scheduleset}{R}
\newcommand{\cohdamessage}{m}
\newcommand{\strategy}{s}
\newcommand{\strategyset}{S}
\newcommand{\loadstate}{L}
\newcommand{\power}{P}
\newcommand{\loadstaterepr}{G_L}
\newcommand{\powerrepr}{G}
\newcommand{\peakcost}{p^{\rm cost}}
\newcommand{\energymarket}{E}
\newcommand{\threshold}{\phi}
\newcommand{\relativethreshold}{\phi_r}

\newcommand{\scheduleindex}{g}
\newcommand{\agentindex}{a}
\newcommand{\timeindex}{t}
\newcommand{\intervalindex}{i}
\newcommand{\coupledderindex}{j}

\newcommand{\coupledderlength}{n^{\rm der}}
\newcommand{\schedulelength}{n^{\rm schedule}_{a}}
\newcommand{\agentlength}{n^{\rm agent}}
\newcommand{\intervallength}{n^{\rm interval}}

\newcommand{\bestcurrent}{\textit{best}_c}
\newcommand{\bestknown}{\textit{best}_k}
\newcommand{\bestlastfitness}{\textit{fitness}^{\rm last}}
\newcommand{\population}{V_P}
\newcommand{\parents}{V_E}
\newcommand{\solutionvar}{V_S}
\newcommand{\generation}{V_G}
\newcommand{\besthistory}{V_R}
\newcommand{\restartcounter}{\textit{restart}^{\rm count}}

\newcommand{\mutationlowerbound}{b_{\rm lower}}
\newcommand{\mutationupperbound}{b_{\rm upper}}
\newcommand{\mutationbound}{b}
\newcommand{\mutationloadstate}{l}
\newcommand{\mutationloadstatenew}{l^{\rm new}}
\newcommand{\mutationstrategynew}{s^{\rm new}}
\newcommand{\mutationtuple}{t_{\rm G_L}}
\newcommand{\mutationtuplenew}{t^{\rm new}_{G_L}}

\newcommand{\mutationdiscoveryindex}{i_m}

\newcommand{\recsearchrange}{r_n}
\newcommand{\recsearchmid}{r_m}
\newcommand{\recsearchstart}{r_s}
\newcommand{\recsearchend}{r_e}
\newcommand{\reccuttingpoints}{R_I}
\newcommand{\recindex}{i_r}

\definecolor{lime}{HTML}{A6CE39}
\DeclareRobustCommand{\orcidicon}{
	\begin{tikzpicture}
	\draw[lime, fill=lime] (0,0) 
	circle [radius=0.16] 
	node[white] {{\fontfamily{qag}\selectfont \tiny ID}};
	\draw[white, fill=white] (-0.0625,0.095) 
	circle [radius=0.007];
	\end{tikzpicture}
	\hspace{-2mm}
}
\foreach \x in {A, ..., Z}{\expandafter\xdef\csname orcid\x\endcsname{\noexpand\href{https://orcid.org/\csname orcidauthor\x\endcsname}
			{\noexpand\orcidicon}}
}

\title{A Multi-Criteria Metaheuristic Algorithm for Distributed Optimization of Electric Energy Storage}

\usepackage{xwatermark}
\newwatermark[firstpage,color=gray!90,angle=0,scale=0.28, xpos=0in,ypos=-5in]{*correspondence: \texttt{rico.schrage@uni-oldenburg.de}}

\usepackage{authblk}

\author[1\thanks{\tt{rico.schrage@uni-oldenburg.de}}]{Rico Schrage\orcidA{}}
\author[1]{Paul Hendrik Tiemann\orcidB{}}
\author[1]{Astrid Nieße\orcidC{}}

\affil[1]{Digitalized Energy Systems Group, Carl von Ossietzky University Oldenburg}

\begin{document}

\twocolumn[ 
  \begin{@twocolumnfalse} 
  
\maketitle

\begin{abstract}
    The distributed schedule optimization of energy storage constitutes a challenge. Such algorithms often expect an input set containing all feasible schedules and, therefore, require searching the schedule space efficiently. However, it is hardly possible to accomplish this with energy storage due to its high flexibility. In this paper, the problem is introduced in detail and addressed by a metaheuristic algorithm, which generates a preselection of schedules. Two contributions are presented to achieve this goal: First, an extension for a distributed schedule optimization allowing a simultaneous optimization is developed. Second, an evolutionary algorithm is designed to generate optimized schedules with respect to multiple criteria. It is shown that the presented approach is suitable to schedule electric energy storage in actual households and industries with different generator and storage types.
\end{abstract}
\vspace{0.35cm}

  \end{@twocolumnfalse} 
] 



\section{Introduction}
In future power grids, distributed energy resources (DERs) such as solar panels and wind turbines lead to a weather-dependent power supply. To balance this, energy storage is of major importance  \cite{roberts2011role}. In order to use multiple (small) DERs combinedly for different purposes, it is established to aggregate them to virtual power plants (VPP) \cite{bitsch2002virtuelle,niee2015fully}. In this context, distributed algorithms have been discussed to protect private data, such as local constraints, and ensure scalability. Because many approaches \cite{landaburu2006optimal,li2010coordination,hinrichs2017distributed} rely on provided operational schedule (OS) sets and energy storage can utilize much flexibility, searching the solution space is not trivial. In this work, the Combinatorial Optimization Heuristic for Distributed Agents (COHDA) will be used as an example of a fully distributed scheduling approach to address the flexibility dispatch problem in distributedly controlled VPP \cite{hinrichs2017distributed}.
This involves designing and integrating an evolutionary multi-criteria algorithm as a parallel process for each storage to allow the usage of electric energy storage for different local purposes while globally fulfilling different goals. In this paper we will consider \textit{arbitrage}, \textit{local supply-demand matching (local SDM)} and \textit{peak shaving}. With this concept, the presented algorithm realizes a multi-purpose operation of electrical energy storage \cite{TERLOUW2019356,tiemann2020electrical}.

The research methodology used here is similar to the SGAE approach described in the work of Nieße et al. \cite{niesse2014sgae}. Therefore, the work has been done in an iterative process of requirements engineering, designing, prototyping, and simulative evaluation, with each step connected to a software artifact.

This paper is structured as follows. In the next section, we motivate the research gap in more detail based on related work. After that, the system setting is introduced. Then, some preliminaries are explained. The consecutive section presents the methodology for integrating energy storage using a local optimization process into a global optimization heuristic. The following section will explain how the local optimization process works with the inclusion of private constraints. Finally, the results of the simulations of the whole system will be discussed, and a conclusion will be drawn.%
\section{Related Work}%
The energy dispatch problem has been addressed with central approaches like linear programming (LP) or dynamic programming (DP) \cite{nottrott2013energy}. In \cite{ikeda2015metaheuristic} the authors compare the metaheuristics particle swarm optimization (PSO), genetic algorithms, and cuckoo search (CS) to solve this problem. They also showed that CS was able to find a semi-optimal solution 135 times as fast as a dynamic programming algorithm. The general problem with central approaches in our context is that it is not sufficient to cover different stakeholders of DERs due to privacy aspects. Further, LP and DP are not suitable for Pareto optimization algorithmic-wise.
While some approaches for the distributed energy scheduling problem with energy storage have been presented \cite{logenthiran2011multi,logenthiran2010multi,silva2012integrated,en13153921}, none of them realizes the integration of a local objective. Besides, using well-known distributed mechanisms like distributed gradient descent (DGD) \cite{ram2009distributed} or the alternating direction method of multipliers (ADMM) \cite{wei2012distributed} may generally not be possible due to constraining mathematical requirements for the objective function. In this terms, combinatorial methods \cite{landaburu2006optimal,li2010coordination,hinrichs2017distributed} can provide better flexibility and privacy advantages.

In \cite{bremerdecoder2013}, the authors introduced a decentralized support-vector-machine-based concept, which values privacy, but is unable to integrate energy storage due to the sampling-based approach, thus leading to a repeatedly needed time-consuming training phase. Furthermore, in \cite{bremer2019evo}, the example heuristic COHDA used in this work is extended to a multi-objective distributed algorithm using the evolutionary metaheuristic. Although the approach seems similar due to the usage of an evolutionary algorithm, the presented concept differs w.r.t local constraints: As they have to be published following the approach from Bremer et al., it is not applicable in a multi-stakeholder setting because of privacy aspects. In \cite{bremer2020steering}, which is about the inclusion and optimal integration of local objectives in COHDA, an evolutionary algorithm was used to determine an optimal balance between a local and a global objective. This is useful to integrate local objectives but does not reflect the specific requirements resulting from highly flexible energy storage.

To be able to perform a Pareto optimization without weighting, metaheuristic algorithms are the default choice \cite{mochapter}. Moreover, there exists plenty of algorithms for this task, e.g., \cite{marichelvam2013discrete, FADAEE20123364, mirjalili2016multi, liao2018comparative, deb2002fast}. We tested some of these without success. We conclude that none of them could be feasible to be used in this context, as the valid solution space of energy storage optimization problems is sparse, and due to the state of energy storage, the dependencies of energy values in different time intervals are complex. 

To sum up, state of the art lacks algorithms that can integrate energy storage in fully distributed energy scheduling approaches having the following abilities:
\begin{itemize}
  \item[-] Preserving private data of different stakeholders, and
  \item[-] efficiently searching solutions considering local objectives, thus allowing for multi-purpose usage of energy storage.
\end{itemize}

The first gap will be tackled by developing a novel extension for COHDA, allowing a local schedule optimization with respect to the local objective and the objective of some coalition. For the second, we will present GABHYME, an algorithm based on the evolutionary metaheuristic with novel recombination, mutation, and repairing mechanisms to perform a multi-objective optimization. 
\section{General system setting}
\label{sec:system_setting}
Before starting with the technical details, this section will introduce the type of system setting and its challenge.

The system we are looking at consists of several participants using energy storage. An agent represents each participant, and we assume it is economically rational to team up for a VPP to fulfill some product. In this work, we will exemplarily consider active power supply. To jointly fulfill a product, the participants in a VPP must negotiate during which interval and which participant provides how much power. Additionally, we assume that single participants try to share a minimal amount of information about themselves. As a consequence, privacy-preserving distributed optimization is necessary to aggregate power supply schedules for fulfilling a specific target product. As reasoned in the previous sections, COHDA complies with these requirements and is a fit for that setting. However, in the case of energy storage, COHDA is hardly useable due to the enormous flexibility and its need to express these in a concrete schedule set. Furthermore, especially energy storages have a strong incentive to utilize remaining flexibility besides the distributed aggregation. 

That said, we define a simplified example scenario to explain the system setting further. Assume there is one pumped storage plant (PSP), one household with a solar power plant (SPP) and battery storage, and a medium-sized industrial demand with a battery. Each is represented by an agent, and they teamed up intending to fulfill some given product target schedule. Besides, every agent has a different local objective. For example, the household storage optimizes local demand matching, the industry battery performs peak shaving, and the PSP uses its capacity to realize a profit with variations in the energy market (arbitrage). The local objectives must be considered when negotiating the VPP's objective (the global objective) to reach all these goals. As a result, at least two objectives should be considered when generating the schedules for each energy resource.
\section{Model foundation}

To be applicable to different energy storage technologies, an abstract and technology-independent model is needed. 
The purpose of the model is to describe the state of charge (SoC) over time. 
So we need a mapping charged/discharged power to SoC dependent on the time, respectively the interval number. 

The abstract energy storage model from \cite{tiemann2020electrical} is used in this work.
It comprises four different attributes of energy storage:
\begin{itemize}
    \item[-] power limits for charging/discharging
    \item[-] capacity
    \item[-] charge/discharge-efficiency
    \item[-] self-discharge.
\end{itemize}
In this work, the storage management equation from \cite{tiemann2020electrical} will be used with minor notational adaptions.
It is defining the storage power $P(t)$, depends on the set power $P_\timeindex^{\rm set}$ at time $\timeindex$ and is calculated as follows:
\begin{equation}\label{eq:constraints_storage}
    P(t) =
    \begin{cases}
        P^{\rm max}_{\rm ch} & \text{for } 0 \leq L(\timeindex) < 1 \land P_\timeindex^{\rm set} > P^{\rm max}_{\rm ch} \\
        P_t^{\rm set} & \text{for } 0 \leq L(\timeindex) < 1 \land P^{\rm max}_{\rm ch} \geq P_\timeindex^{\rm set} > 0 \\
        P_t^{\rm set} & \text{for } 0 < L(\timeindex) \leq 1 \land 0 > P_\timeindex^{\rm set} \geq -P^{\rm max}_{\rm dis} \\
        -P^{\rm max}_{\rm dis} & \text{for } 0 < L(\timeindex) \leq 1 \land P_\timeindex^{\rm set} < -P^{\rm max}_{\rm dis} \\
        0 & \text{else}
    \end{cases}
\end{equation}
Hereby, $P^{\rm max}_{\rm ch/dis}$ are constants for the upper charging/discharging limit and $\timeindex$ is the time. The state of energy of the storage $L(\timeindex)$ is expressed as a differential equation:
\begin{equation*}
    C_{\rm E}\frac{\text{d}L(\timeindex)}{\text{d}\timeindex}=-\frac{C_{\rm E}\cdot L(\timeindex)}{\tau} + P(\timeindex)\cdot
    \begin{cases}
        \eta_{\rm ch} \text{ if } P(t)>0 \text{ (charge)}\\
        \frac{1}{\eta_{\rm dis}} \text{ if } P(\timeindex)>0 \text{ (discharge)}
    \end{cases}
\end{equation*}
Therein $C_{\rm E}$ is the capacity, $L$ is the state of charge, $\tau$ is the self-discharge time constant,
$P(\timeindex)$ is the power charged/discharged at time $\timeindex$, and $P^{\rm max}_\text{ch/dis}$ are the charging/discharging efficiencies of the storage. Instead of $t$, an interval number $i$ will be used in this work. Each interval lasts a defined number of minutes (e.g., $15$). Therefore $\tau$ is chosen accordingly. For the sake of brevity, we will omit the formalization of the time mappings.

\section{Basic algorithm and extensions}
In the subsequent sections, the COHDA heuristic \cite{hinrichs2017distributed}, which has been used as basis for the presented work, is introduced, and the extension done by this work is presented.

\subsection{COHDA}

COHDA is a distributed optimization heuristic developed with a focus on energy scheduling using a multi-agent system. An agent, in this context, is a unit that can sense the environment and react accordingly. In its simplest form, an agent represents one DER (other variants are possible, including sub-aggregations, but will not be discussed here for reasons of brevity). All agents are connected using an arbitrary connected topology. Hinrichs describes the problem as $\mathcal{MC}-\mathcal{COP}$ (Multiple Choice Combinatorial Optimization Problem), which is about selecting a set of solutions from a solution space, minimizing the distance of the sum to a target solution. In the energy scheduling application of COHDA, the target solution is an OS \cite{hinrichs2013cohda}, which is an ordered set of active power values describing at which interval how much power should be aggregated. The solution candidate is a cluster schedule (CS).

Every agent, denoted by the index $\agentindex$, defines a set of feasible OS $\scheduleset_{\rm \agentindex}$ with $0 \leq \agentindex < \agentlength$, where $\agentlength$ is the number of agents. Thereby it is not specified how this set has to be generated. The heuristic is executed by each agent in three different steps: 
\begin{enumerate}
  \item \texttt{(update)} The agent receives the current solution of its neighbours and eventually updates its own if their solution is better.
  \item \texttt{(choose)} Eventually, the agent adapts its own choice to optimize the solution.
  \item \texttt{(send)} If the agent finds a better solution with respect to the global optimization target, it will send it to all neighbours. 
\end{enumerate}
These three steps will get executed by every agent until no better solutions are found. It is proven that COHDA will always converge to a solution. Details on prerequisites and further characteristics of the heuristic are given in \cite{hinrichs2017distributed}. For the work presented here, we reduced the presentation of COHDA to the necessary details.

\subsection{Extension}
\label{subsec:extension}

\begin{figure}[htbp]
  \centering
  \includegraphics[width=1\columnwidth]{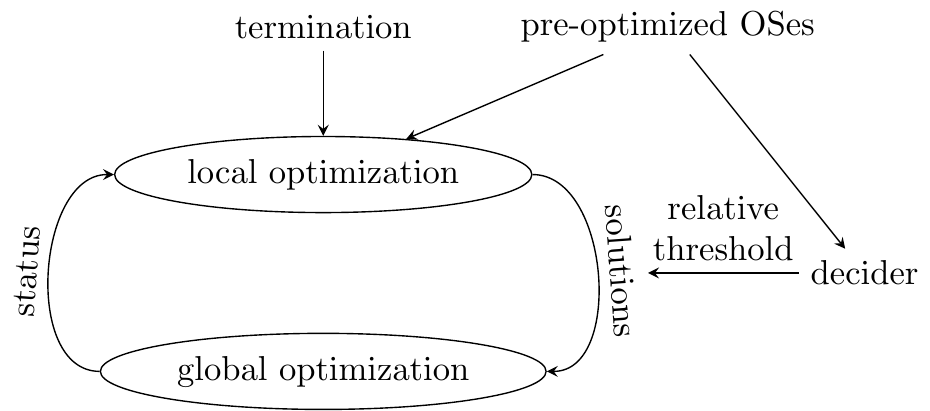}
  \caption{Local optimization and global optimization}
  \label{fig:cohda_system}
\end{figure}

At this point, the necessary extensions and modifications of COHDA for multi-criteria storage operation are introduced. The work presented here aims to establish a parallel local optimization process running in the background at every agent, which participates in the schedule optimization with COHDA. This concept is visualized in figure \ref{fig:cohda_system}. The extension addresses the following three challenges:
\begin{enumerate}
  \item Local optimization: how to generate the operational schedules?
  \item Global optimization: how to integrate the local optimization?
  \item Termination: how to handle termination if a parallel local optimization exists?
\end{enumerate}
The first part directly tackles the problem of efficient solution space searching and protecting private data. The other two are necessary to integrate the approach in the fully distributed heuristic COHDA.
\paragraph{Local optimization} A local optimization is running in parallel to the global optimization, continuously generating new high-quality OS due to a local and the global COHDA objective. The optimization is explained in detail in section \textit{\nameref{sec:local_optimization}}.
\paragraph{Global optimization} All generated OS which are locally acceptable are added to the set of feasible OS $\scheduleset_{\rm \agentindex}$. To determine whether an OS is acceptable, the decider is introduced. The decider uses a threshold, which is defined using the pre-optimization. The fitness of the best pre-optimized solution is multiplied by the relative threshold to determine the minimum fitness for new operational schedules. The decider will allow adding an OS if the schedule's local fitness is above this absolute threshold $\threshold$. Therefore, it contributes to resolving conflicting objectives. After a solution has been picked, only the global fitness will matter. The pre-optimization is equivalent to the local optimization but uses the local objective only. 

Another fundamental difference is the start indication for the three COHDA steps: \texttt{update,\,choose,\,send}. Only starting this step-chain with receiving a message is impossible because $\scheduleset_{\rm \agentindex}$ is not an immutable set anymore due to continuously adding new OS generated by the local optimization. Consequently, the three steps also start with changing the set $\scheduleset_{\rm \agentindex}$. 
\paragraph{Termination} Lastly, the termination criterium has to be changed. COHDA will terminate when no agent chooses an improving OS, but the local optimization continuously generates new OS, which might be better. To handle this, the local optimization uses a Kalman-filter \cite{kalman1960new} to determine whether significantly better OS with regard to a local criteria are continuously getting found. When the improvement becomes too slow, the generation process will terminate. As a consequence, the global optimization must not stop before every local optimization is finished. In practice, one should relax this to avoid very long negotiations.
\newline
\newline
These modifications result in the overall procedure, outlined in figure~\ref{fig:cohda_system} and algorithm~\ref{alg:agent_behaviour_new}. Therein $\cohdamessage$ is the first message received. It is assumed that the agents are already connected to each other according to their topology. The influence of the concrete topology is neglected here; the interested reader can find more information about this in \cite{hollytopo}. 
\begin{algorithm}[htpb]
  \caption{Schematic overall procedure per agent $\agentindex$}
  \label{alg:agent_behaviour_new}
\begin{algorithmic}[1]
      \Procedure{handleFirstMessage}{$\cohdamessage$}
        \State $\scheduleset_{\rm \agentindex}$ = \textsc{startLocalOptimization(}$\cohdamessage$\textsc{)} \Comment{pre-optimization}
        \State \textsc{globalOptimization(}$\cohdamessage,\,\scheduleset_{\rm \agentindex}$\textsc{)} \Comment{init global optimization}
        \While{\textsc{isLocalOptimizationNotDone()}} \Comment{termination}
          \State $\scheduleset_{\rm \agentindex}$ = \textsc{getNewOperationalSchedules()}
          \If{\textsc{newMessageOrSchedulesAvailable(}$\scheduleset_{\rm \agentindex}$\textsc{)}}
            \State \textsc{globalOptimization(}$\cohdamessage,\,\scheduleset_{\rm \agentindex}$\textsc{)}\Comment{global optimization}
          \EndIf
        \EndWhile
        \State \textsc{stop()}
        \State \textsc{postResult()}
      \EndProcedure
\end{algorithmic}
\end{algorithm}

Because the local optimization surely terminates at some point, this extension does not modify any property of COHDA relevant to the convergence and termination proofs from Hinrichs in \cite{hinrichs2017distributed}. As a result, the proofs are still valid for the extended version.
\section{Local Optimization}\label{sec:local_optimization}
In this section, the local optimization responsible for generating the OS is described. First, a solution representation is introduced, second, the objectives and constraints are defined, and last, the solving procedure is explained.

\begin{figure}[htpb]
  \centering
  \includegraphics[width=1\columnwidth]{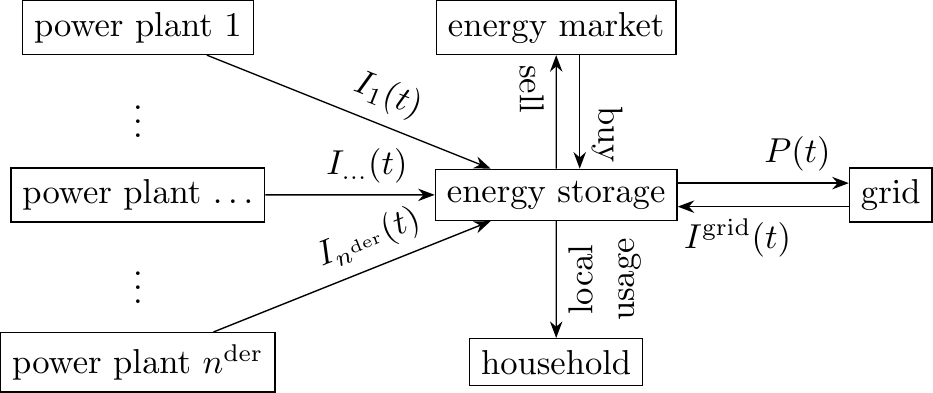}
  \caption[StorageModelKontext]{Energy storage and its environment}
  \label{fig:energy_storage_mode_environment}
\end{figure}

In figure \ref{fig:energy_storage_mode_environment}, the environment of the storage is depicted. For the objective \textit{local SDM}, it is assumed that the storage belongs to a coupled power plant. Also, there is an energy market of which we have perfect knowledge.
\subsection{Objectives}
In the local optimization, a multi-criteria maximation problem with respect to the global objective and some local objectives is solved. Therefore a solution to the problem is defined as follows.
\begin{equation}\label{eq:strategy_eq}
  \begin{split}
      \powerrepr &= \gSet{(\strategy_1,\,\power(1)),\,(\strategy_{\dots},\,\power(\dots)),\,(\strategy_{\rm \intervallength},\,\power(\intervallength))} \\
      \loadstaterepr &= \gSet{(\strategy_1,\,\loadstate(1)),\,(\strategy_{\dots},\,\loadstate(\dots)),\,(\strategy_{\rm \intervallength},\,L(\intervallength))}\\
      \strategyset &= \gSet{\text{buy},\,\text{sell},\,\text{charge},\,\text{discharge},\,\text{local SDM}}\\
      \text{with } &0 < \intervalindex \leq \intervallength \text{, } \strategy_{\rm \intervalindex} \in \strategyset \text{ and } \intervallength\in\mathbb{N}.
  \end{split}
\end{equation}
$\powerrepr$ is the power representation consisting of tuples of a strategy and the active power.
$\loadstaterepr$ is the load state representation, where the power is expressed with the energy storage's charge state. The variable $\intervallength$ is the number of intervals. Finally, $\strategyset$ denotes the set of strategies. A strategy determines what has to be done with the paired power value and is necessary to implement different objectives.
Our objective is to create operational schedules which fit the global optimization with COHDA and optimize with respect to the local objective. The following equation is the first objective.
\begin{equation}\label{eq:cohda_ziel}
  \begin{split}
      &\underset{x_{\rm\agentindex,\scheduleindex}}{\text{max}}~\left(-\norm{\targetvar - \sum_{\agentindex=1}^{\agentlength}\sum_{\scheduleindex=1}^{\schedulelength} (\schedulevar_{\rm \agentindex,\scheduleindex}\cdot x_{\rm \agentindex,\scheduleindex})}_1\right)\\
      \text{with } &\sum_{\scheduleindex=1}^{\schedulelength}x_{{\rm \agentindex,\scheduleindex}} = 1\\
      &x_{{\rm \agentindex,\scheduleindex}}\in\gSet{0,\,1},~\agentindex=1,\,\dots,\,\agentlength,~\scheduleindex=1,\,\dots,\,\schedulelength.
  \end{split}
\end{equation}
Therein the variable $\targetvar$ is the target OS of COHDA, which is a set of active power, $\choosevar_{\agentindex,\scheduleindex}$ is a binary decision variable for the choice of the $\scheduleindex$'th OS of the agent with index $\agentindex$, $\schedulevar_{\rm \agentindex,\scheduleindex}$ is the $\scheduleindex$'th schedule of an agent with index $\agentindex$. Therefore $\schedulelength$ is the number of OS of an agent with index $\agentindex$, and $\agentlength$ is the number of agents.

As a next step, the local objectives follow.
\begin{equation}
  \label{eq:arbitrage}
  \underset{\powersell,~\powerbuy}{\text{max}}~\sum_{\intervalindex=1}^{\intervallength}\left(\marketdischargevar(\intervalindex)\cdot \powersell(\intervalindex) - \marketchargevar(\intervalindex)\cdot \powerbuy(\intervalindex)\right)
\end{equation}
\begin{equation}
  \label{eq:own_consumption}
  \underset{\powerlocal}{\text{max}}~\sum_{\intervalindex=1}^{\intervallength}\left(\marketchargevar(\intervalindex) \cdot \left(\text{max}(\householdprofile_{\intervalindex},\,0) - \abs{\powerlocal(\intervalindex) - \text{max}(\householdprofile_{\rm \intervalindex},\,0)}\right)\right)
\end{equation}
\begin{equation}\label{eq:peak_shaving}
    \begin{split}
      &\underset{\dischargestorage}{\text{max}}~\left(\Delta\householdprofile_p \cdot \peakcost\right)\\
      \text{with } &\Delta\householdprofile_p = \left(\text{max}(\dischargestorage) - \text{max}(\householdprofile+\dischargestorage)\right)
    \end{split}
\end{equation}
\begin{equation}
  \begin{split}
      \marketchargevar,\,\marketdischargevar&\colon \mathbb{N}_0 \rightarrow \mathbb{R} \\
      \energymarket &= (\marketchargevar,\,\marketdischargevar).\\
  \end{split}
\end{equation}
Equation (\ref{eq:arbitrage}) defines the objective \textit{arbitrage}, where it is about earning money with the fluctuation of the energy market. 
The mapping $\marketchargevar$ is the cost to buy energy at interval $\intervalindex$, $\marketdischargevar$ describes the money to earn when selling energy in $\intervalindex$. The functions $\powersell(\intervalindex)$ and $\powerbuy(\intervalindex)$ represent the energy to sell/buy in $\intervalindex$. Furthermore, the equation (\ref{eq:own_consumption}) describes the money saved when using power for the local supply-demand matching (local SDM) instead of buying it from the energy market. Here, $\powerlocal(\intervalindex)$ is the energy planned for the local SDM in the interval $\intervalindex$, $\householdprofile_{\rm \intervalindex}$ is power consumption in the interval $\intervalindex$ of a household/industry. The last objective is represented by equation (\ref{eq:peak_shaving}), which is about \textit{peak shaving} to reduce costs. The variable $\householdprofile$ is hereby the load profile of the household/industry, $\dischargestorage$ is the power discharge profile of the energy storage, $\Delta\householdprofile_p$ is the peak energy demand reduction amount done by the storage, and $\peakcost$ is the cost constant for peak energy, given by the energy vendor.
\subsection{Constraints}
If the storage is coupled with a generator, charging is constrained to the generated output. Consequently for all mappings, representing the OS of the coupled DER, $\current_{\rm \coupledderindex}\colon \mathbb{N} \rightarrow \mathbb{R} \text{ with \coupledderindex } = 1,\,\dots,\,\coupledderlength$ the following condition holds.
\begin{equation*}
    \forall (\strategy_\intervalindex,\,P(\intervalindex))\in\powerrepr\colon~\strategy_\intervalindex=\text{charge} \ \Rightarrow \power(\intervalindex) \leq I_{\coupledderindex}(\intervalindex) \text{ with } 0\leq \intervalindex < \intervallength.
\end{equation*}
Moreover, the strategy has to be consistent with the set power, meaning if the strategy is \enquote{buy}, it is impossible to choose $\power(\intervalindex)<0$.
\begin{equation*}
  \begin{split}
      \forall (\strategy_\intervalindex,\,\power(\intervalindex))\in \powerrepr\colon \power(\intervalindex) > 0 &\Leftrightarrow \strategy_\intervalindex \in \gSet{\rm buy,~\rm charge}\\
      \forall (\strategy_\intervalindex,\,\power(\intervalindex))\in \powerrepr\colon \power(\intervalindex) < 0 &\Leftrightarrow \strategy_\intervalindex \in \gSet{\rm sell,~\rm discharge,~\rm local\,SDM}\\
      \text{with } 0 \leq \intervalindex < \intervallength
  \end{split}
\end{equation*}
Additionally, the constraints, defined by the energy model are applied by definition of the OS (see equation (\ref{eq:constraints_storage}) and (\ref{eq:strategy_eq})).
\subsection{GABHYME}
In this section the metaheuristic algorithm \textbf{Ga}ussian \textbf{B}rush \textbf{Hy}brid \textbf{M}emetic \textbf{E}volutionary (GABHYME) is presented. It can solve the previously introduced multi-criteria problem and, therefore, will be used to generate the OS in the local optimization. 
GABHYME is based on the evolutionary metaheuristic \cite{back1996evolutionary} and can be characterized as ($\populationsize + \generationsize$)-Evolutionary Algorithm (EA). It combines a novel heuristically cut-point recombination with a hybrid mutation operator, consisting of local search and gaussian random-based search.
The algorithm has five parameters: planning horizon $\intervallength$, population size $\populationsize$, number of generations $\numberofgenerations$, generation size $\generationsize$, number of parents $\numberofparents$. The chosen approach to set these parameters optimally is presented in the appendix. The flow is written as pseudocode in algorithm \ref{alg:evo_base}. The next paragraphs will introduce the methods used one by one.
\begin{algorithm}[htpb]
  \caption{GABHYME}
  \label{alg:evo_base}
  \begin{algorithmic}[1]
        \Procedure{Evo}{$\intervallength$, $\populationsize$, $\numberofgenerations$, $\generationsize$, $\numberofparents$}
            \State $\besthistory = \emptyset$
            \State $\population = \Call{initPop}{\populationsize,\,\intervallength}$
            \State $\bestcurrent,\,\bestknown,\,\bestlastfitness = \text{None}$
            \State $\restartcounter = 0$
            \For{$\numberofgenerations \text{ times}$}
                \State $\generation = \emptyset$
                \For{$\generationsize \text{ times}$}
                    \State $\parents = \Call{selectParents}{\population,\,\numberofparents}$
                    \State $\solutionvar = \Call{recombine}{\parents,\,\numberofparents}$
                    \State $\solutionvar = \Call{mutate}{\solutionvar}$
                    \State $\generation = \generation \cup \solutionvar$
                \EndFor 
                \State $\population = \textsc{select}(\population \cup \generation,\,\populationsize)$
                \State $\bestcurrent = $ choose solution of $\population$ with maximal fitness
                \If{$\bestknown$ is None \textbf{or} $\textsc{fit}(\bestknown) < \textsc{fit}(\bestcurrent)$}
                    \State $\besthistory = \besthistory \cup \bestcurrent$
                    \State $\bestknown=\bestcurrent$
                \EndIf
                \If{$\restartcounter > \frac{\numberofgenerations}{10}$} \Comment{population restart}
                    \If{$\Call{fit}{\bestcurrent}-\bestlastfitness \leq 0.01$}
                        \State $r = \text{int}(\frac{\populationsize}{2})$
                        \State $\population = \population[0:r] \cup \Call{initPop}{\mu - r,\,\intervallength}$
                    \EndIf
                    \State $\restartcounter = 0$
                    \State $\bestlastfitness = \Call{fit}{\bestcurrent}$
                \EndIf
                \State $\restartcounter=\restartcounter+1$
            \EndFor
            \State \Return $\besthistory$ \phantom{$min(k_i)$}
        \EndProcedure
  \end{algorithmic}
\end{algorithm} 
\subsubsection{Fitness-function}
A fitness function is needed to rate solutions due to the objectives and constraints. In the algorithm, the objective functions of the previous section are used for that. To integrate the constraints, a penalty function is necessary. We define it so that the fitness is set to $-\infty$ when a constraint violation has been detected. This approach is called death-penalty \cite{michalewicz1995genetic}.
\subsubsection{Solution-repairing}
Although most operations try to preserve the validity of the solution, it can not be guaranteed (for example, when the recombination gets executed). To avoid breaking solutions with high fitness values, a repairing routine will be used, which is able to repair the constraints  \textit{limits}, \textit{consistency}, and \textit{satisfiability}. When violating the limits of the storage, a simple clipping routine will get executed (analog to equation (\ref{eq:constraints_storage})). If the consistency is violated, a random allowed strategy in $\strategyset$ will be used. Also, the power will be clipped to the storage maximal power limits.
\subsubsection{Self-adaptive mutation step size}
In this algorithm a concept called \textit{self adaptive mutation step size} is used \cite{schwefel1981numerical}. The idea is to optimize the mutation step size $\stepsize$ together with the solution. So every solution is paired with its mutation step size.
\begin{equation}\label{eq:sigma_init_update}
\begin{aligned}
  \stepsize &= e^{0{.}22\cdot \mathcal{N}(0,\,1)}\\
  \stepsize &= \stepsize \cdot e^{0{.}22\cdot \mathcal{N}(0,\,1)}
\end{aligned}
\end{equation}
When a solution mutates the step size mutates as well, using the equations (\ref{eq:sigma_init_update}) for initializing and updating $\stepsize$. The advantage of this approach is to enable fine control of mutation by the solution itself, following the idea that solutions with high fitness values will be created with high fitness mutation step sizes. However, because of the load state and the high restrictions for the state, the problem results in a cluttered and sparse solution space, and every solution has other requirements for the step size. The value $0.22$ as constant is used because of a recommendation in \cite{mutationrate}.
\subsubsection{Mutation}
The mutation is a hybrid memetic operator integrating the sampling of values from a normal distribution (referred to by \textit{gaussian brush}). It consists of two parts: the random-based explorative mutation and the local search approach. 
The latter is responsible for the exploitation and is based on the local search heuristic. For one element in $\loadstaterepr$, the load state will be step-wise adjusted within a valid interval, and it is checked whether the new solution outperforms the old one. While $\mutationdiscoveryindex$ is the number of the step, the equation (\ref{eq:mutation_local_search_step_calc}) calculates the load state. The variable $\mutationlowerbound$ is the lower, and $\mutationupperbound$ is the upper boundary. Furthermore, $l$ is the load state to be altered, $\mutationbound$ is the bound to be considered, and $\mutationloadstatenew$ is the new load state to be tested.
\begin{equation}\label{eq:mutation_local_search_step_calc}
  \begin{split}
      &\mutationloadstatenew = \mutationloadstate + \Call{sigmoid}{\mutationdiscoveryindex}\cdot (\mutationbound-\mutationloadstate)\\
      \text{with } &0\leq \mutationdiscoveryindex<5 \text{ and } \mutationbound\in\gSet{\mutationlowerbound,\,\mutationupperbound}\\
      &\Call{sigmoid}{x} = \frac{1}{1+e^{1{,}5x - 5}}
  \end{split}
\end{equation}
\begin{algorithm}[htpb]
  \caption{EA: Mutation}
  \label{alg:mutation}
  \begin{algorithmic}[1]
      \Function{mutate}{\textit{solution}}
          \State $(\stepsize, \powerrepr) = \textit{solution}$
          \State $\stepsize = \stepsize \cdot e^{0.22\cdot \mathcal{N}(0,\,1)}$
          \State $\loadstaterepr =$ state of charge representation of $\powerrepr$
          \If{$\textsc{random()} \geq 0.5$}
              \State $\intervalindex=$ choose random index of tuple in $\loadstaterepr$ \Comment{exploration}
              \State $g=$ gaussian distribution $\mathcal{N}(0,\,0.5)$ with size $\stepsize\cdot\intervallength$
              \State Add $\stepsize\cdot g$ to load state values of $\loadstaterepr$ starting at $\intervalindex$
              \State Set random index of $\loadstaterepr$ to a random strategy
              \State Repair solution $\loadstaterepr$ with regard to the constraints
          \Else
              \State $\mutationtuple=$ choose random tuple of $\loadstaterepr$ \Comment{exploitation}
              \State $\mutationloadstate$ = load state of $t_{\loadstaterepr}$
              \State $(\mutationlowerbound,\,\mutationupperbound)=$ calc. valid S.o.C. range for $\mutationtuple$
              \State $best = \mutationtuple$
              \For{$\mutationdiscoveryindex=0;\,\mutationdiscoveryindex<5;\,\mutationdiscoveryindex=\mutationdiscoveryindex+1$}
                  \For{$\mutationbound$ \textbf{in} $\gSet{\mutationlowerbound, \mutationupperbound$}}
                      \State $\mutationloadstatenew = \mutationloadstate + \Call{sigmoid}{\mutationdiscoveryindex}\cdot (\mutationbound-\mutationloadstate)$
                      \State $\mutationstrategynew = $ choose for $\mutationloadstatenew$ a fitting strategy for $\mutationtuple$
                      \State $\mutationtuplenew = (\mutationstrategynew,\,\mutationloadstatenew)$
                      \State $\loadstaterepr^{\text{new}} =$ original solution $\loadstaterepr$ with $\mutationtuplenew$
                      \If{\Call{fit}{$\loadstaterepr^{\text{new}}$} $>$ \Call{fit}{$\loadstaterepr$}}
                          \State $best = \mutationtuplenew$
                      \EndIf
                  \EndFor
              \EndFor
              \State Replace the tuple in $\loadstaterepr$ with $best$
          \EndIf
          \State \Return $(\stepsize,\,\powerrepr)$ \phantom{$min(k_i)$}
      \EndFunction
  \end{algorithmic}
\end{algorithm}
\subsubsection{Recombination}
The recombination operation has to fulfill two requirements: it should preserve the validity, and high fitness solutions should be able to pass on attributes to following generations. Therefore, a recombination operation is proposed, which recombines an interval of each parent's OS using the most similar cutting point. For example when there are two parents, the first step is to find the values $L_a(i) \in G_{\rm L,a}$ and $L_b(i) \in G_{\rm L,b}$ so that $\abs{L_a(i) - L_b(i)}$ is minimal. This search is done on the interval $[\recsearchstart,~\recsearchend],\text{ with }\recsearchstart=\recsearchmid-\text{int}(\frac{\recsearchrange}{2}),~\recsearchend=\text{max}(\recsearchmid+\frac{\recsearchrange}{2},~\recsearchstart+1),~\recsearchmid=\recsearchrange\cdot \recindex,~\recsearchrange=\text{int}(\frac{\intervallength}{\numberofparents})$. The variable $\recindex$ is the index of the interval (see algorithm \ref{alg:rekombination}), $\numberofparents$ is the number of parents, $L_a(\intervalindex),~L_b(\intervalindex)$ are the load state values for the OS a, b in the interval $\intervalindex$, and $G_{\rm L,a},~G_{\rm L,b}$ are the parent OS a and b.
\begin{algorithm}[htpb]
  \caption{EA: Recombination}
  \label{alg:rekombination}
  \begin{algorithmic}[1]
      \Function{recombine}{$\parents$, $\numberofparents$}
          \State $result = \emptyset$
          \State $R_I = \emptyset$
          \State $\parents = $ state of charge representation of $\parents$
          \For{$\recindex = 1;\,\recindex < \abs{\parents};\,\recindex=\recindex+1$}
              \State $\recsearchrange = \text{int}(\frac{\intervallength}{\numberofparents})$
              \State $\recsearchmid = \recsearchrange \cdot \recindex$
              \State $\recsearchstart = \recsearchmid - \text{int}(\frac{\recsearchrange}{2})$
              \State $\recsearchend = \text{max}(\recsearchmid + \frac{\recsearchrange}{2},\,\recsearchstart + 1)$
              \State $\reccuttingpoints = \reccuttingpoints \cup \text{min}_{\text{index}}\abs{V_{E,\recindex}[\recsearchstart:\recsearchend] - {V}_{E,\recindex-1}[\recsearchstart:\recsearchend]}$
          \EndFor
          \State $result=$ combine parents $\parents$ at the indices $\reccuttingpoints$.
          \State \Return $result$ \phantom{$min(k_i)$}
      \EndFunction
  \end{algorithmic}
\end{algorithm}
\subsubsection{Initialization}   
At the population's initialization step, the most important thing is generating diverse but valid solutions to enable an extensive exploration of the solution space. Therefore, for each solution and interval, a random strategy and fitting amount of power for charging/discharging are chosen. Also, the initial mutation step size is set (see equation \ref{eq:sigma_init_update}).
\subsubsection{Multi-criteria Extension}
There are two typical ways to integrate multiple objectives in metaheuristic optimization: Pareto-optimization and combining the objectives by multiplying some weight or translating it to some common unit. In this work, both approaches are used and evaluated separately.
\paragraph{Pareto}
As we consider, without loss of generality, only two objectives (local and global) at the same time, the NSGA-II selection operator, can be applied \cite{deb2002fast} for Pareto-optimization. NSGA-II has two different phases; first, the solutions get divided into fronts, and second, within each front, they get sorted by the descending crowding distance. The first step is needed to find all non-dominated solutions, while the second introduces a quality measure for those instances. For GABHYME, this can just be seen as the selection operation.
\paragraph{Normalization}
The second approach is normalizing the value ranges of both objectives. Here we will exploit the attribute that the fitness value range of the global objective is known, and the local objective range can be guessed via pre-optimization as described in the previous section.
To implement this approach, the local and global fitness will be normalized as follows:
\begin{equation}
\label{eq:normalization}
  \begin{split}
    g_{\rm norm}(x) &= 1 - \abs{\frac{g(x)}{\norm{\targetvar}_1}}\\
    f_{\rm norm}(x) &= \text{min}\left(\frac{f(x)}{\threshold},\,1\right)
  \end{split}
\end{equation}
Therein $f_{\rm norm}(x)$ is the local fitness normalized with respect to the absolute threshold $\threshold$, while $g_{\rm norm}(x)$ is the global fitness normalized using the norm of the target OS $\targetvar$ for COHDA, $g(x)$ is the global objective (see equation (\ref{eq:cohda_ziel})), $f(x)$ is the local objective.
\subsubsection{Termination}
\label{subsec:termin}
For the termination, GABHYME will use the approach described in \cite{marti2016stopping}, which uses the MGBM (after the authors' surnames) criterium with a special parameterized Kalman-Filter to detect a lack of change in the fitness. For more information, see the original paper \cite{marti2016stopping}.
In general, the work does not depend on any specific termination algorithm, so it is possible to change this choice without side effects on other components.
\section{System example}
To conclude the methodology, we present an example of the overall systems procedure. It is assumed there are three agents (household-, industry- and PSP-agent) configured as introduced in section \ref{sec:system_setting}, and GABHYME Pareto will be used as local optimization. 

First, each of the three agents starts a pre-optimization using GABHYME in a single-objective variant. In this pre-optimization, only the local objective of each agent matters: local SDM eq. (\ref{eq:own_consumption}), arbitrage eq. (\ref{eq:arbitrage}), and peak shaving eq. (\ref{eq:peak_shaving}). The fitness of the best solution $f_b$ will determine the actual threshold $\threshold = \relativethreshold\cdot f_b$ to determine if a schedule is feasible. Here, $\relativethreshold$ is the relative threshold. After the pre-optimization is finished, the global COHDA negotiation starts, and every agent has at least an initial schedule set with a length of one. At this point also, the parallel local optimization (e.g., GABHYME with NSGA-II) will start to generate fitting schedules with respect to two objectives, the local and the global. Every time new non-dominated solutions are found, they will be passed to the COHDA negotiation and inserted in the schedule set if their local fitness exceeds the threshold. This process runs until the local optimization terminates (see \ref{subsec:termin}) and an ordinary COHDA optimization remains with fixed size schedule sets.
\section{Evaluation}
\begin{figure}[htbp]
  \centering
  \includegraphics[width=0.8\columnwidth]{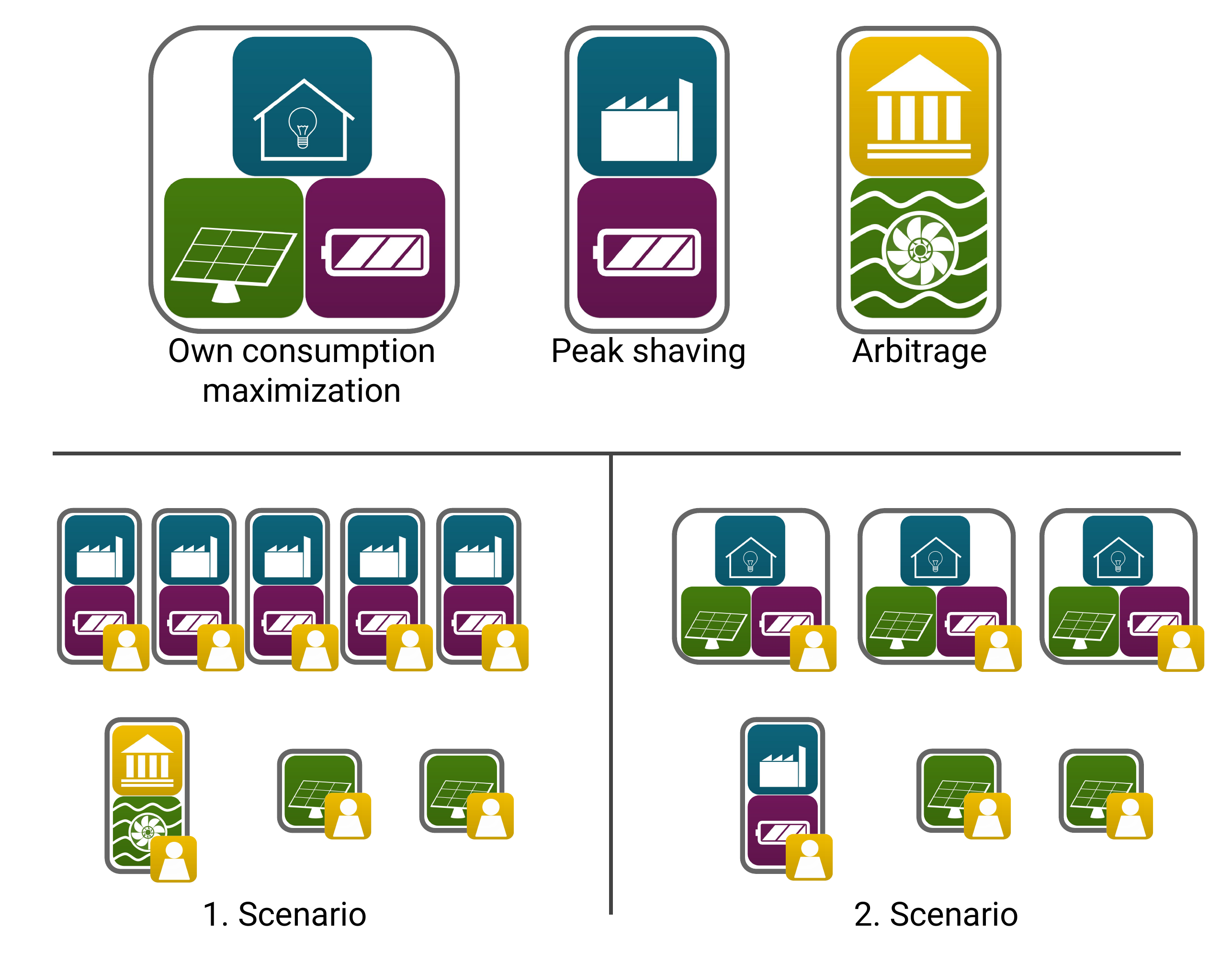}
  \caption{Evaluation Overview: The top depicts local test setups, including their objectives; the bottom shows the two evaluation scenarios in which the different agents with their local setup are displayed.}\label{fig:eval_overview}
\end{figure}

In this section, the system will be evaluated using two different scenarios as the local objectives may appear on different network levels. The first scenario represents a negotiation on the transmission level and the second on distribution level. An overview is shown in figure~\ref{fig:eval_overview}. Using GABHYME, each scenario will get executed twice, once with the normalization and once with the Pareto method (referred to by GABHYME-N and GABHYME-P).

Additionally, we evaluate two different baseline methods to show the effectiveness compared to typical schedule generation strategies. First, we apply an ordinary evolutionary algorithm instead of GABHYME. Second, we evaluate a random sampling approach with the raw COHDA heuristic. The EA uses standard techniques: a self-adaption controlled random mutation operator \cite{schwefel1981numerical} with basic solution repairing, no recombination due to the high risk of breaking the solution, and the NSGA-II selection operator \cite{deb2002fast}. So it mainly differs from GABHYME in terms of the mutation operator, the recombination operator, and the loss of the ability to restart the population. Two variants are applied, similar to GABHYME, the normalization, and the Pareto method. In the following, we refer to them as EA-N and EA-P. The sampling approach naively randomly generates $10\,000$ valid schedules, which COHDA will use to find a suitable CS. Consequently, that method does not use the full extension presented in section~\ref{subsec:extension}.

In the evaluation, we will consider the robustness and effectiveness of the approach. For robustness, the median, the average, standard deviation, and outliers are considered. The effectiveness will be measured with the fulfillment rate of the target OS. This rate can be calculated using $g_{\rm norm}$ (see equation (\ref{eq:normalization})). Regarding the quality of the solutions, we will calculate a relative measure, referred to as local quality (LQ):
\begin{equation}
    LQ_{\rm m,\agentindex} = \frac{f_{\rm local}(G_{a,m})}{\underset{\rm x}{max}(f_{\rm local}(G_{a,x}))}
\end{equation}
$LQ_{\rm m,\agentindex}$ is the relative local solution quality of a single method $m$ for an agent $a$. The average of this metric over all agents is defined as $LQ_m$ or short LQ-value.
\subsection{Local test setup}
\begin{figure*}[htbp]
  \centering
  \begin{subfigure}{0.29\textwidth}
    \includegraphics[width=1\textwidth]{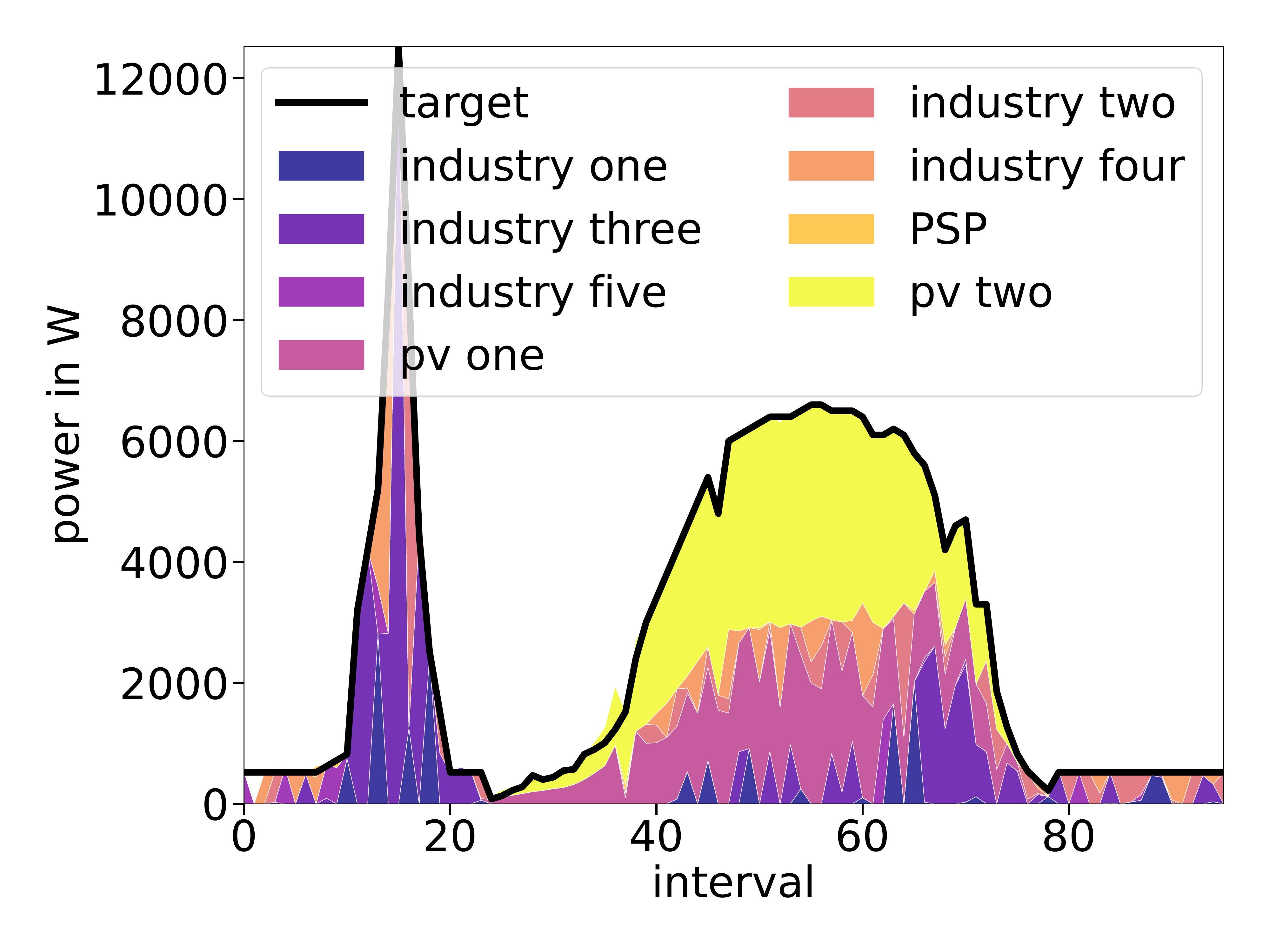}
    \subcaption{Cluster schedule: GABHYME-N}\label{subfig:evo_norm_large_cs}
  \end{subfigure}
  \begin{subfigure}{0.29\textwidth}
    \includegraphics[width=1\textwidth]{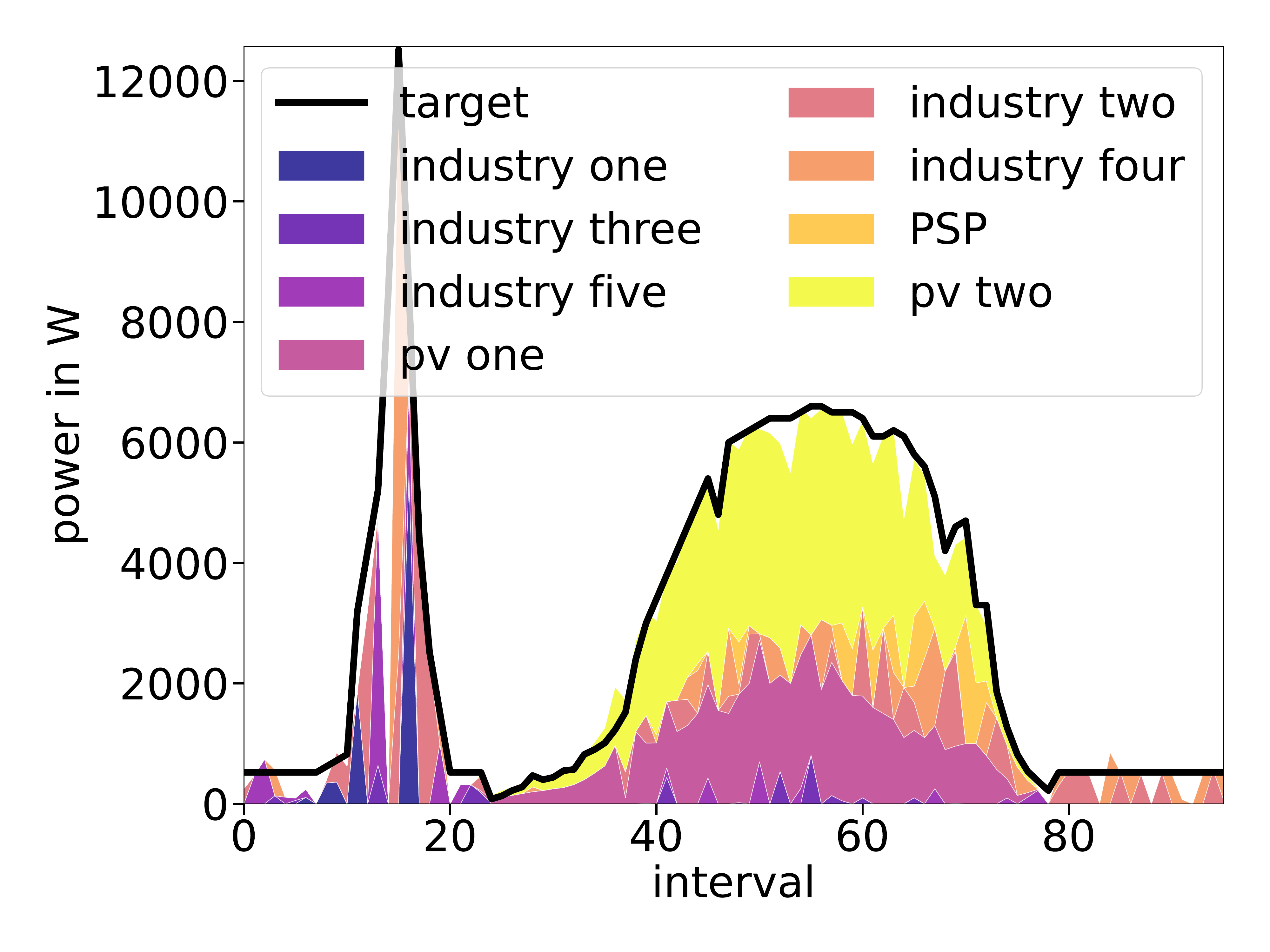}
    \subcaption{Cluster schedule: GABHYME-P}\label{subfig:evo_pareto_large_cs}
  \end{subfigure}
  \begin{subfigure}{0.40\textwidth}
    \includegraphics[width=1\textwidth]{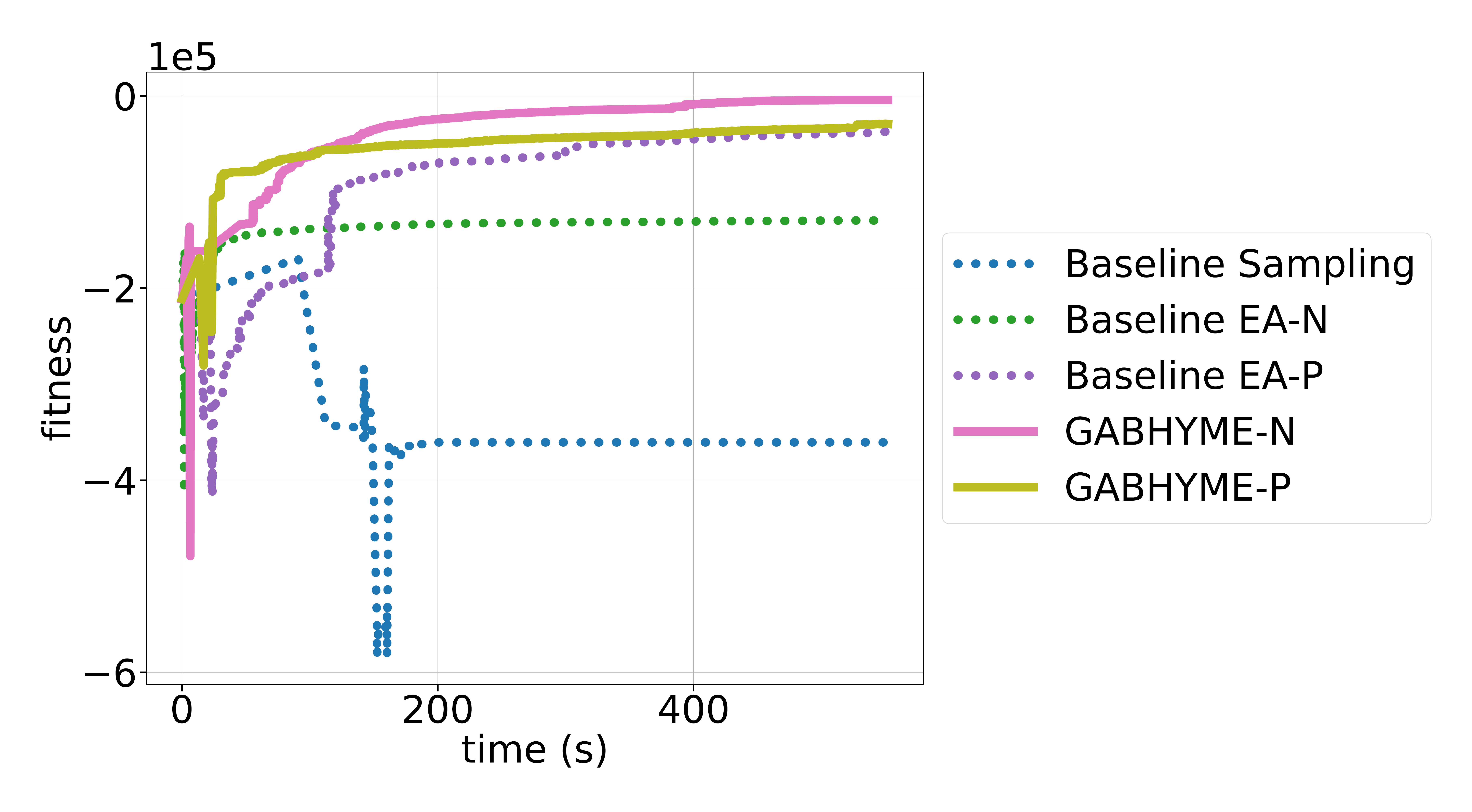}
    \subcaption{System performance}\label{subfig:large_fitness_system} 
  \end{subfigure}
  \begin{subfigure}{0.38\textwidth}
    \includegraphics[width=1\textwidth]{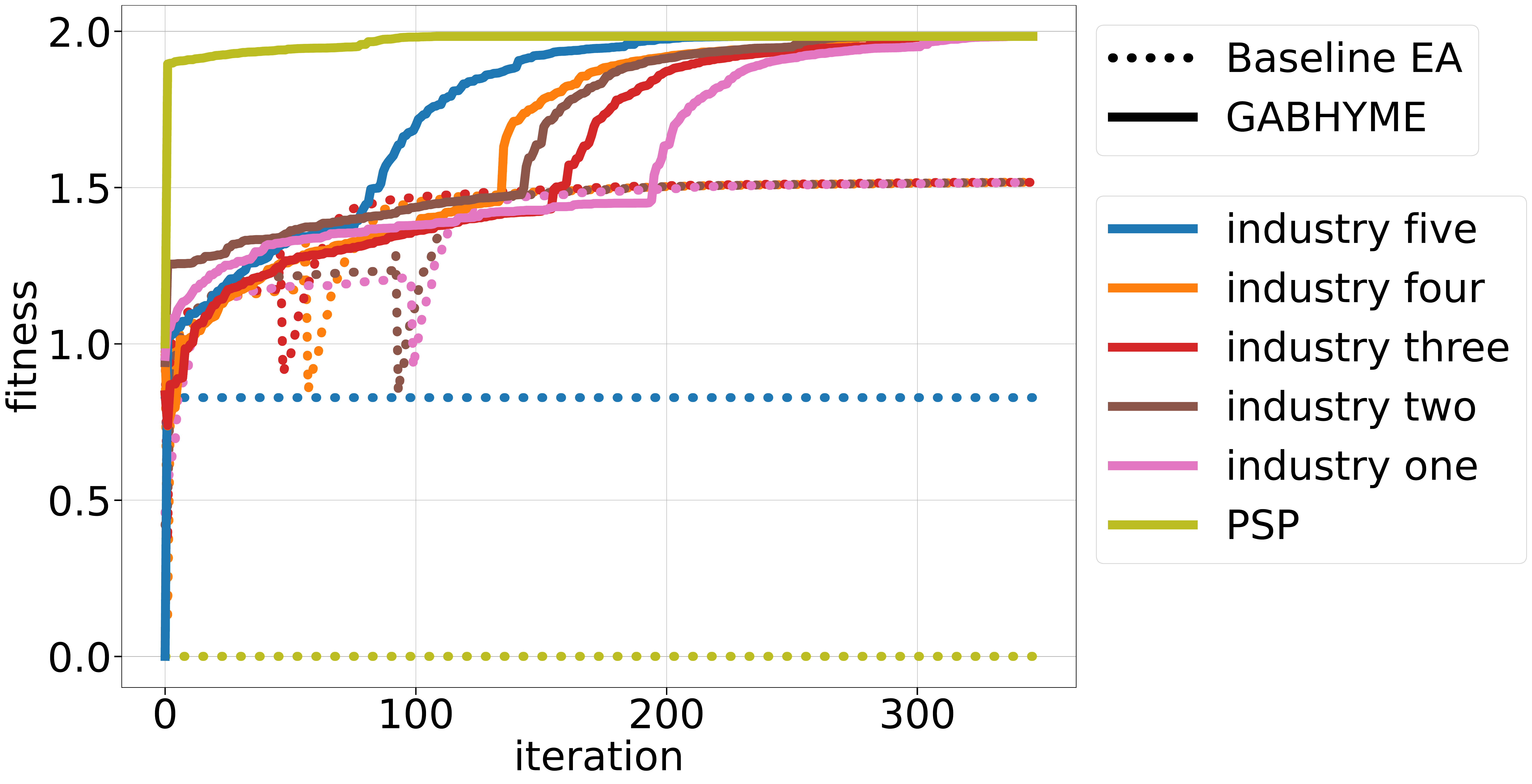}
    \subcaption{Fitness history of the normalization methods}\label{subfig:large_fitness_individuals} 
  \end{subfigure}
  \begin{subfigure}{0.31\textwidth}
    \includegraphics[width=1\textwidth]{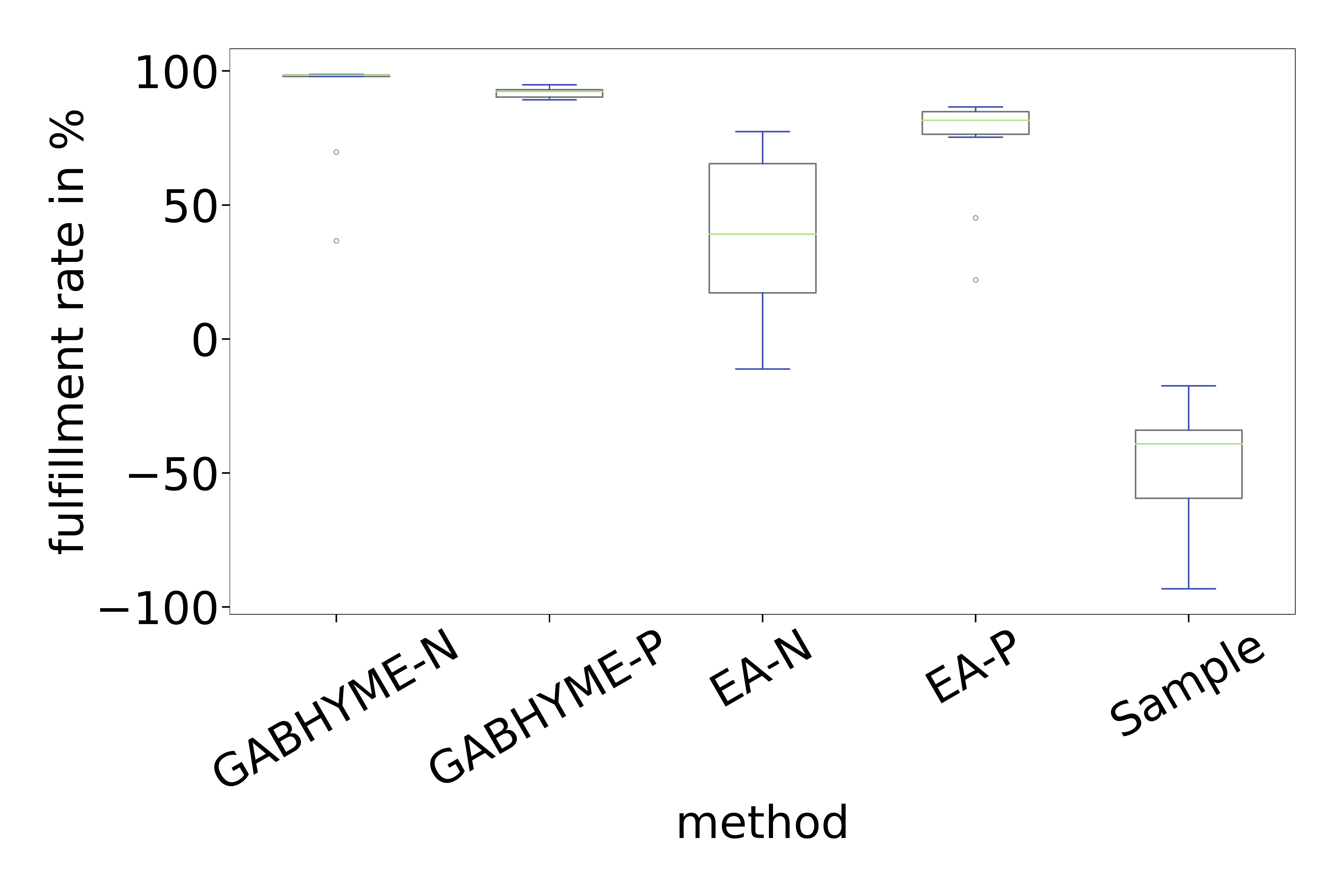}
    \subcaption{Robustness}\label{subfig:robust_cohda_large} 
  \end{subfigure}
  \caption{Overall result -- 1st Scenario}\label{fig:first_scenario_overall}
\end{figure*}
The objective functions for \textit{peak shaving}, \textit{local SDM} and \textit{arbitrage} imply three different local setups. The data is based on real data from households, industries, and the energy market. The storage model will be parameterized using specifications of real-world storage. The planning horizon is $96$ using $15$ minute intervals. The storage parameters are shown in table \ref{tab:all_storages}.
\subsubsection{\textit{Arbitrage}}
This test case will use the following technical data, based on the PSP Kirchentellinsfurt from table~\ref{tab:all_storages}.
Furthermore, the interval 30.03.2018 12:00 AM --- 31.03.2018 12:00 AM in the region DE/AT/LU is chosen as market data \cite{smard2019market}.
\subsubsection{\textit{Peak shaving}}
For \textit{peak shaving} a price of $16.49$ Cent/kWh is assumed (average price for industries in 2018 with an electrical consumption of $2000$ --- $20.000$ MWh \cite{eurostat2019preise}.
Two load profiles will be used, one based on a small industry and the other on big industry. The data is part of the data set used by Tiemann et al. \cite{tiemann2020electrical}. 
\subsubsection{\textit{Local SDM}} 
For \textit{local SDM}, a storage with a coupled solar power plant is assumed as part of a household. As consumer electricity price, the average 29.47 Cent/kWh of 2018 is used. The storage parameters are based on a technical product sheet of the RUSOL GmbH \cite{rusol2019energy}.
A profile of the solar power plant in Kronberg is used \cite{brod2018kronberg}. The household profile used is part of the data set of Tjaden et al. \cite{tjaden2015reprasentative}.
\subsection{Scenarios}
The extension will be evaluated using two scenarios, as presented in figure \ref{fig:eval_overview}. The evolutionary methods will use the parameters listed in table \ref{tab:alg_param}.

The first scenario consists of eight agents, five representing big industries applying a battery and the objective \textit{peak shaving}. Furthermore, there is a PSP agent and two agents controlling solar power plants. The local test cases were described above. Each industry uses a different load profile. The capacity of the PSP will be scaled down to $10\%$. As a communication topology, a complete graph is used. The target OS $\targetvar$ is depicted in figure \ref{fig:first_scenario_overall}\subref{subfig:evo_norm_large_cs}.

The second scenario consists of six agents: one representing a small industry applying a battery and the objective \textit{peak shaving}, three more controlling batteries coupled to solar power plants, and households. The last two are predefined generation profiles of SPPs. The local test cases were described above. Furthermore, a complete graph is used as a topology. The target OS is depicted in figure \ref{fig:second_scenario_overall}\subref{subfig:evo_norm_small_cs}.

Besides the single solution quality of the negotiation, it is crucial to determine the robustness of the approach because of its large basis on randomness. To accomplish this, the simulation scenarios have been executed ten times.
\subsection{Results}
\begin{figure*}[htbp]
  \centering
  \begin{subfigure}{0.29\textwidth}
    \includegraphics[width=1\textwidth]{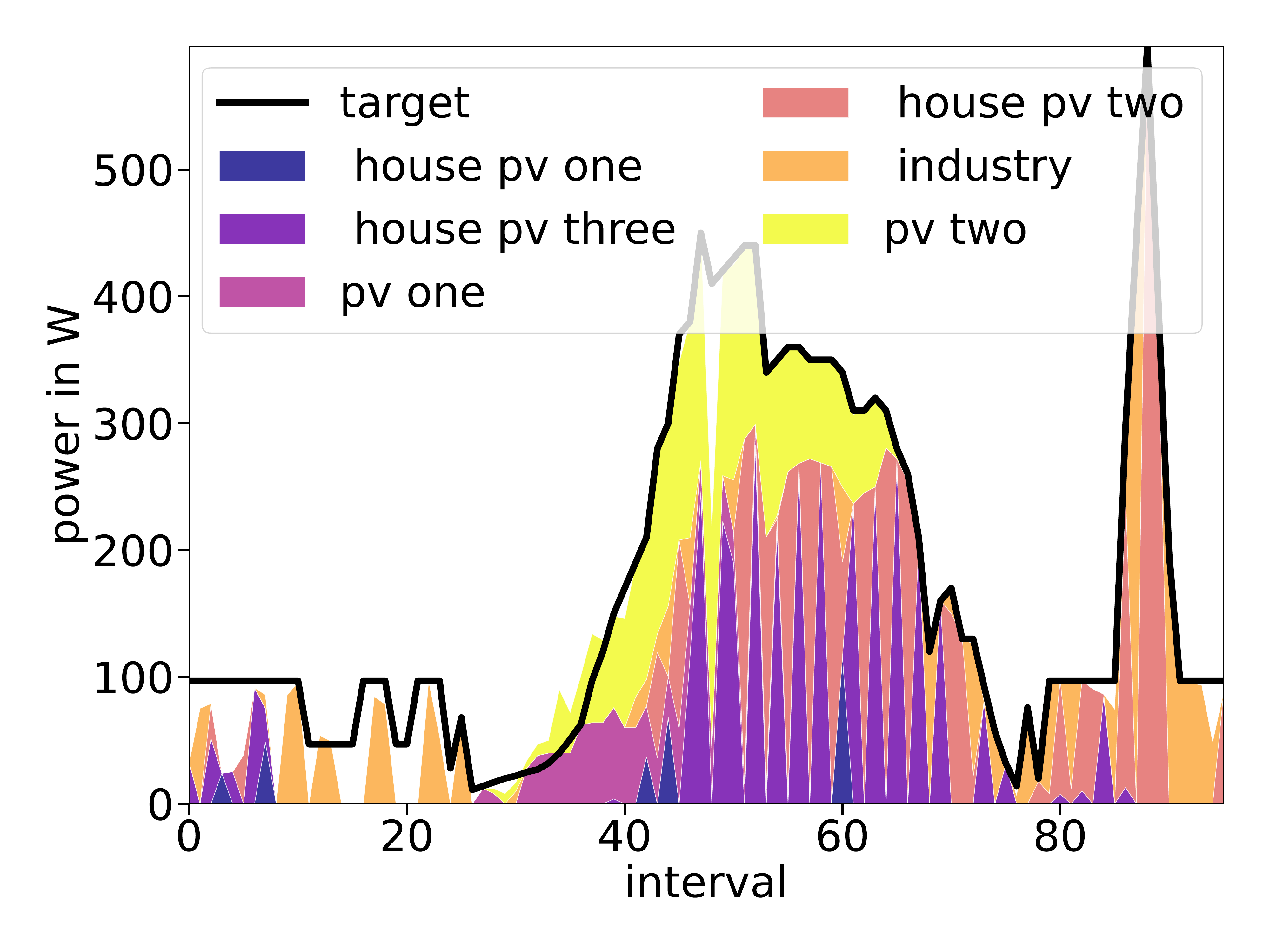}
    \subcaption{Cluster schedule GABHYME-N}\label{subfig:evo_norm_small_cs}
  \end{subfigure}
  \begin{subfigure}{0.29\textwidth}
    \includegraphics[width=1\textwidth]{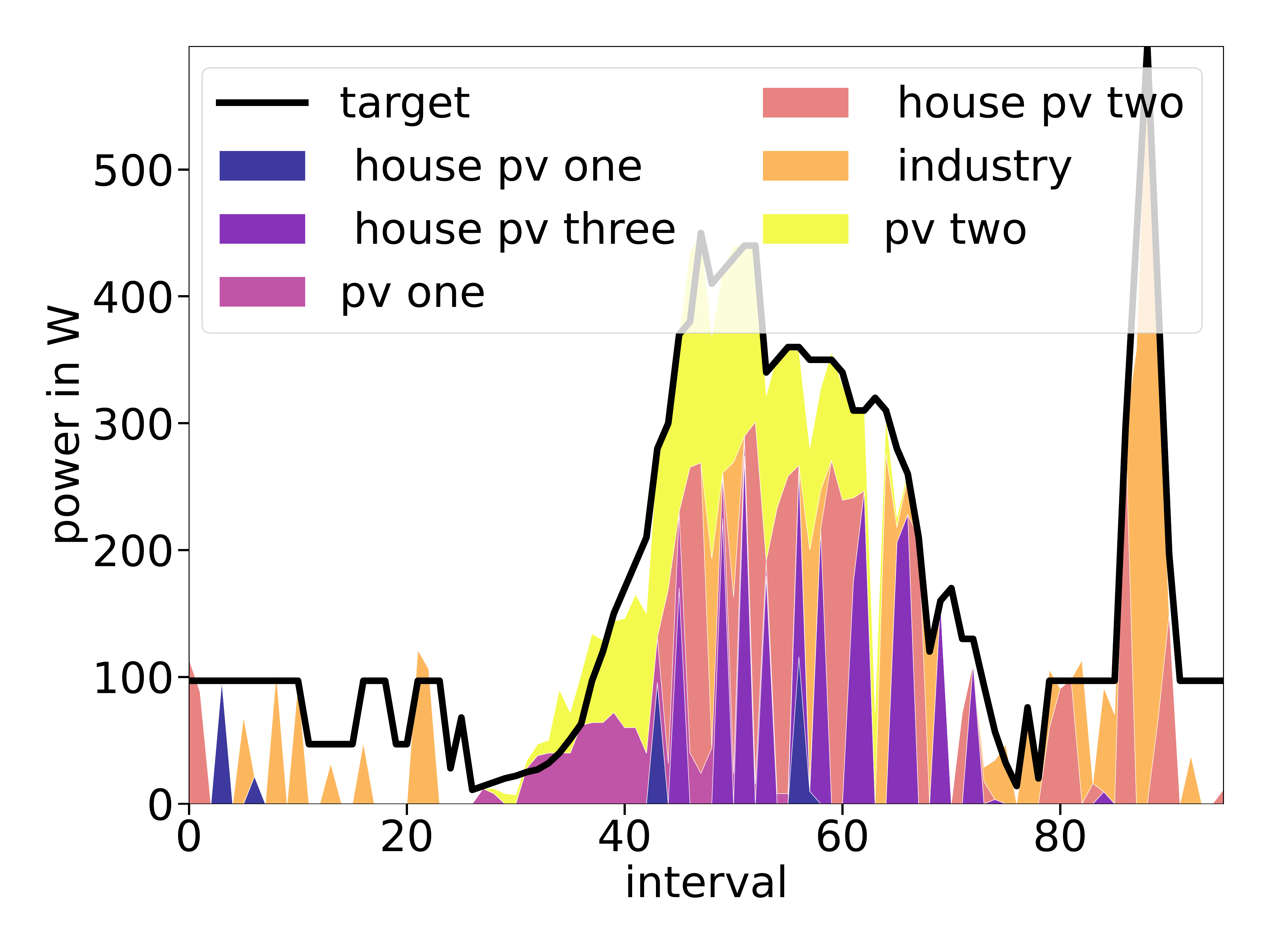}
    \subcaption{Cluster schedule GABHYME-P}\label{subfig:evo_pareto_small_cs}
  \end{subfigure}
  \begin{subfigure}{0.40\textwidth}
    \includegraphics[width=1\textwidth]{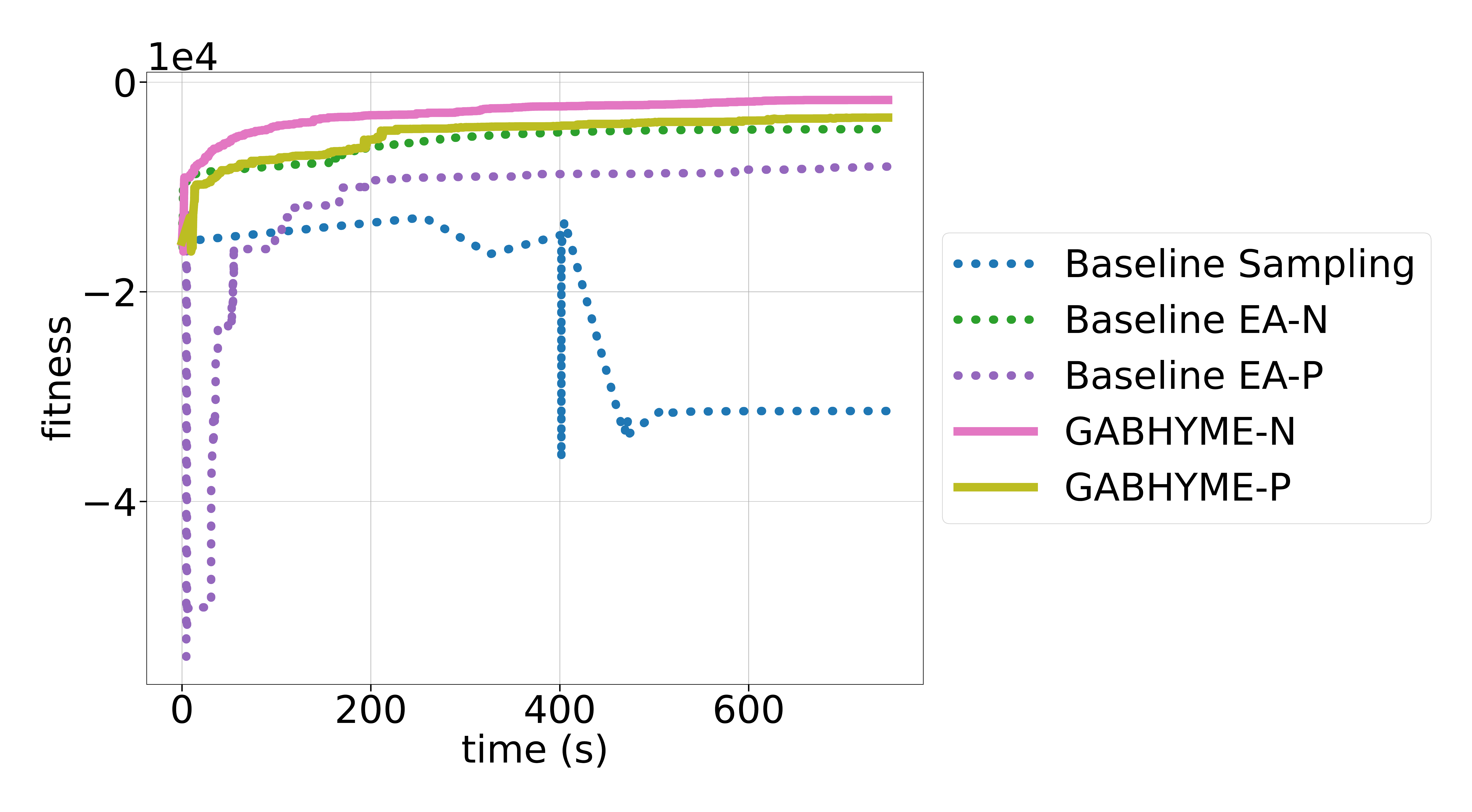}
    \subcaption{System performance}\label{subfig:small_fitness_system} 
  \end{subfigure}
  \begin{subfigure}{0.38\textwidth}
    \includegraphics[width=1\textwidth]{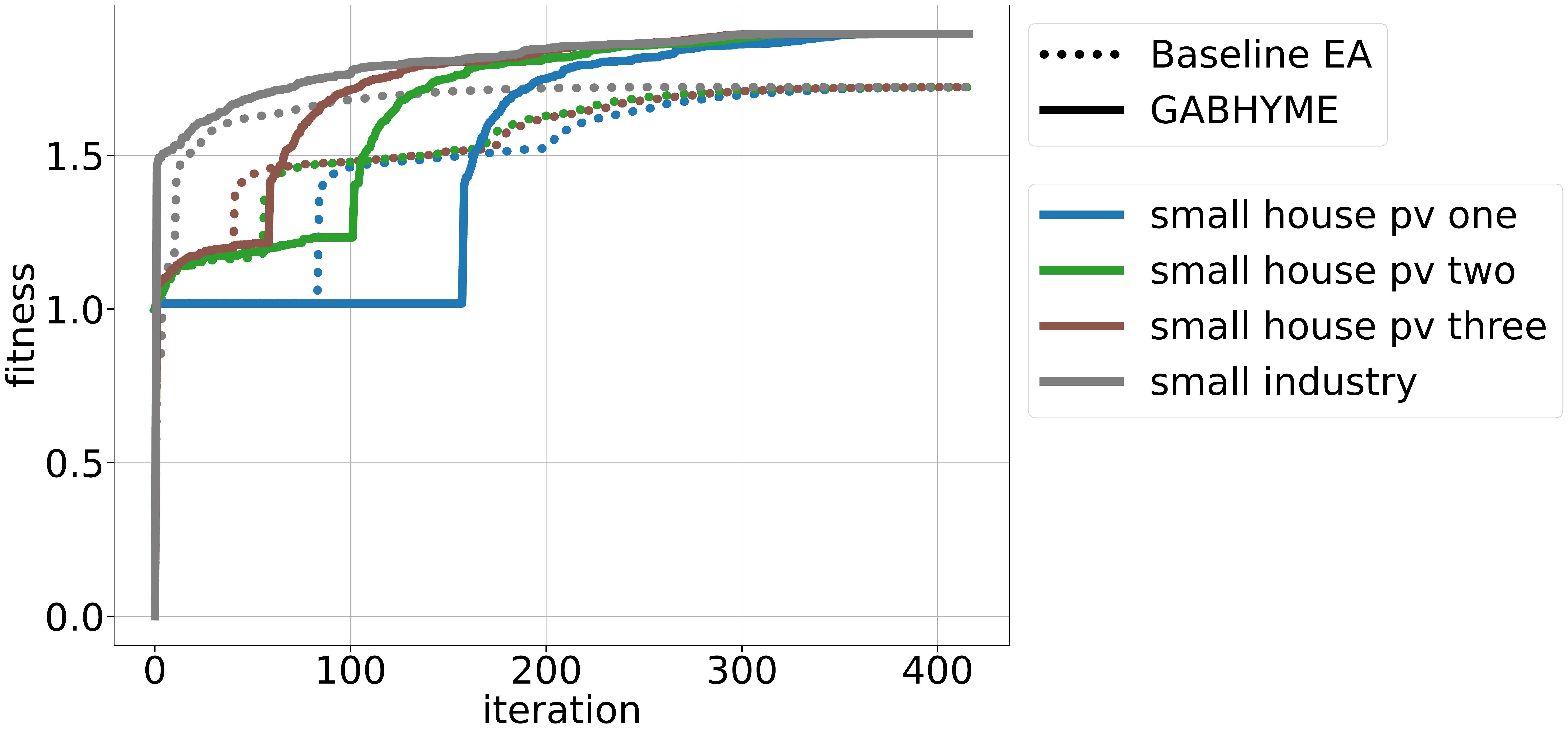}
    \subcaption{Fitness history of the normalization methods}\label{subfig:small_fitness_individuals} 
  \end{subfigure}
  \begin{subfigure}{0.31\textwidth}
    \includegraphics[width=1\textwidth]{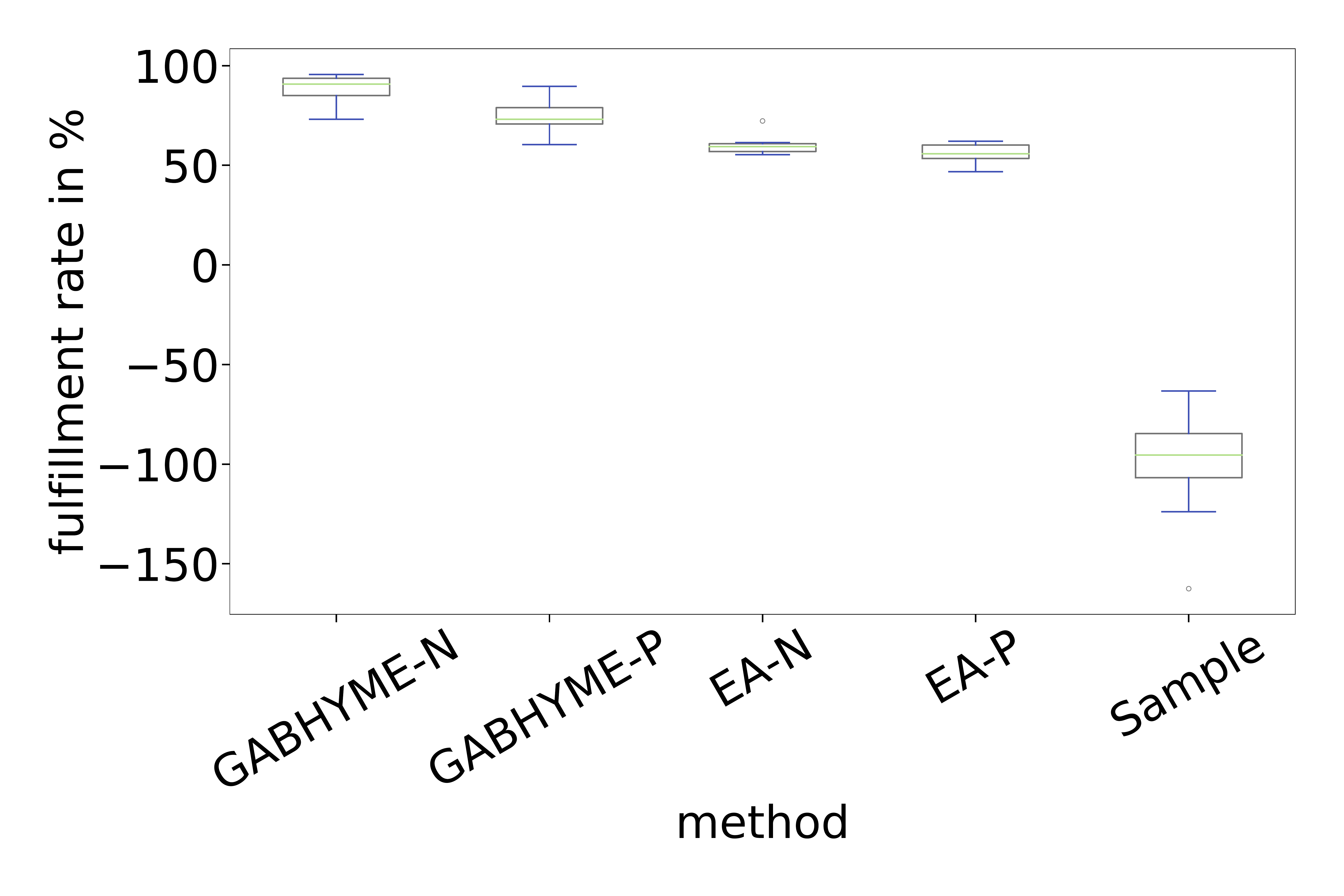}
    \subcaption{Robustness}\label{subfig:robust_cohda_small} 
  \end{subfigure}
  \caption{Overall result -- 2nd Scenario}\label{fig:second_scenario_overall}
\end{figure*}

The numerical results are shown in table \ref{tab:all_num_results}, the corresponding boxplots are depicted in figures \ref{fig:first_scenario_overall}\subref{subfig:robust_cohda_large} and \ref{fig:second_scenario_overall}\subref{subfig:robust_cohda_small}. Besides, there are several figures showing representative single runs of the evaluated methods (figures \ref{fig:first_scenario_overall}a-d and \ref{fig:second_scenario_overall}a-d). Hereby, a and b show the cluster schedules resulting from GABHYME, and c and d are the fitness histories of the global objective, respectively, the normalized combined local and global fitness. The local solutions are shown in the appendix (figure \ref{fig:first_norm_solutions}, \ref{fig:first_industry_one_to_four_pareto}, \ref{fig:second_solutions_norm}, and \ref{fig:second_solutions_pareto}).
\paragraph{1st Scenario}
At first look, the target schedule was fulfilled by GABHYME pretty closely, especially GABHYME-N has a high median compared to the other algorithms. Further, considering the evolutionary methods, the histories of COHDA's objective function (see figure \ref{fig:first_scenario_overall}\subref{subfig:large_fitness_system}) have a typical logarithmic shape, except for the first 50 seconds, where several performance jumps happen. These can also be seen in \ref{fig:first_scenario_overall}\subref{subfig:large_fitness_individuals}, especially with the industry agents. The reason for this could be a slow reaction to the flexibility of the other agents due to the partly incomplete knowledge about them. Also, it is often a sign of escaping a local extreme point, especially considering the cluttered solution space. Compared to the baseline, it is clear that GABHYME dominates in terms of the defined metrics. The EA-N stagnates quickly, mainly caused by two agents who cannot find suitable schedules within their environment. This is also not a one-time effect, EA-N is consistently unstable, as shown in \ref{fig:first_scenario_overall}\subref{subfig:robust_cohda_large}. EA-P shows a way better and more robust result but is also slightly off regarding median and average. Moreover, as repeatedly discussed, the sampling approach is insufficient for energy storage, most likely due to the large solution space.

For GABHYME-P, the chosen solutions and their Pareto fronts are shown in figure \ref{fig:second_solutions_pareto}. Looking at the LQ-values, it generates the best local solutions among the methods while obtaining a competitive fulfillment rate.

While GABHYME-N has two severe outliers and therefore is less robust than most of the other methods, the biggest problem with GABHYME-P is the precision. A reason can be an insufficient mutation control or too few iterations, respectively a too aggressive stopping criterium. The fitness curve also supports the latter point.
\paragraph{2nd Scenario}
The CS of GABHYME-N largely matches the target OS, starting at interval 20. However, the prior intervals are much harder to fulfill because the batteries have to coordinate their initial energy (10\% load). In this scenario, as seen in the first scenario, a similar difference exists between the baseline methods and GABHYME regarding the fulfillment rate. Nevertheless, in this scenario, EA-N is more robust. Consequently, the difference is much less severe.

For GABHYME-N, optimization success started very late when looking at the solution fitness histories. One reason is that COHDA needs some ramp-up time before everybody knows each other. Another reason might be that the single local objectives in this scenario are relatively easy to maximize, which leads to high fitness pre-optimization solutions, making it hard to find feasible solutions in multi-criteria optimization. The baseline EA methods show similar but less severe behavior.

The target OS of GABHYME-P was fulfilled with an average rate of around 74\%. This is a low rate compared to the normalization approach. There are many slots where the target cannot be reached, especially since the first and last 20 slots are far from the target. As the generators are mostly SPPs, it makes sense that it is especially problematic to manage the slots where no generation is achieved. However, GABHYME-P shows a clear improvement to the baseline EA-P. The sampling approach proves again not to be feasible.
\paragraph{Robustness}
The result of the first test case with normalization is robust, with minimal fluctuations regarding the fulfillment rate. However, two outliers are massively underperforming (see figure \ref{fig:first_scenario_overall}\subref{subfig:robust_cohda_large}) The Pareto variant has a bit more fluctuation but no outliers. Excluding outliers, the robustness of GABHYME in the second test case (see figure \ref{fig:second_scenario_overall}\subref{subfig:robust_cohda_small}) is way worse with respect to the solution distribution. However, the standard deviation of GABHYME and the baseline EA is lower overall. The Pareto approach is, by all means, less robust.

From what we could observe, the available flexibility and the number of generators mainly (and therefore agents) influence the robustness because higher flexibility and more agents make it easier to reach the target values without losing too much local value from a single agent's perspective. This means, the robustness of the approach would highly depend on the scenario and how to handle multiple objectives. However, this is just a hypothesis from the two scenarios, and further work is needed to analyze and justify this more thoroughly.
%
%
%
\begin{table*}
\caption{Median $\tilde{x}$, average $\overline{x}$ standard deviation $\sigma$ of the fulfillment rates, and the relative local quality (LQ) of the methods EA, GABHYME, and Sampling in the scenarios 1. and 2.}\label{tab:all_num_results}
\centering
  \begin{tabular}{cx{2cm}cccc}
  \toprule 

  scenario & method & $\overline{x}$ & $\tilde{x}$ & $\sigma$ & LQ \\

  \midrule

  \multirow{5}{*}{1.} & GABHYME-N & 0.8936 & \textbf{0.9842} & 0.1955 & 0.4385 \\
  & GABHYME-P & \textbf{0.9192} & 0.9237 & \textbf{0.0171} & \textbf{0.9361} \\
  & EA-N & 0.3850 & 0.3910 & 0.2896 & 0.4430 \\
  & EA-P & 0.7267 & 0.8160 & 0.2047 & 0.7452 \\
  & Sampling & -0.4683 & -0.3915 & 0.2123 & -0.3123 \\
  \midrule
  \multirow{5}{*}{2.} & GABHYME-N & \textbf{0.8776} & \textbf{0.9070} & 0.0762 & 0.4865 \\
  & GABHYME-P & 0.7450 & 0.7312 & 0.0801 & \textbf{0.7860} \\
  & EA-N & 0.5994 & 0.5936 & \textbf{0.0458} & 0.3865 \\
  & EA-P & 0.5858 & 0.5576 & 0.0468 & 0.7159 \\
  & Sampling & -0.9990 & -0.9547 & 0.2640 & 0.1170 \\

  \bottomrule

\end{tabular}
\end{table*}
\section{Conclusion}
In this work, an extension for distributed schedule optimization has been proposed, which is capable of integrating energy storage. A multi-criteria algorithm has been developed, whose responsibility is to preselect high-fitness OS for storage. Besides the objective of the chosen basic heuristic COHDA, three different local objectives were tested in this work: peak shaving, local SDM, and arbitrage. Two approaches to integrating the local objective have been presented and evaluated in a comprehensive system test. In this test, the robustness and effectiveness of the approaches have been evaluated. Due to the reduction of the solution space and the number of possible OS, the extension is capable of integrating storage in the energy scheduling process with distributed schedule optimization algorithms like COHDA. The evaluation shows that the approach leads to solutions with high fitness values concerning the global objective and at least profitable solutions regarding the defined model from a local perspective. Furthermore, the approach preserves all relevant characteristics of COHDA, like the termination and convergence \cite{hinrichs2017distributed}.  

On the problem side, it should be noted that the solutions are locally far from the optimal solution concerning fitness. One problem of the algorithm is the scaling regarding the planning horizon, which is due to the need to choose a strategy, not optimal. It is an option to switch the algorithmic approach completely, for example, by switching the metaheuristic, combining, or even replacing it with other heuristics. Besides the EA, the GWO \cite{mirjalili2014grey} and the FireFly \cite{yang2009firefly} metaheuristics were applied to the presented scenarios. However, both got dominated by the EA pretty heavily regarding the solution quality. So they were excluded from this paper. In conclusion, one general problem regarding metaheuristics is the lack of formal performance guarantees. To solve this, metaheuristic algorithms would have to be omitted entirely, leading to a loss of flexibility in terms of defining the fitness function and the necessity to determine some suitable weights for different objectives.
\section{Future Work}
Many scenarios with energy storage could be investigated to check the general approach. For example, demand-side management could be seen as a household with energy storage and different local constraints.

Besides defining more, mainly financial objectives, it will also be necessary to include ecological objectives or at least some cost for disadvantageous ecological OS.

The most important thing to do is to improve the algorithms for solving the multi-criteria problem. In general, the critical part is the mutation itself. In this work, an approach has been used, which is meant to scale with the planning horizon, but also should be feasible for exploitation. The main problem is the balance between exploration and exploitation operators. Therefore, we implemented a random choice, which might be improvable.

Lastly, we did not consider system scalability in our evaluation, as we did not determine this as a major issue of the extension itself. However, in future work, it has to be shown that the approach scales as expected.

\normalsize
\bibliography{main}
\appendix
\section{Appendix}
\label{ch:appendix}
\FloatBarrier
\begin{figure*}[b]
  \centering
  \begin{subfigure}{0.24\textwidth}
      \centering
      \includegraphics[width=1\columnwidth]{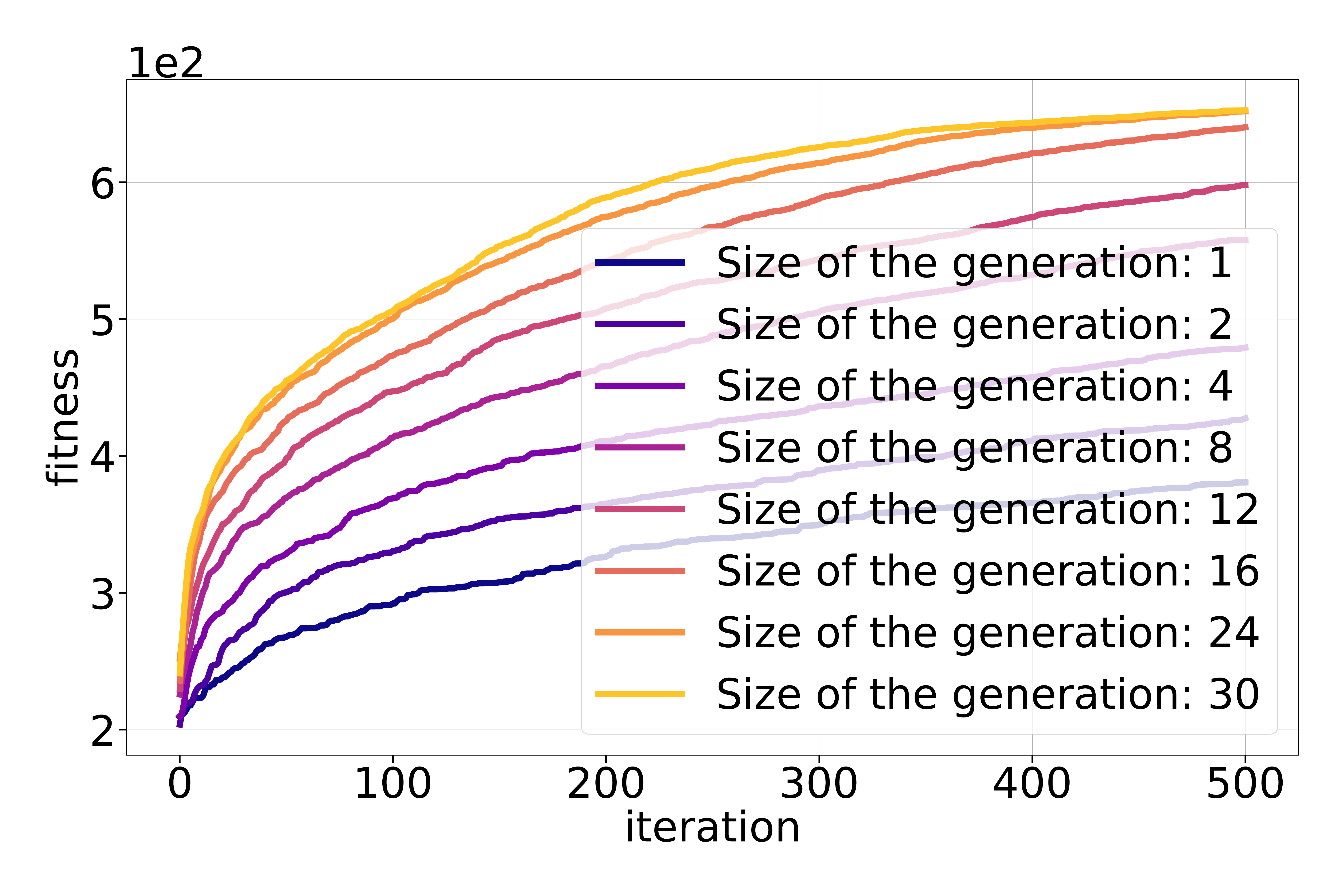}
      \caption{Peak shaving -- fitness}
      \label{subfig:evo_param_gensize_pc_fit}
  \end{subfigure}%
  \begin{subfigure}{0.24\textwidth}
      \centering
      \includegraphics[width=1\columnwidth]{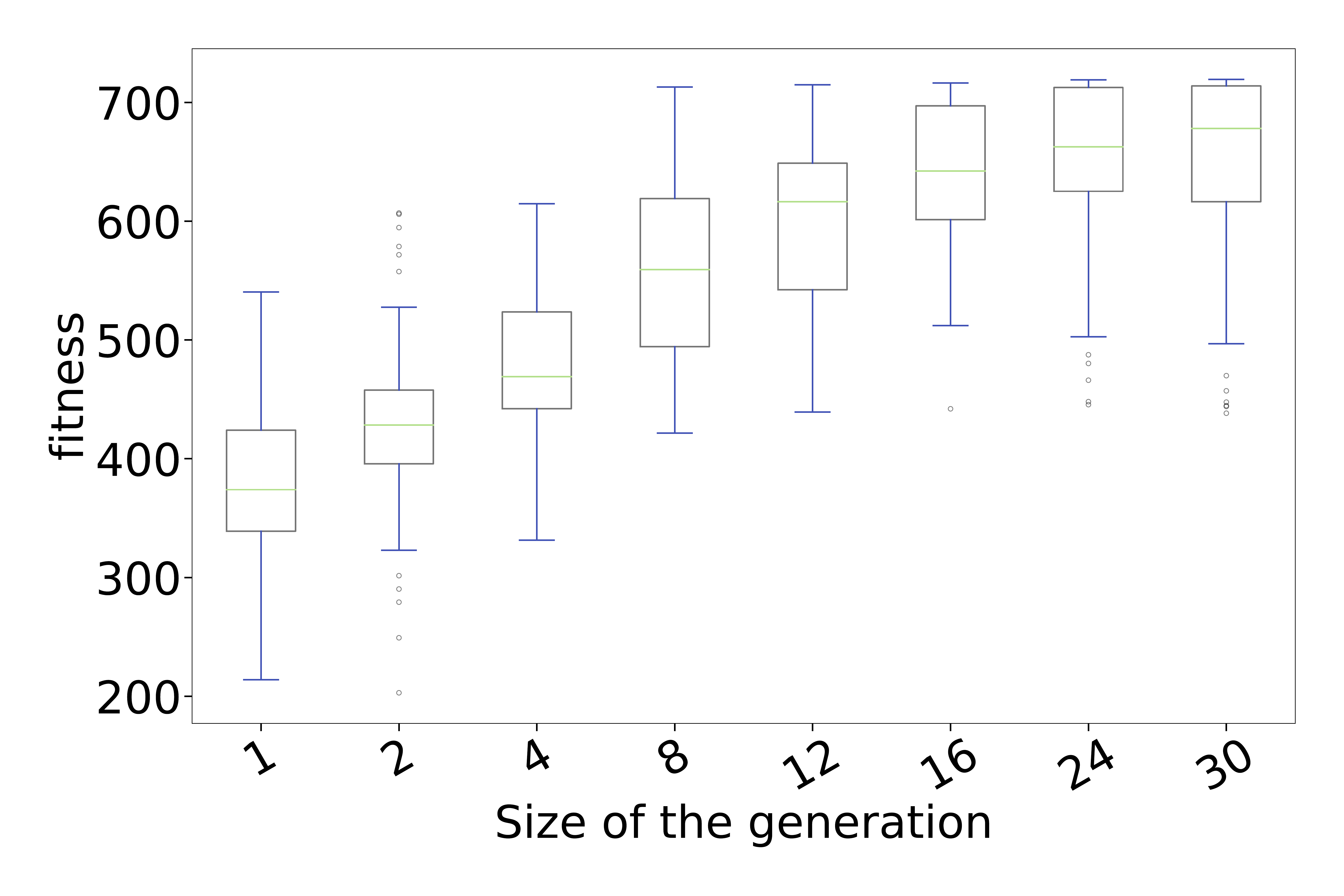}
      \caption{Peak shaving -- robustness}
      \label{subfig:evo_param_gensize_pc_rob}
  \end{subfigure}
  \begin{subfigure}{0.24\textwidth}
      \centering
      \includegraphics[width=1\columnwidth]{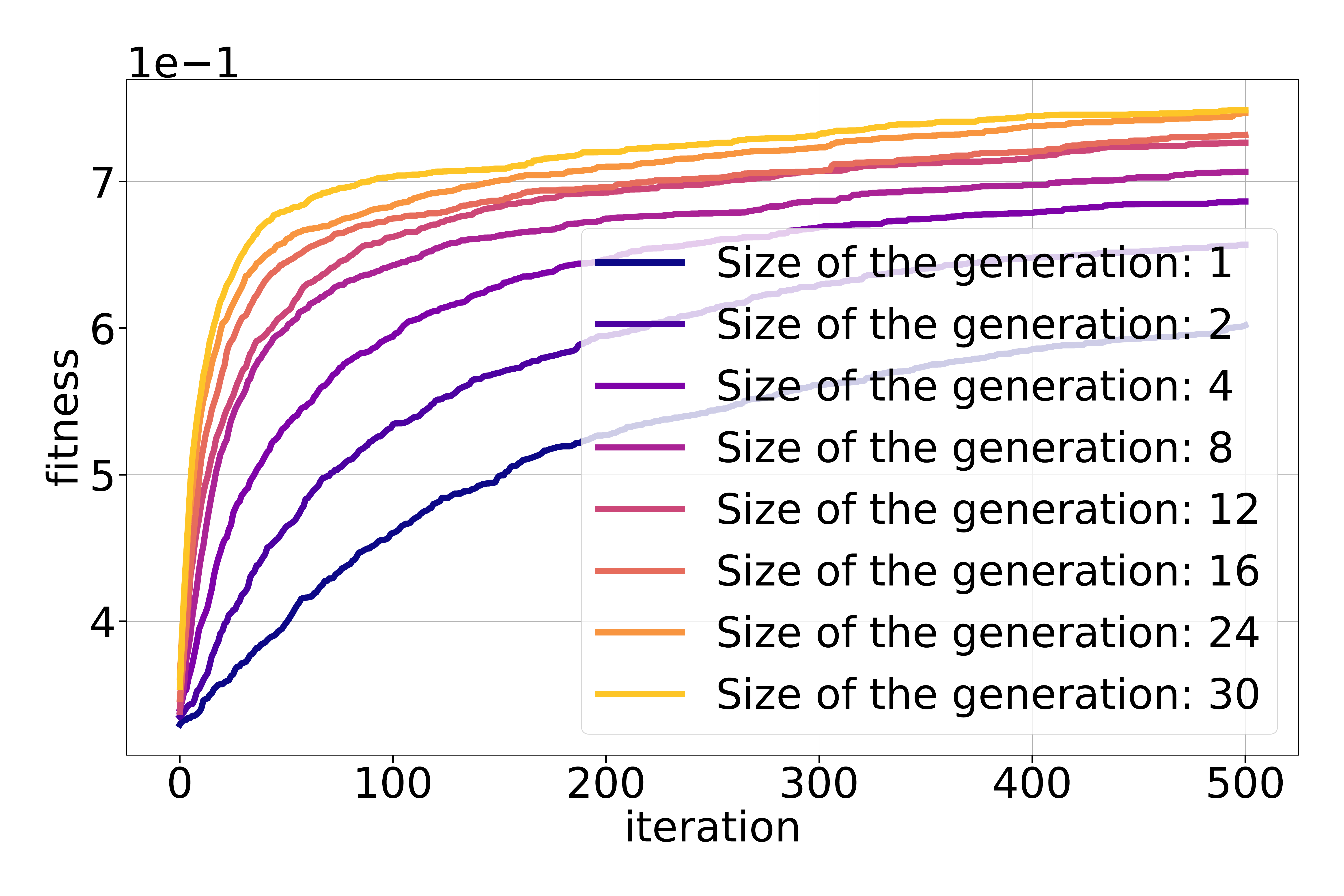}
      \caption{Local SDM -- fitness}
      \label{subfig:evo_param_gensize_oc_fit}
  \end{subfigure}%
  \begin{subfigure}{0.24\textwidth}
      \centering
      \includegraphics[width=1\columnwidth]{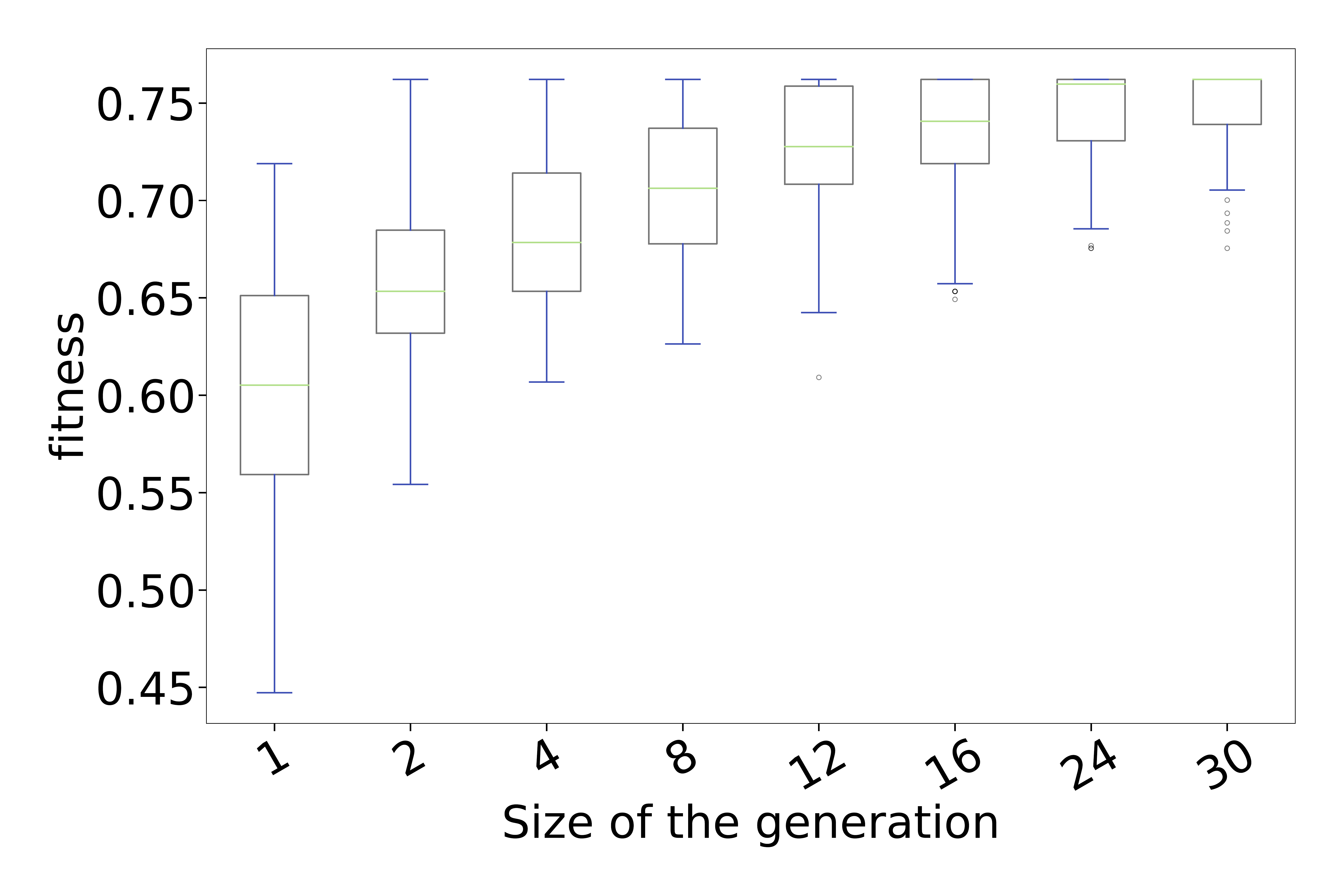}
      \caption{Local SDM -- robustness}
      \label{subfig:evo_param_gensize_oc_rob}
  \end{subfigure}
    \begin{subfigure}{0.24\textwidth}
        \centering
        \includegraphics[width=1\columnwidth]{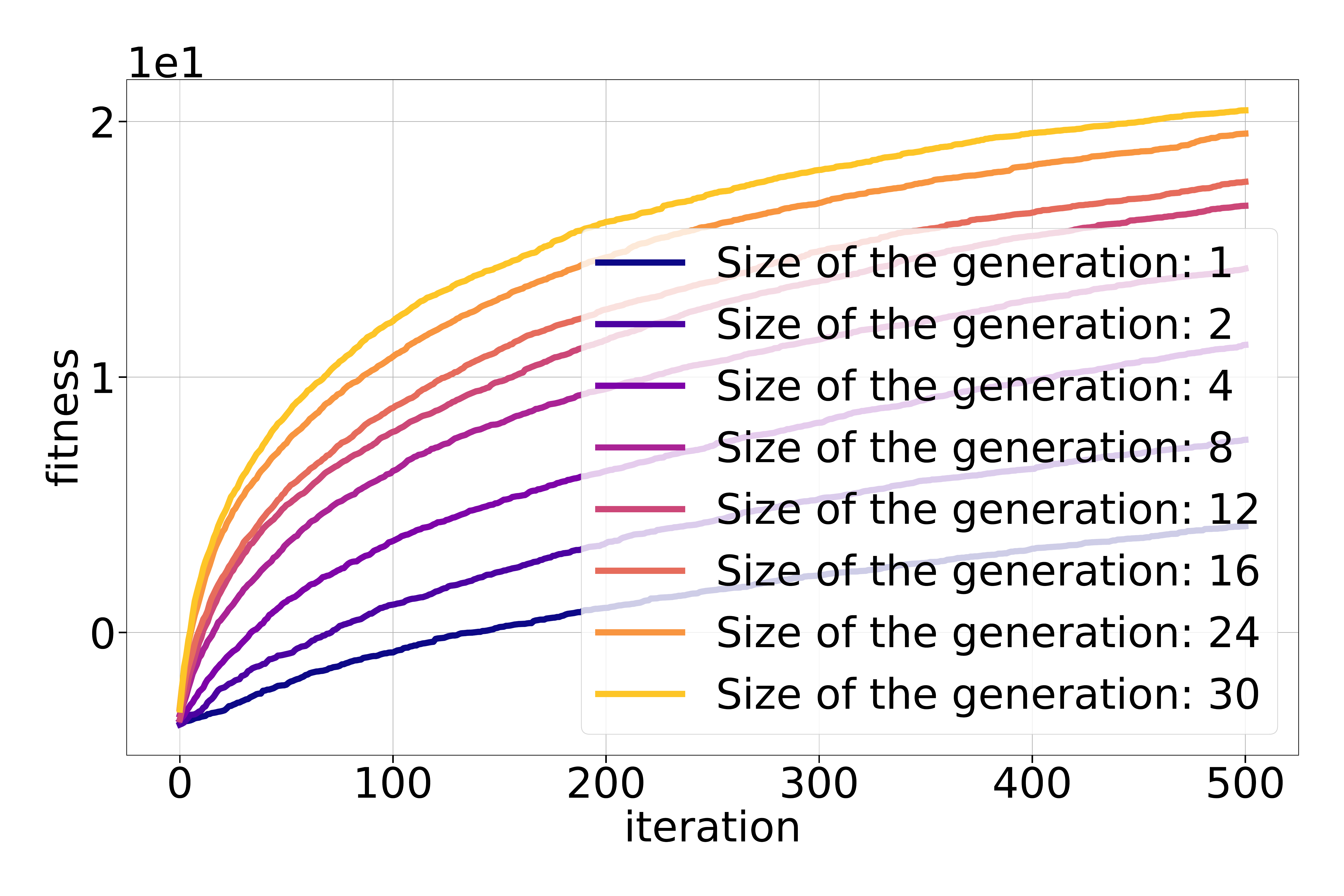}
        \caption{Arbitrage -- fitness}
        \label{subfig:evo_param_gensize_arbitrage_fit}
    \end{subfigure}%
    \begin{subfigure}{0.24\textwidth}
        \centering
        \includegraphics[width=1\columnwidth]{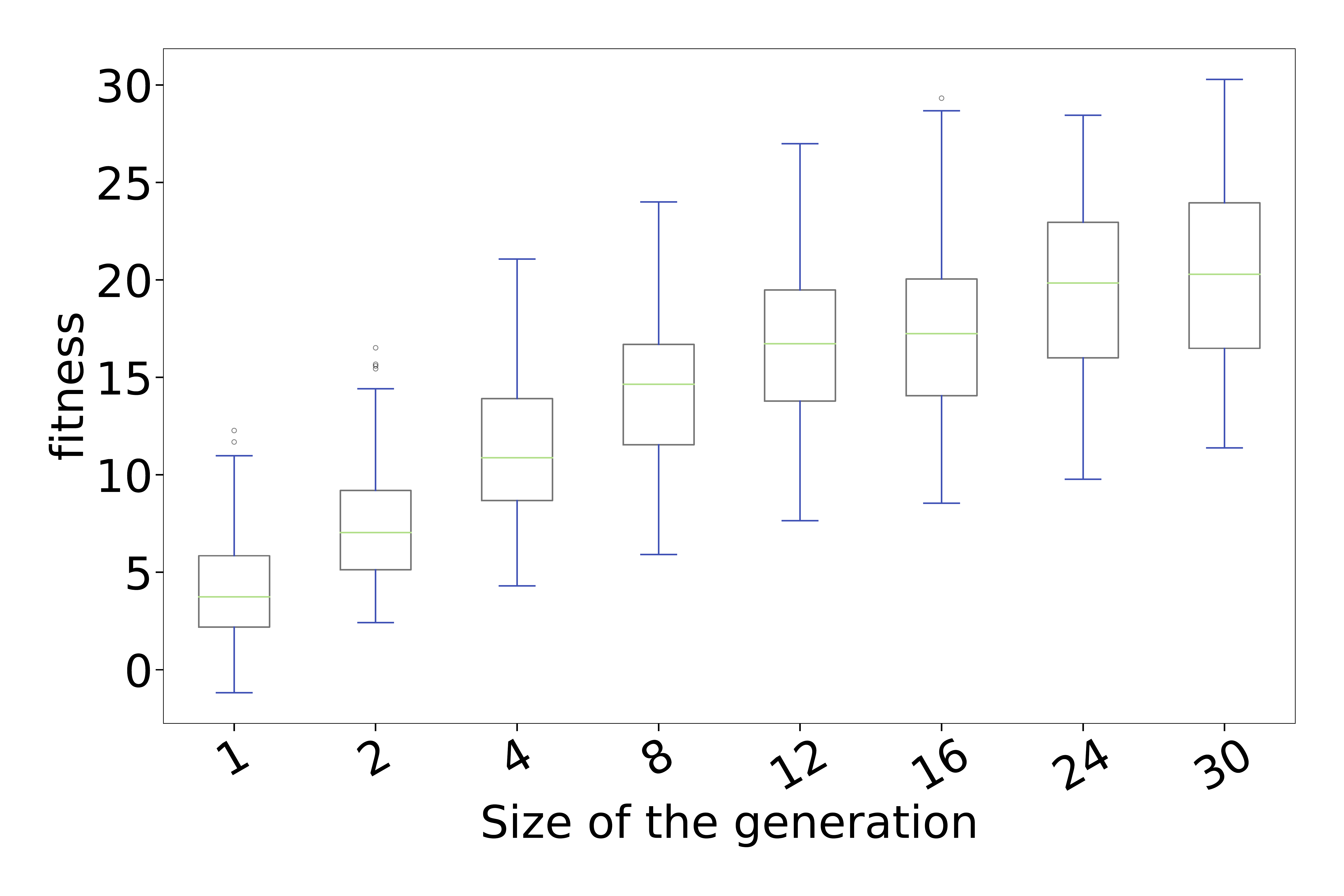}
        \caption{Arbitrage -- robustness}
        \label{subfig:evo_param_gensize_arbitrage_rob}
    \end{subfigure}
  \caption{Size of the generation}
  \label{fig:evo_param_gensize}
\end{figure*}
\begin{figure*}[b]
    \centering
    \begin{subfigure}{0.24\textwidth}
        \centering
        \includegraphics[width=1\textwidth]{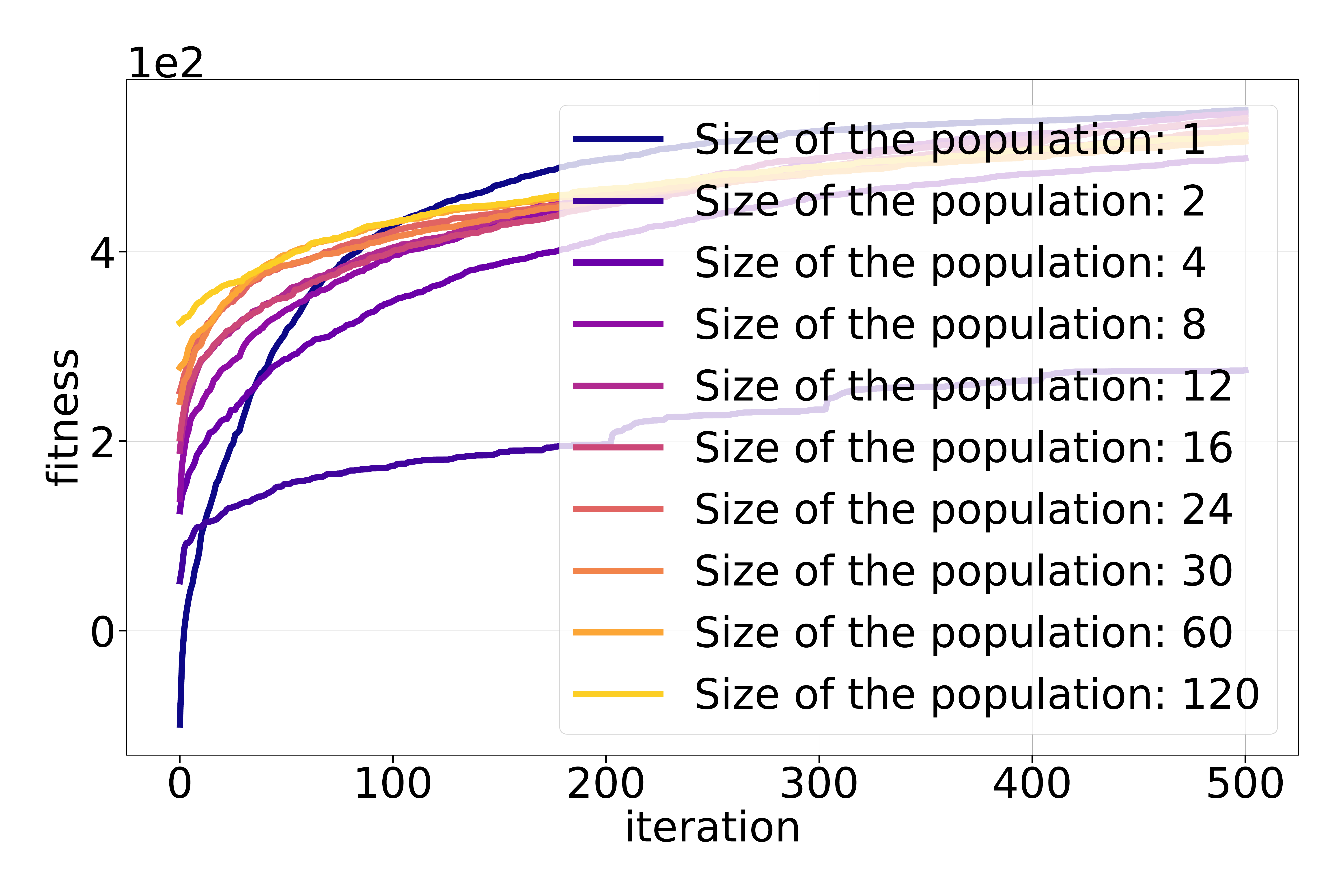}
        \caption{Peak shaving -- fitness}
        \label{subfig:evo_param_popsize_pc_fit}
    \end{subfigure}%
    \begin{subfigure}{0.24\textwidth}
        \centering
        \includegraphics[width=1\textwidth]{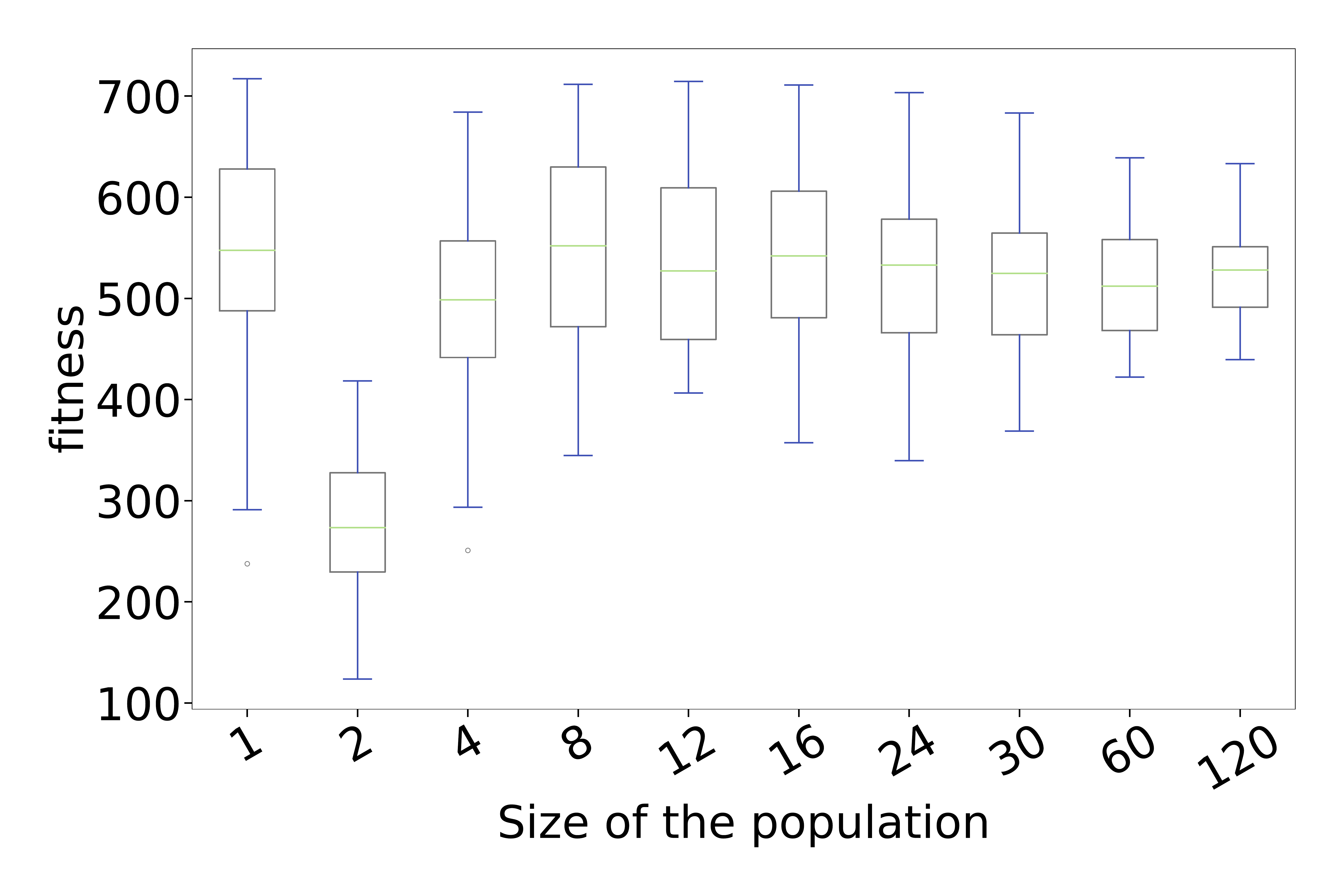}
        \caption{Peak shaving -- robustness}
        \label{subfig:evo_param_popsize_pc_rob}
    \end{subfigure}
    \begin{subfigure}{0.24\textwidth}
        \centering
        \includegraphics[width=1\textwidth]{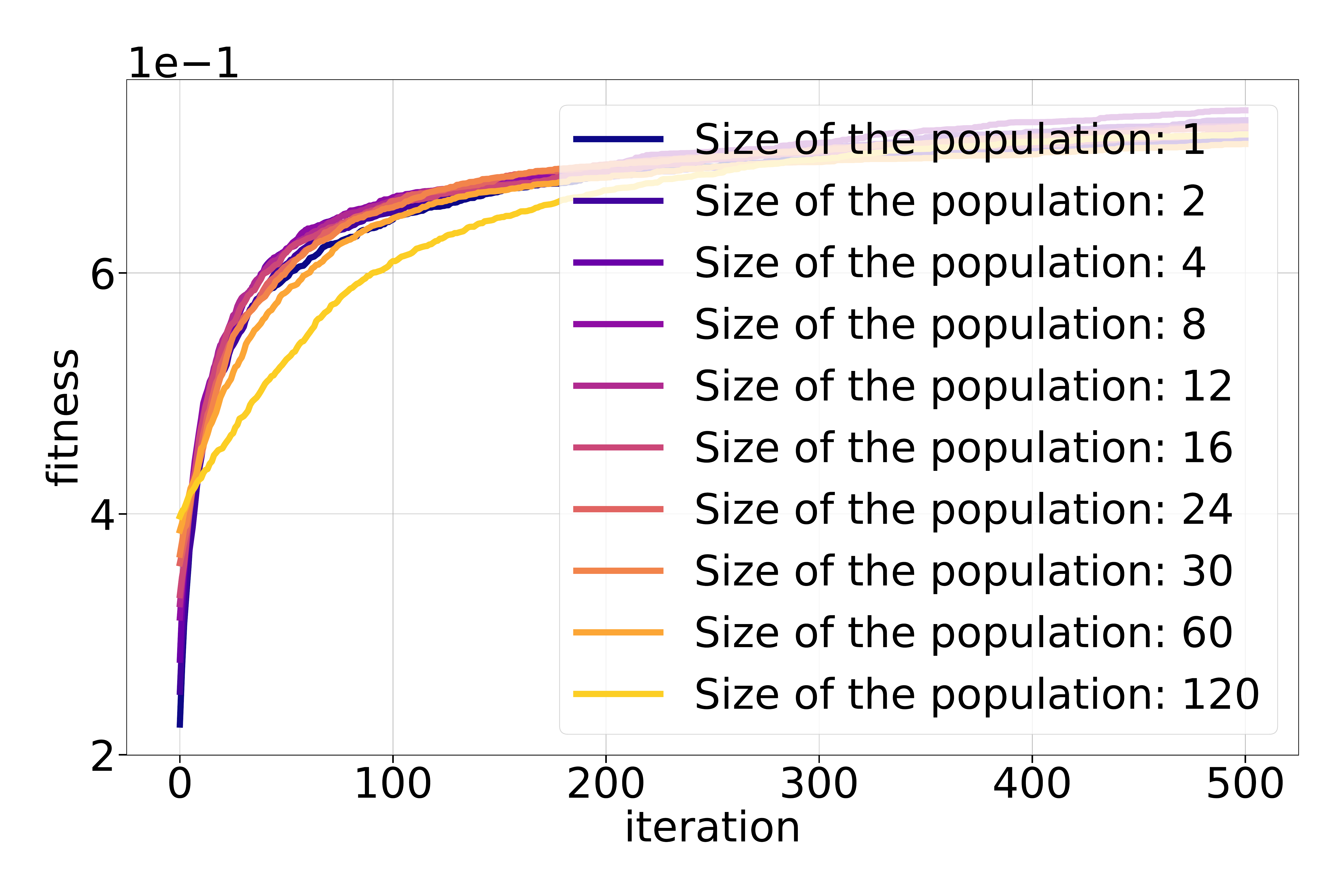}
        \caption{Local SDM -- fitness}
        \label{subfig:evo_param_popsize_oc_fit}
    \end{subfigure}%
    \begin{subfigure}{0.24\textwidth}
        \centering
        \includegraphics[width=1\textwidth]{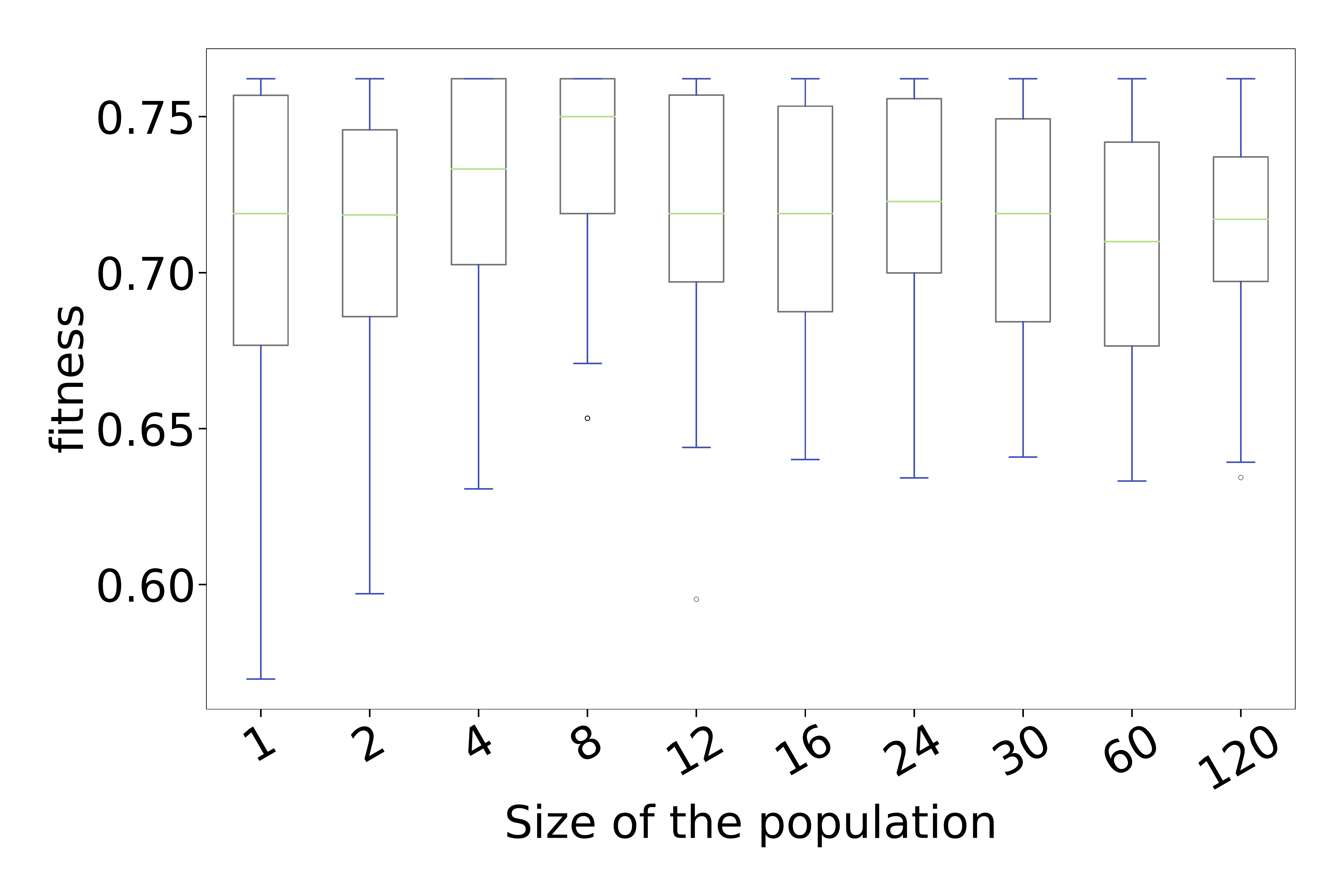}
        \caption{Local SDM -- robustness}
        \label{subfig:evo_param_popsize_oc_rob}
    \end{subfigure}
    \begin{subfigure}{0.24\textwidth}
        \centering
        \includegraphics[width=1\textwidth]{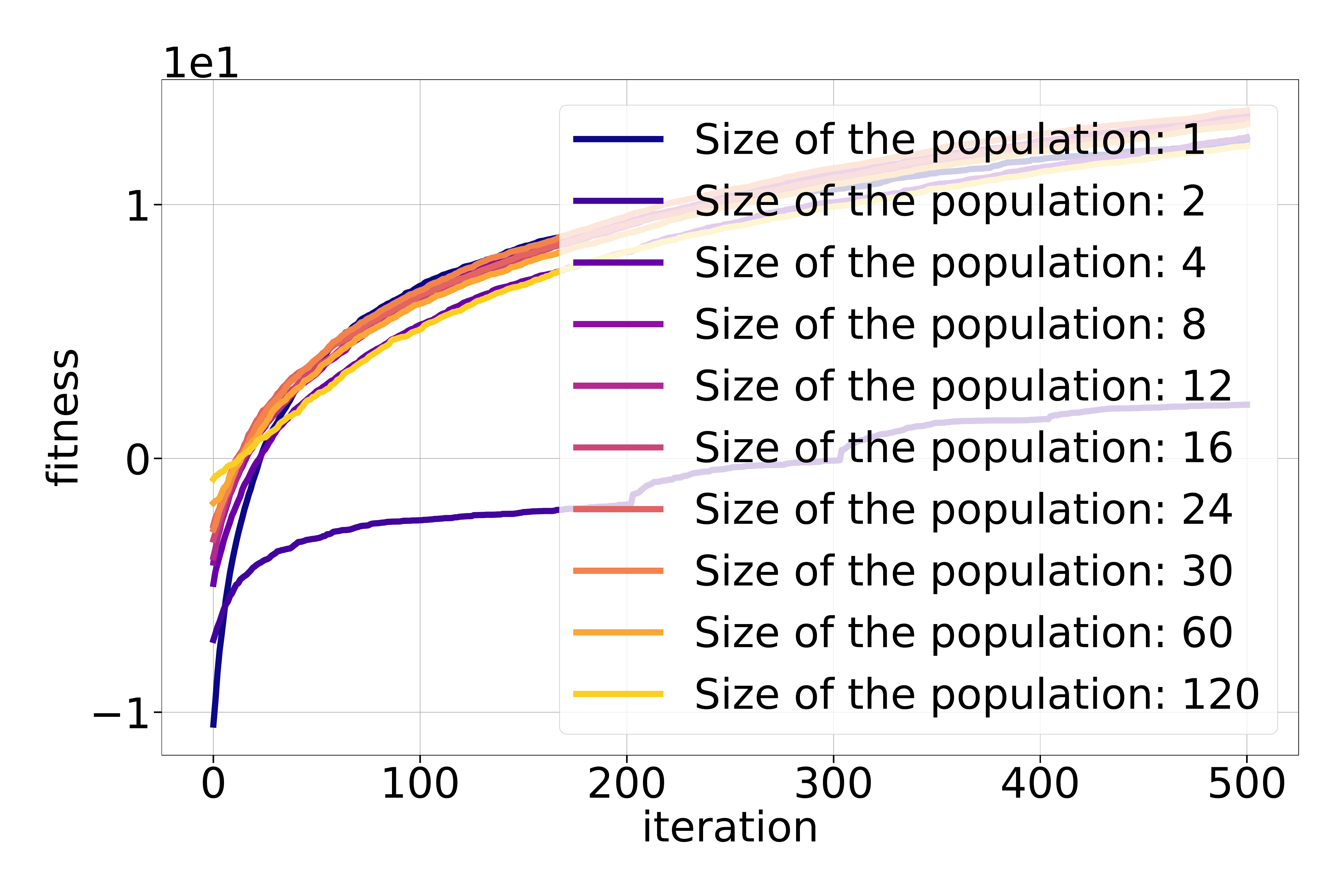}
        \caption{Arbitrage -- fitness}
        \label{subfig:evo_param_popsize_arbitrage_fit}
    \end{subfigure}%
    \begin{subfigure}{0.24\textwidth}
        \centering
        \includegraphics[width=1\textwidth]{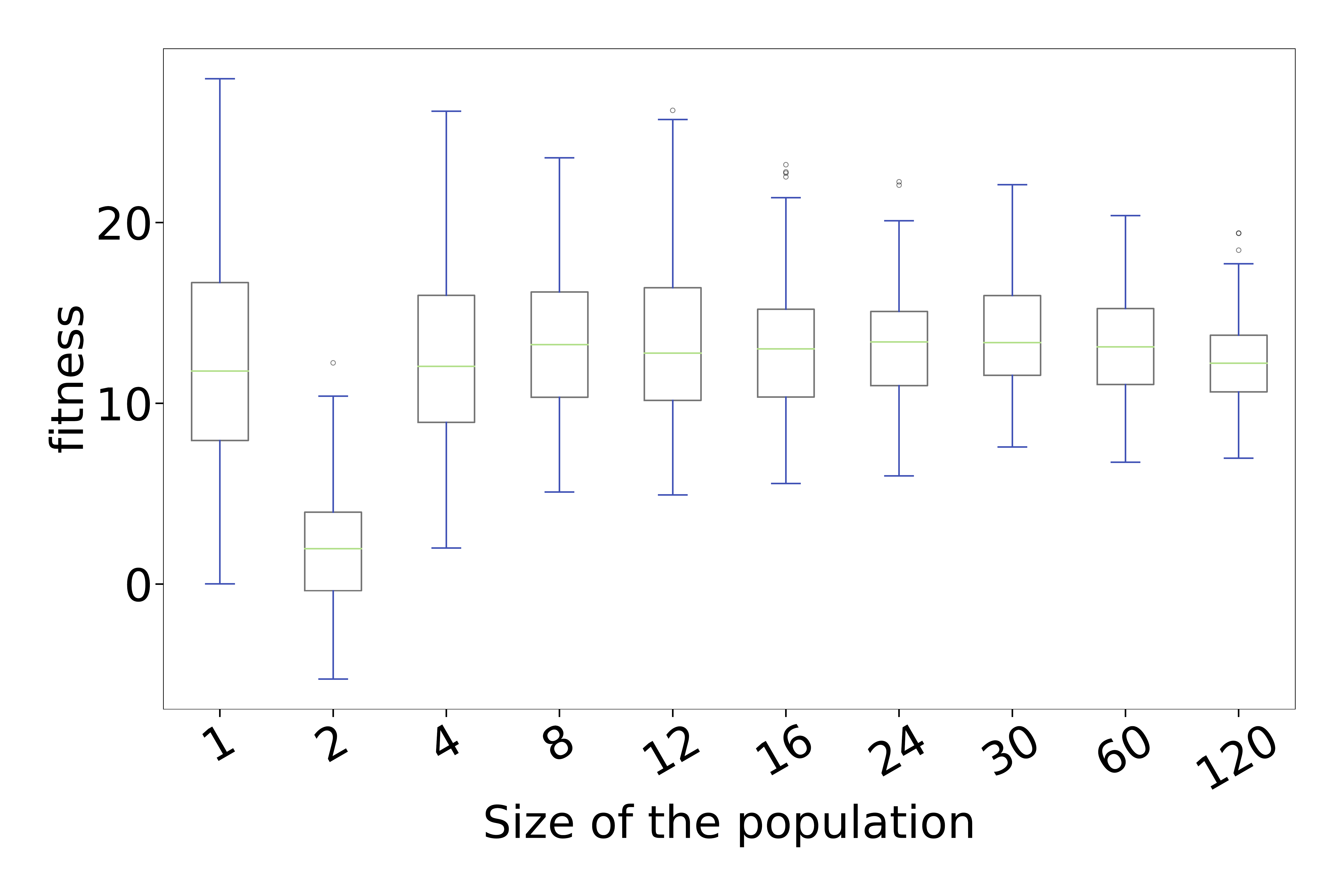}
        \caption{Arbitrage -- robustness}
        \label{subfig:evo_param_popsize_arbitrage_rob}
    \end{subfigure}
    \caption{Size of the population}
    \label{fig:evo_param_popsize}
\end{figure*}
\begin{figure*}[htbp]
  \centering
  \begin{subfigure}{0.24\textwidth}
      \centering
      \includegraphics[width=1\textwidth]{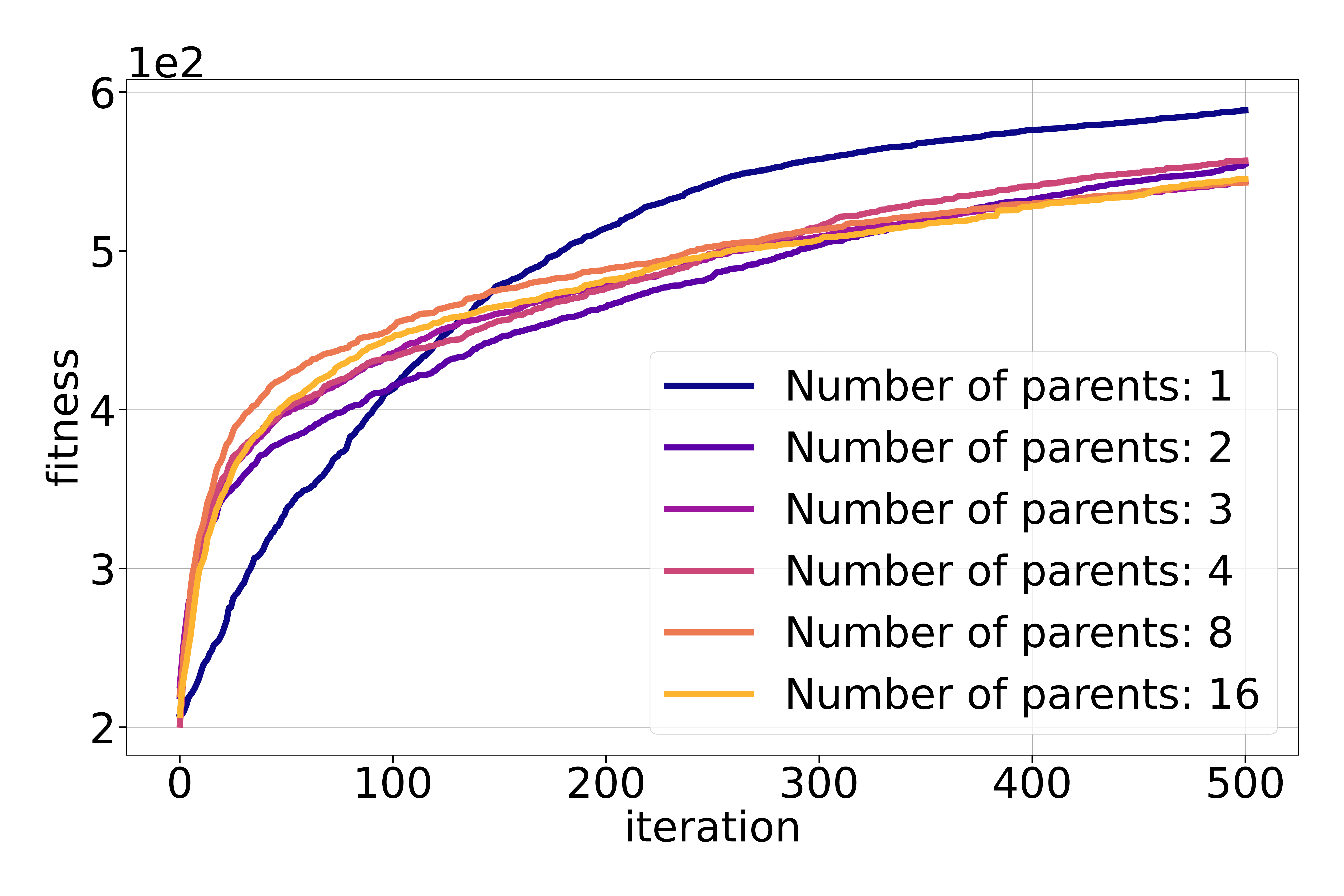}
      \caption{Peak shaving -- fitness}
      \label{subfig:evo_param_parents_pc_fit}
  \end{subfigure}
  \begin{subfigure}{0.24\textwidth}
      \centering
      \includegraphics[width=1\textwidth]{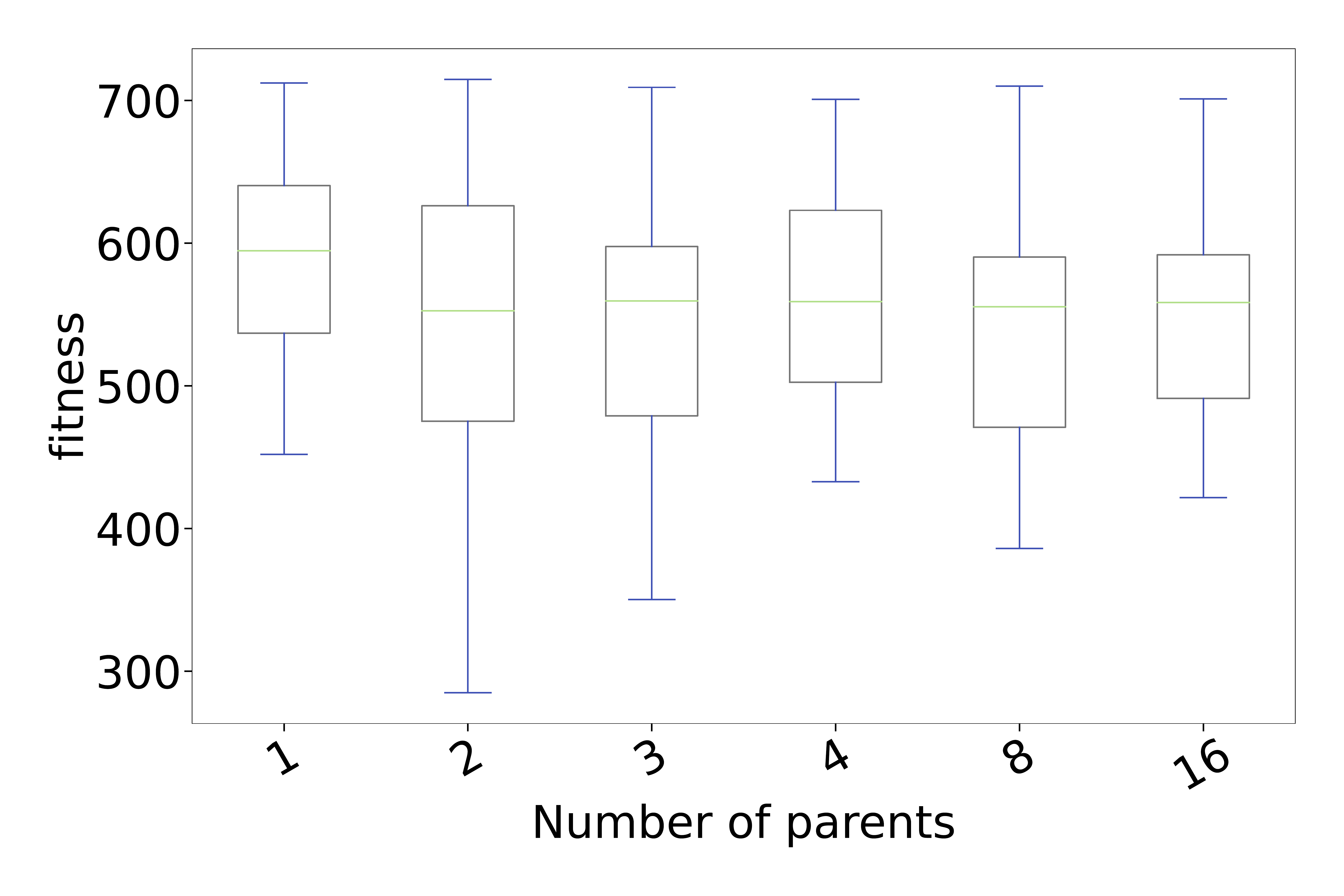}
      \caption{Peak shaving -- robustness}
      \label{subfig:evo_param_pc_parents_rob}
  \end{subfigure}
  \begin{subfigure}{0.24\textwidth}
      \centering
      \includegraphics[width=1\textwidth]{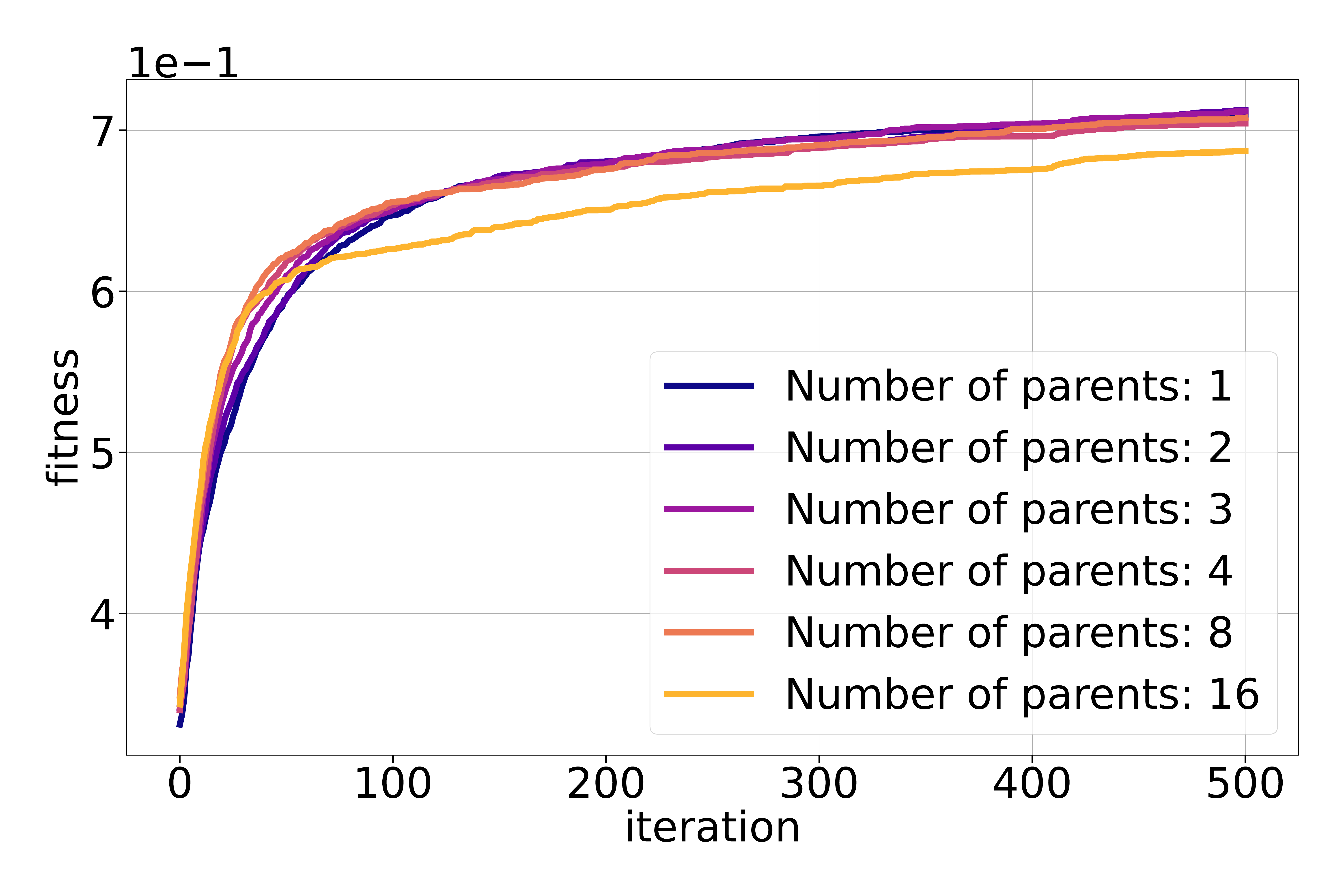}
      \caption{Local SDM -- fitness}
      \label{subfig:evo_param_parents_oc_fit}
  \end{subfigure}
  \begin{subfigure}{0.24\textwidth}
      \centering
      \includegraphics[width=1\textwidth]{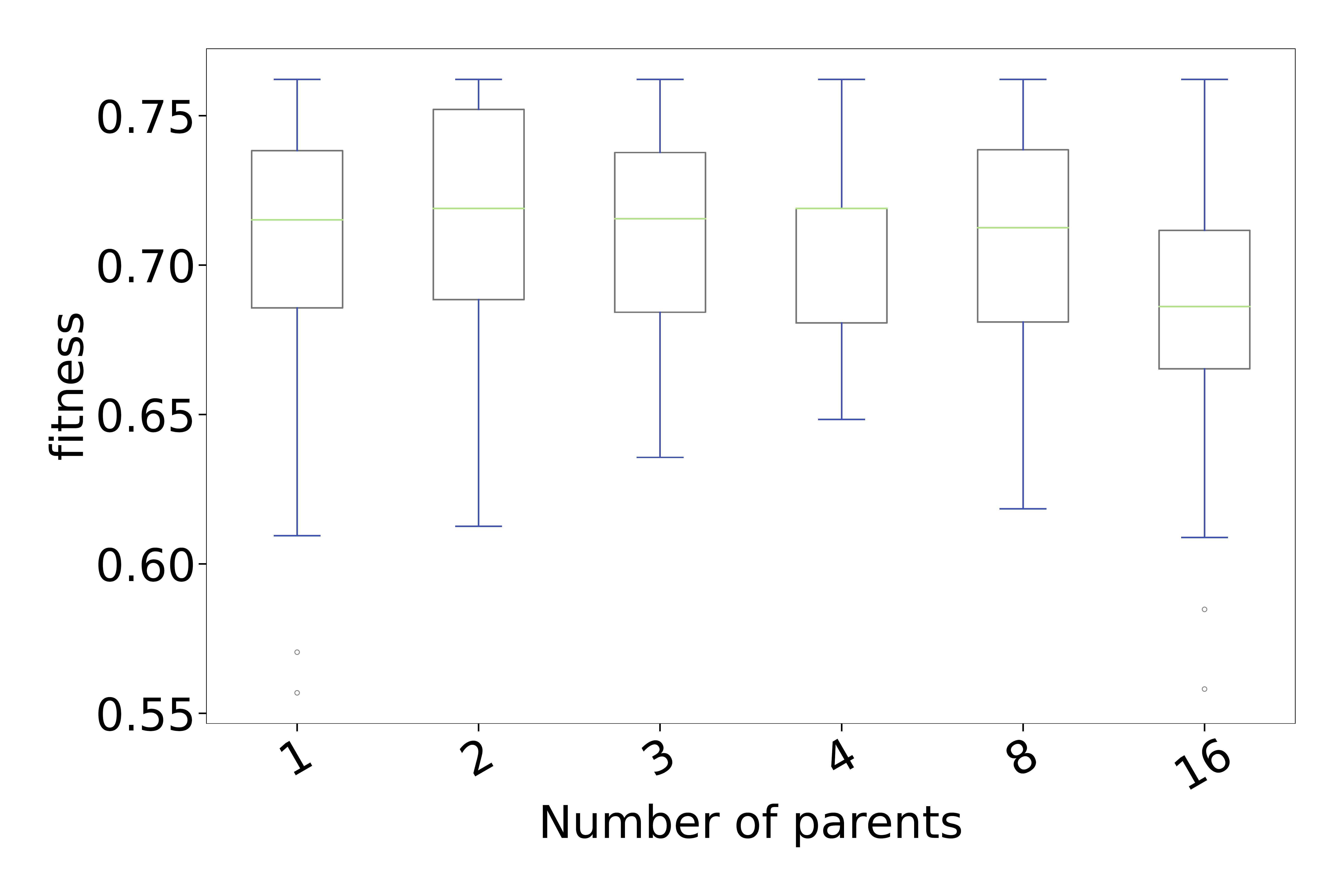}
      \caption{Local SDM -- robustness}
      \label{subfig:evo_param_parents_oc_rob}
  \end{subfigure}
  \begin{subfigure}{0.24\textwidth}
      \centering
      \includegraphics[width=1\textwidth]{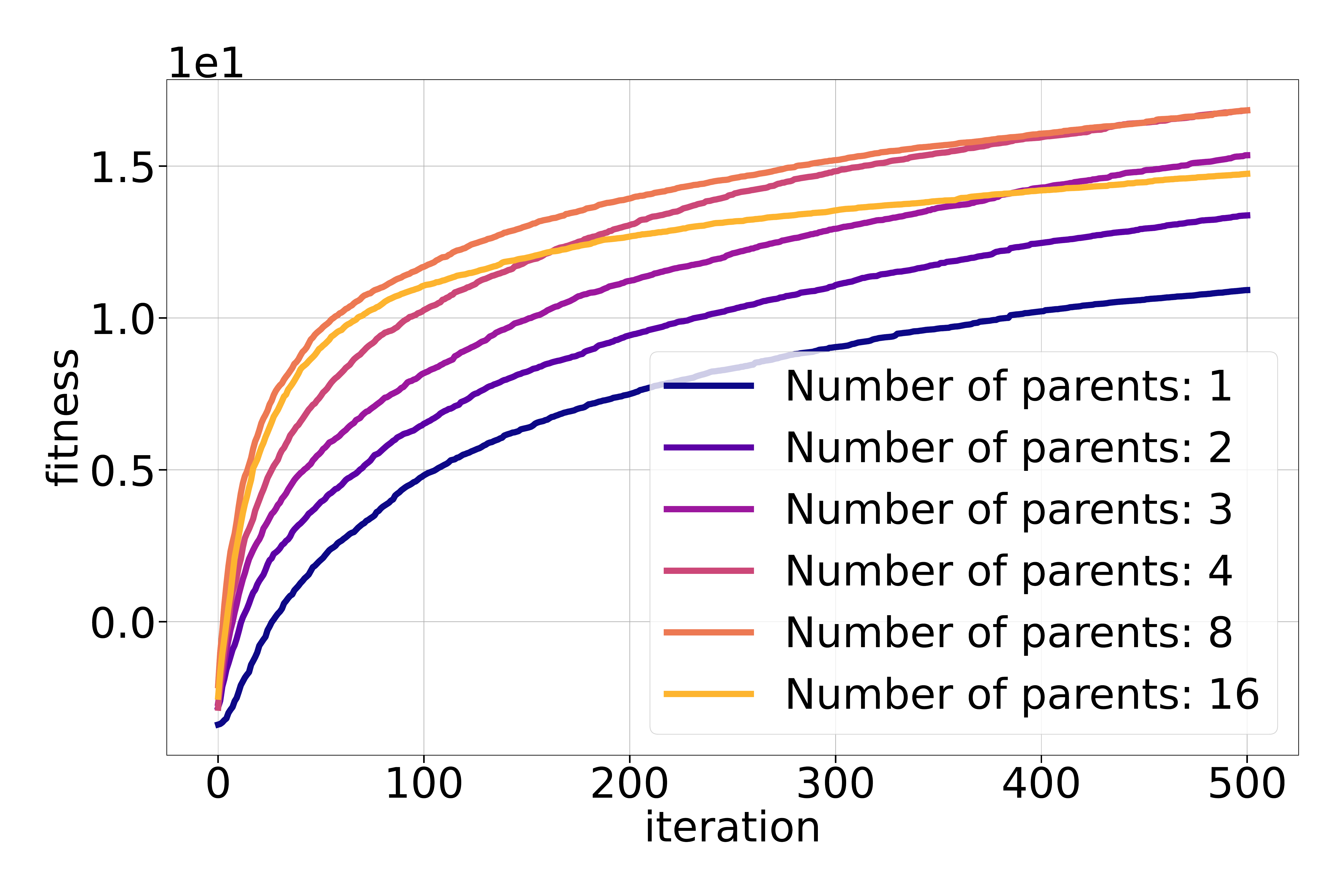}
      \caption{Arbitrage -- fitness}
      \label{subfig:evo_param_parents_arbitrage_fit}
  \end{subfigure}%
  \begin{subfigure}{0.24\textwidth}
      \centering
      \includegraphics[width=1\textwidth]{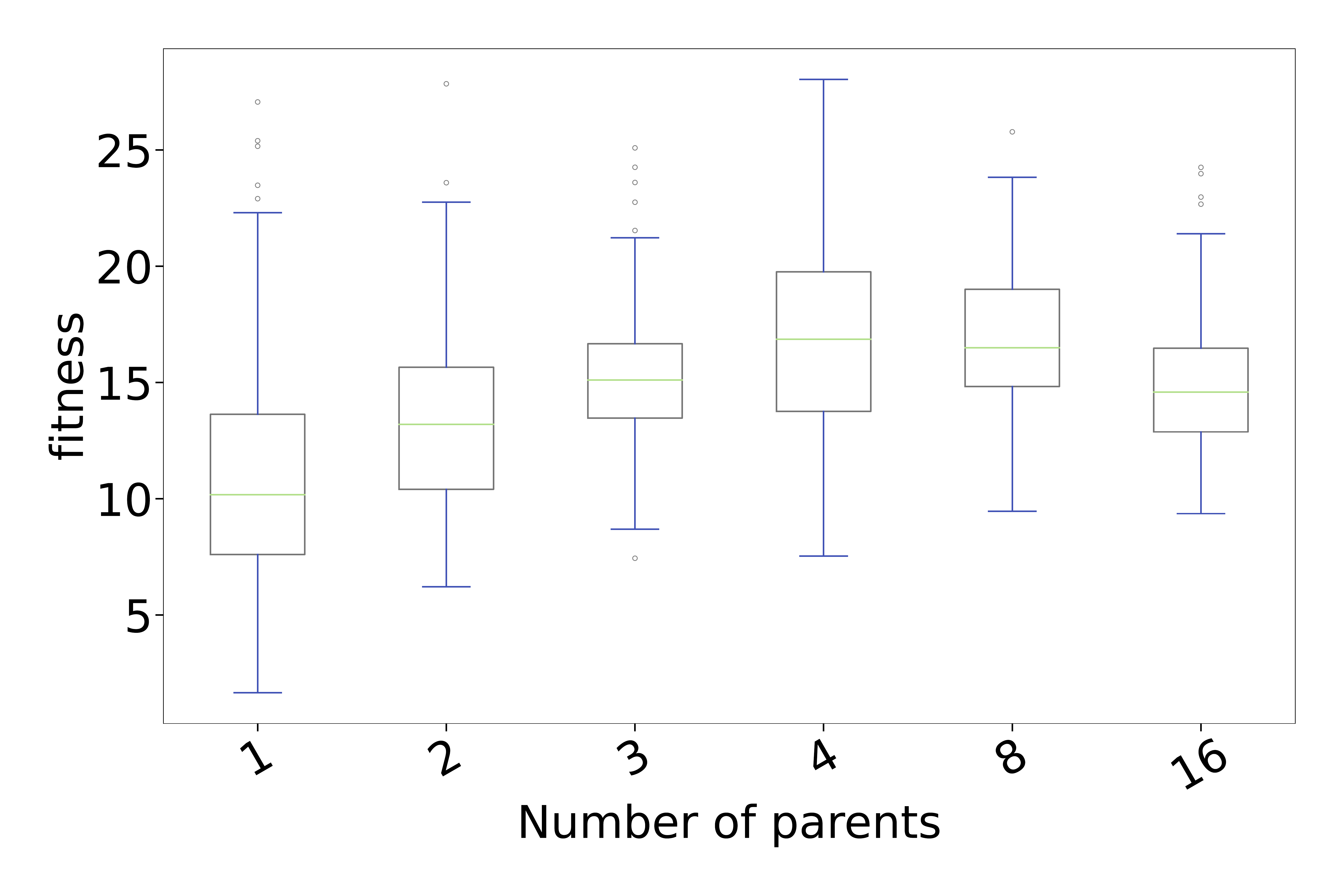}
      \caption{Arbitrage -- robustness}
      \label{subfig:evo_param_parents_arbitrage_rob}
  \end{subfigure}
  \caption{Number of parents}
  \label{fig:evo_param_parents}
\end{figure*}
\begin{table*}
\caption{All storages used in the evaluation.}\label{tab:all_storages}
\centering

  \begin{tabular}{x{4.5cm}ccx{1cm}x{0.7cm}x{1cm}cc}
  \toprule 

  & \multicolumn{2}{c}{efficiency} & & & & \multicolumn{2}{c}{max. power in kW} \\

  \cmidrule{2-3}\cmidrule{7-8}

  name & discharge & charge & initial load in kWh & self discharge in \% & capacity in kWh & charge & discharge \\

  \midrule

  Energy storage model based on the PSP Kirchentellinsfurt \cite{fairenergie2019pumpspeicher}& $\sqrt{0{.}8}$ & $\sqrt{0{.}8}$ & 0 & 0 & 6000 & 1300 & 1300 \\
  Big energy storage for peak shaving \cite{tiemann2020electrical} & $\sqrt{0{.}92}$ & $\sqrt{0{.}92}$ & 17.77 & 0 & 177.7 & 117.6 & 117.6  \\
  Small energy storage for peak shaving \cite{tiemann2020electrical} & $\sqrt{0{.}92}$ & $\sqrt{0{.}92}$ & 3.33 & 0 & 33.3 & 8.25 & 8.25 \\
  Energy storage for local SDM \cite{rusol2019energy} & $\sqrt{0{.}81}$ & $\sqrt{0{.}81}$ & 0.512 & 0 & 5.12 & 1.65 & 2.4 \\

  \bottomrule

\end{tabular}
\end{table*}  
\algparam{30}{4}{30}{8}{1}{24}{8}{2}{30}
\begin{figure*}[htbp]
  \centering
  \begin{subfigure}{0.24\textwidth}
    \includegraphics[width=1\textwidth]{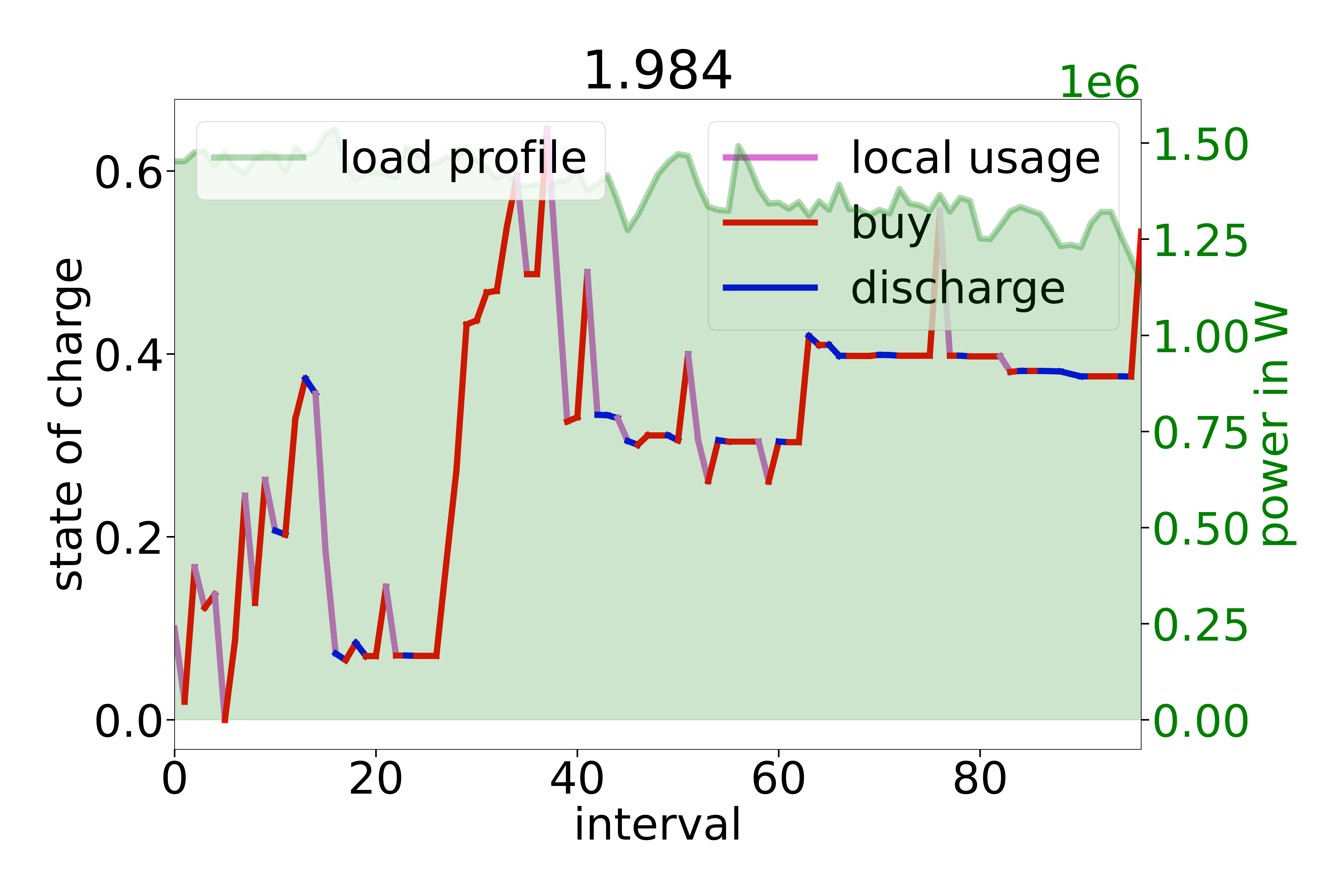}
    \caption{Industry one}
    \label{subfig:first_industry_one_norm}
  \end{subfigure}
  \begin{subfigure}{0.24\textwidth}
    \includegraphics[width=1\textwidth]{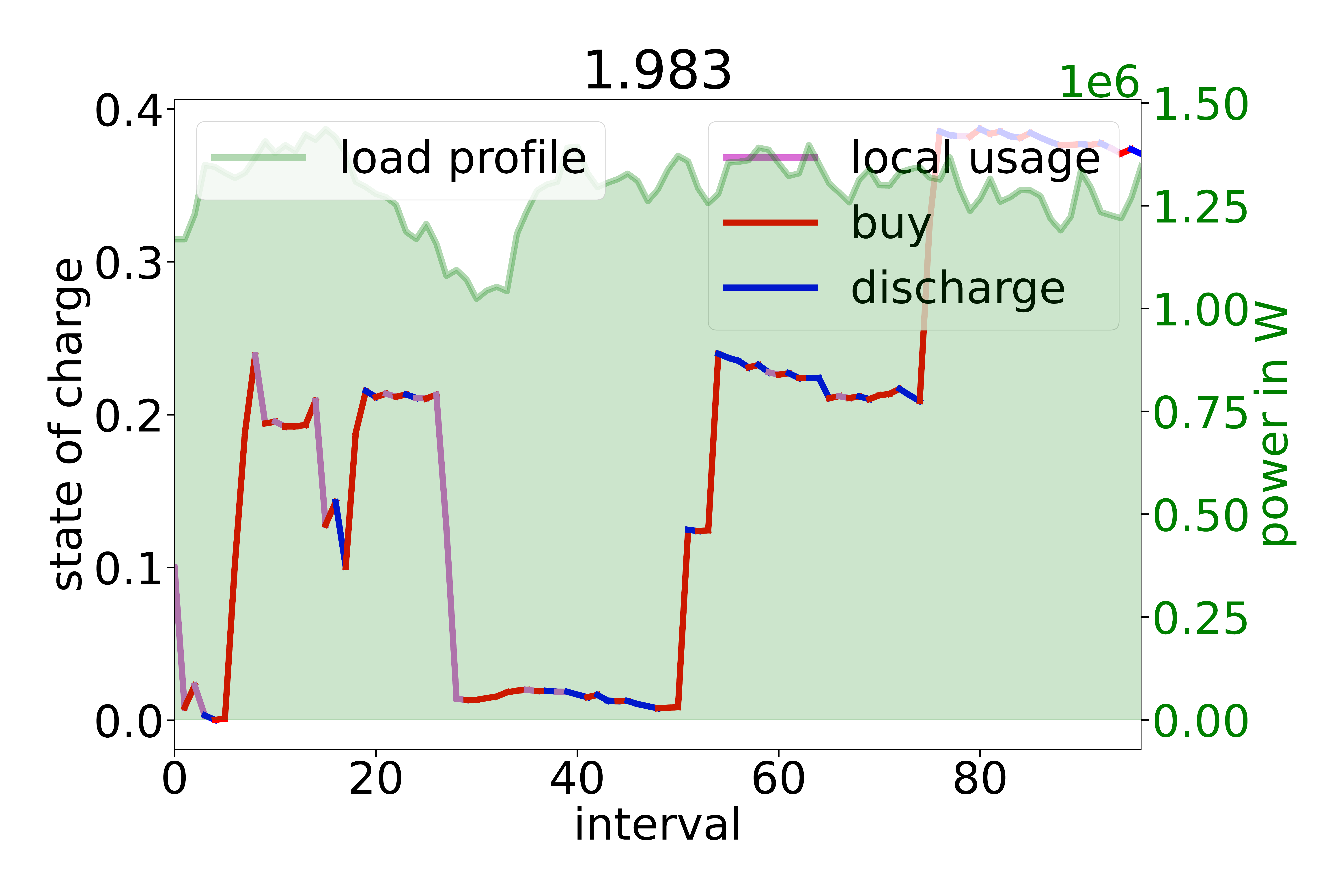}
    \caption{Industry two}
    \label{subfig:first_industry_two_norm}
  \end{subfigure}
  \begin{subfigure}{0.24\textwidth} 
    \includegraphics[width=1\textwidth]{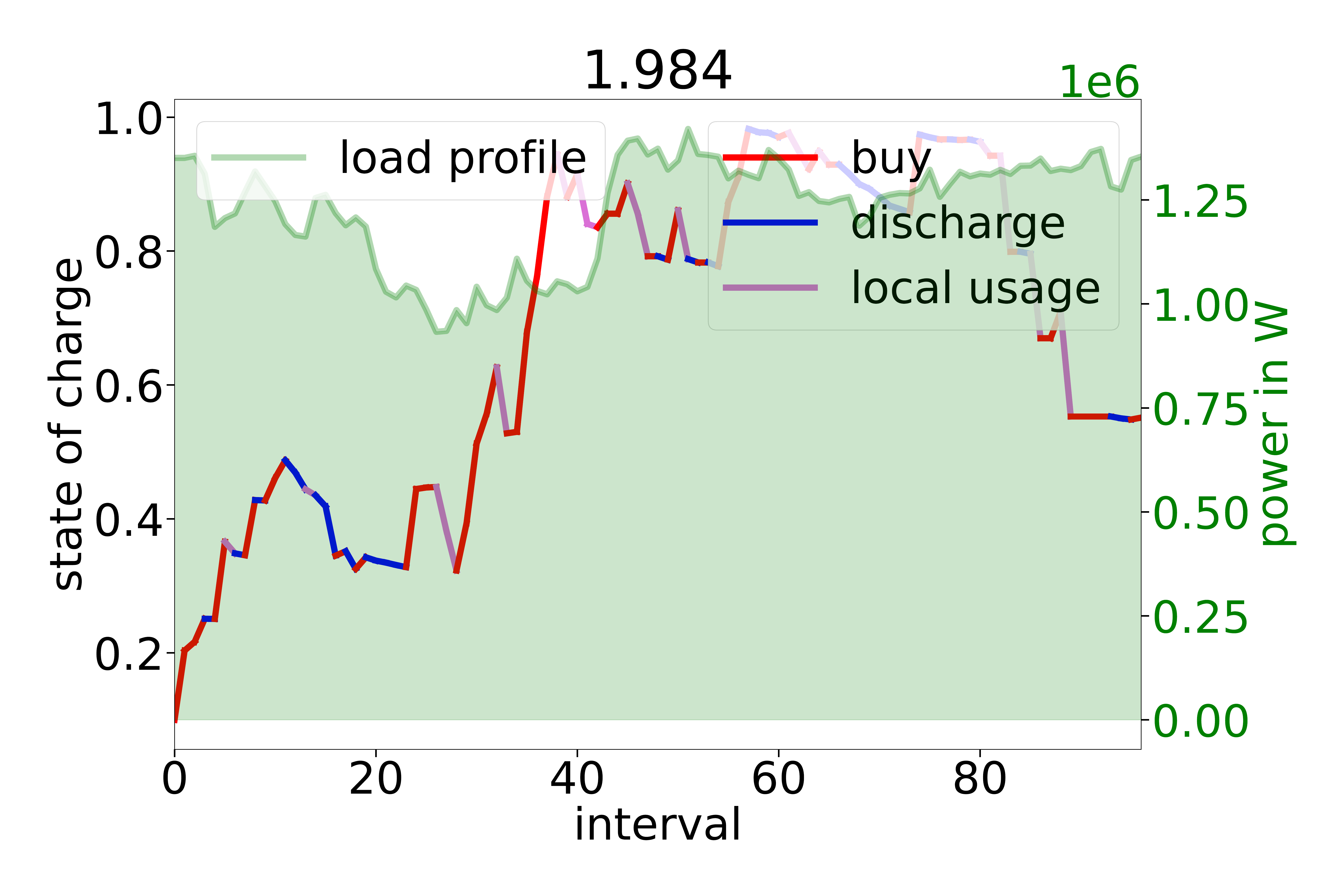}
    \caption{Industry three}
    \label{subfig:first_industry_three_norm}
  \end{subfigure}
  \begin{subfigure}{0.24\textwidth}
    \includegraphics[width=1\textwidth]{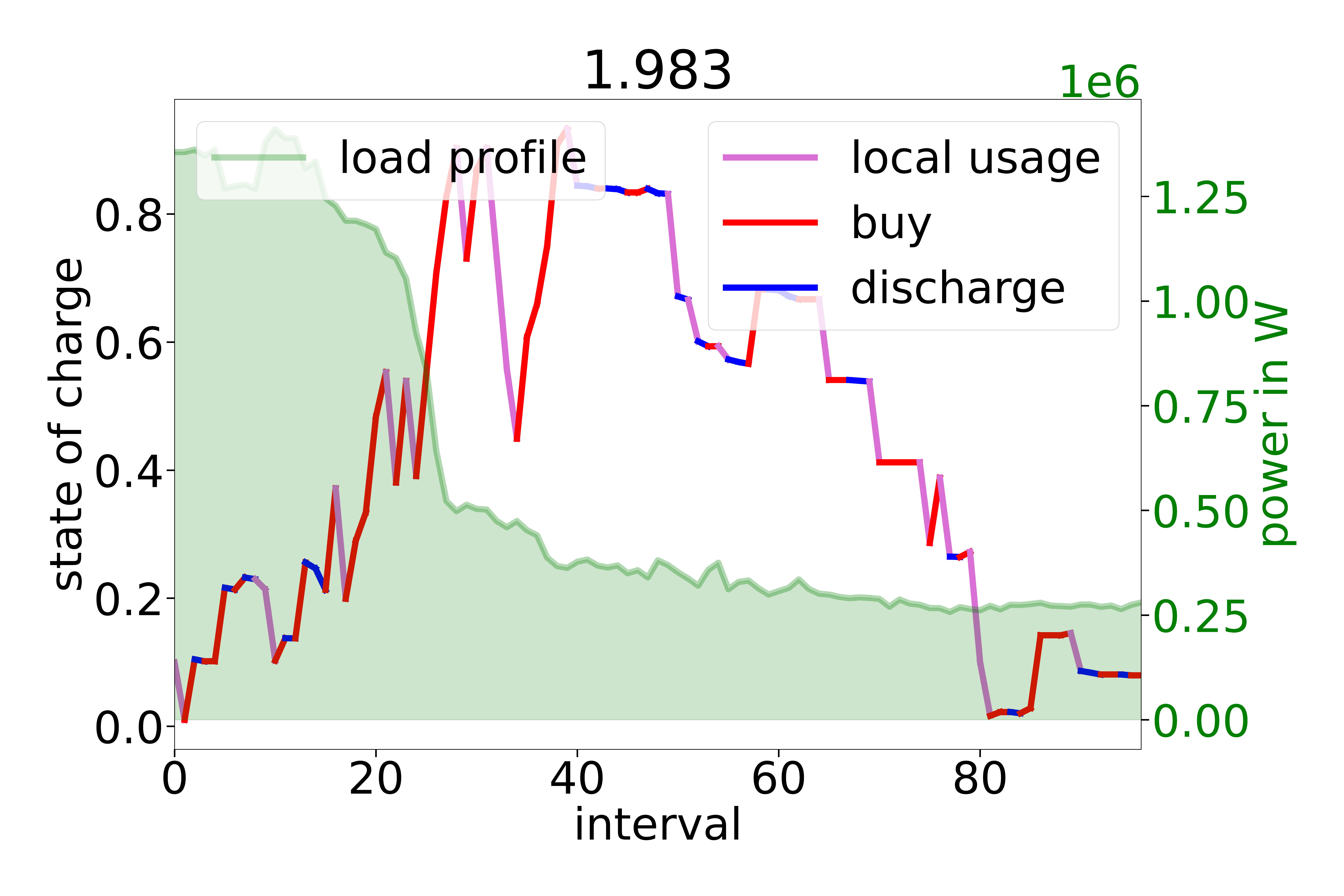}
    \caption{Industry four}
    \label{subfig:first_industry_four_norm}
  \end{subfigure} 
  \begin{subfigure}{0.24\textwidth}
    \centering
    \includegraphics[width=1\textwidth]{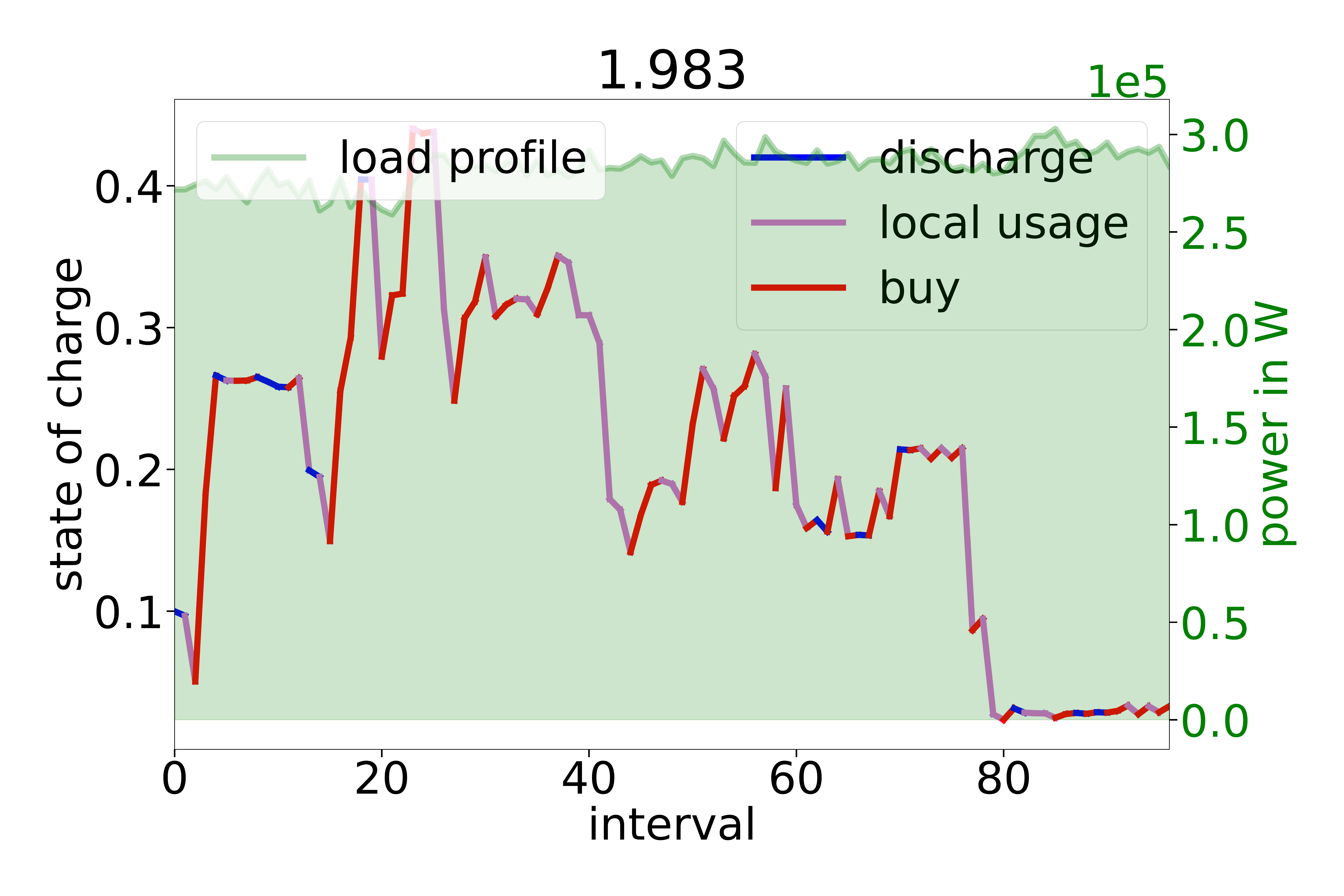}
    \caption{Industry five}
    \label{subfig:first_industry_five_norm}
  \end{subfigure}
  \begin{subfigure}{0.24\textwidth}
    \centering
    \includegraphics[width=1\textwidth]{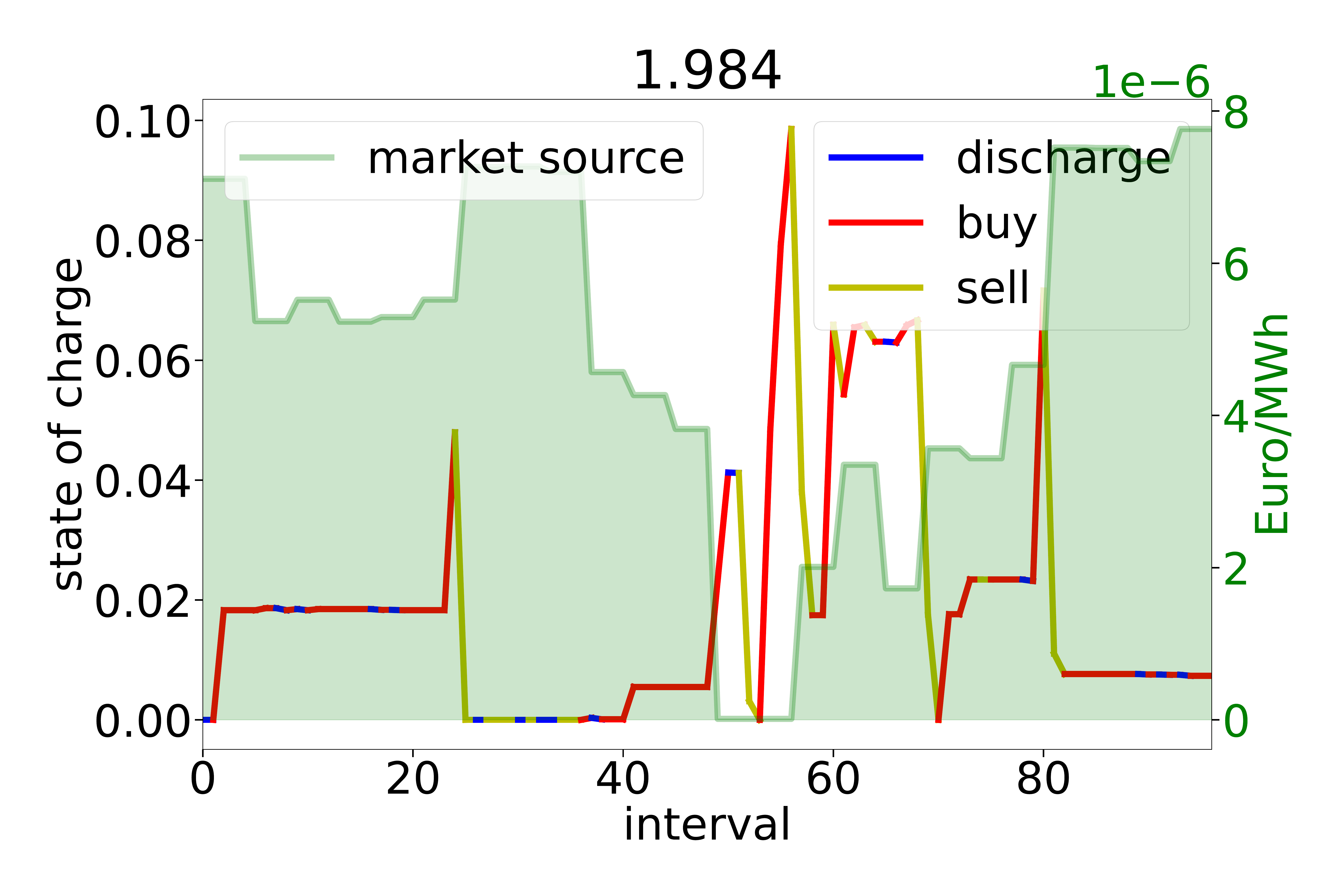}
    \caption{PSP}
    \label{subfig:first_psp_norm}
  \end{subfigure}
  \caption{Locally chosen OS -- 1. Scenario with normalization}
  \label{fig:first_norm_solutions}
\end{figure*}
\begin{figure*}[htbp]
  \centering
  \begin{subfigure}{0.24\textwidth}
    \centering
    \includegraphics[width=1\textwidth]{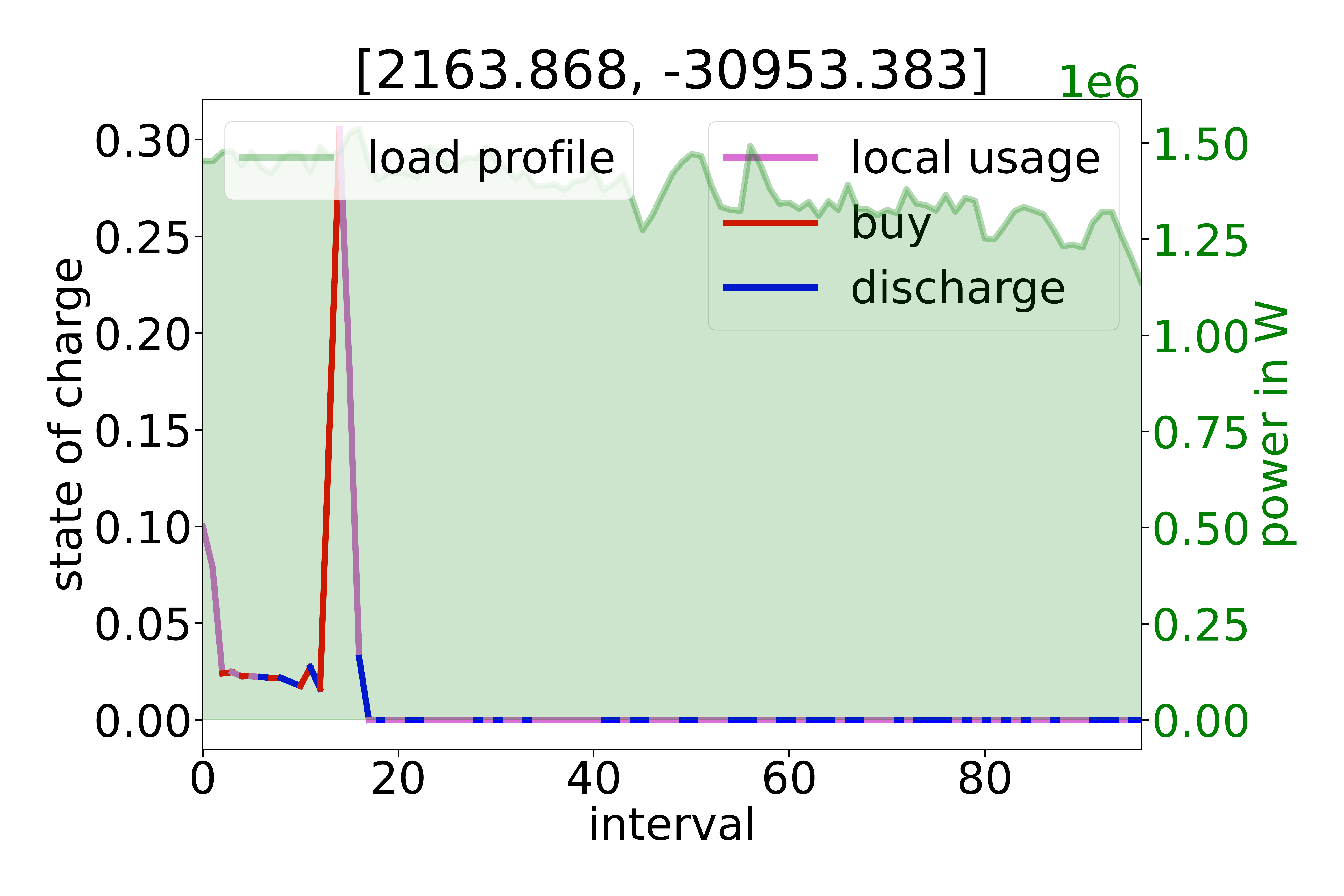}
    \caption{Industry one: chosen OS}
    \label{subfig:first_industry_one_pareto}
  \end{subfigure}
  \begin{subfigure}{0.24\textwidth}
    \centering
    \includegraphics[width=1\textwidth]{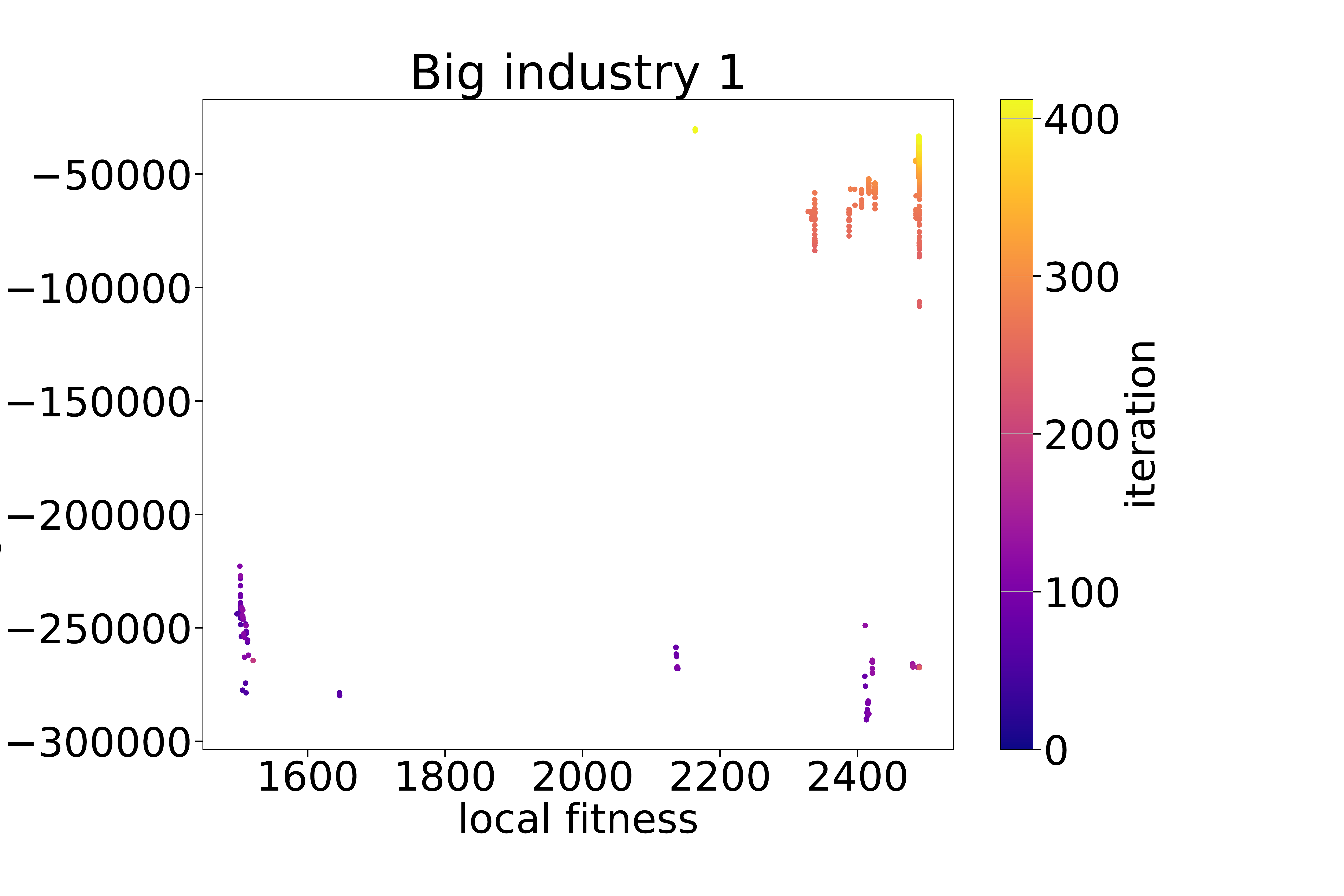}
    \caption{Industry one: Pareto front}
    \label{subfig:first_industry_one_pareto_fit}
  \end{subfigure}
  \begin{subfigure}{0.24\textwidth}
    \centering
    \includegraphics[width=1\textwidth]{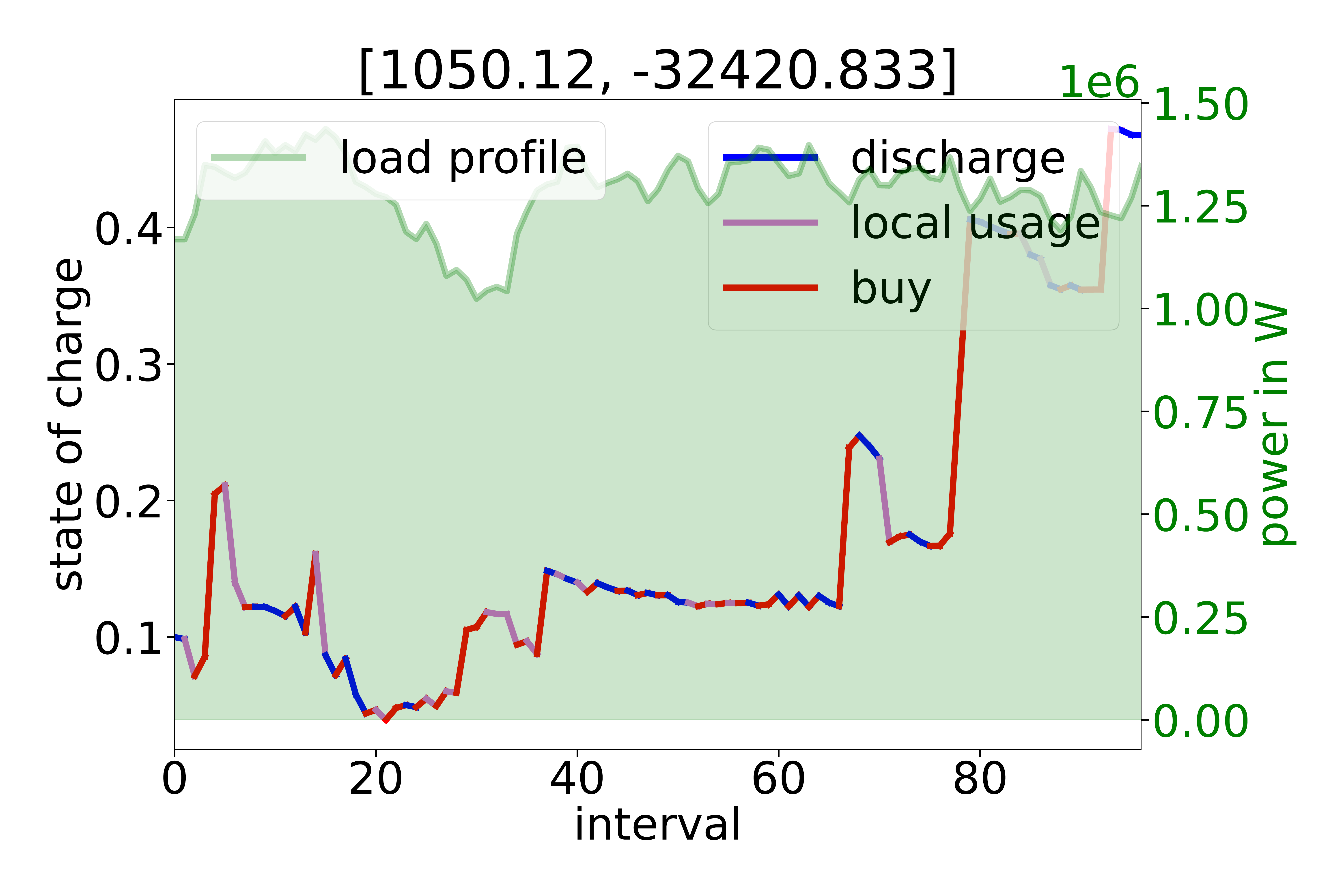}
    \caption{Industry two: chosen OS}
    \label{subfig:first_industry_two_pareto}
  \end{subfigure}
  \begin{subfigure}{0.24\textwidth}
    \centering
    \includegraphics[width=1\textwidth]{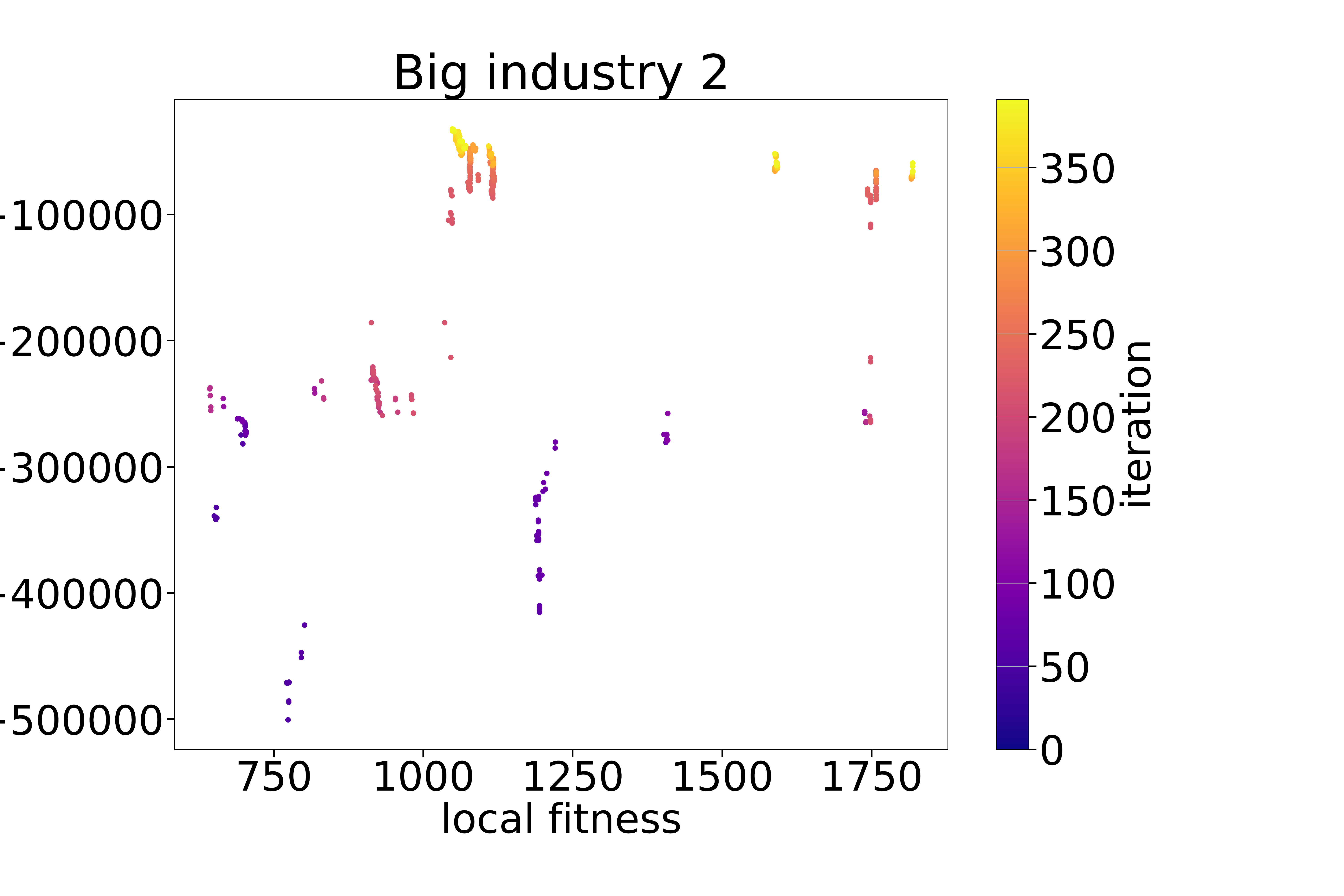}
    \caption{Industry two: Pareto front}
    \label{subfig:first_industry_two_pareto_fit}
  \end{subfigure}
  \begin{subfigure}{0.24\textwidth} 
    \centering
    \includegraphics[width=1\textwidth]{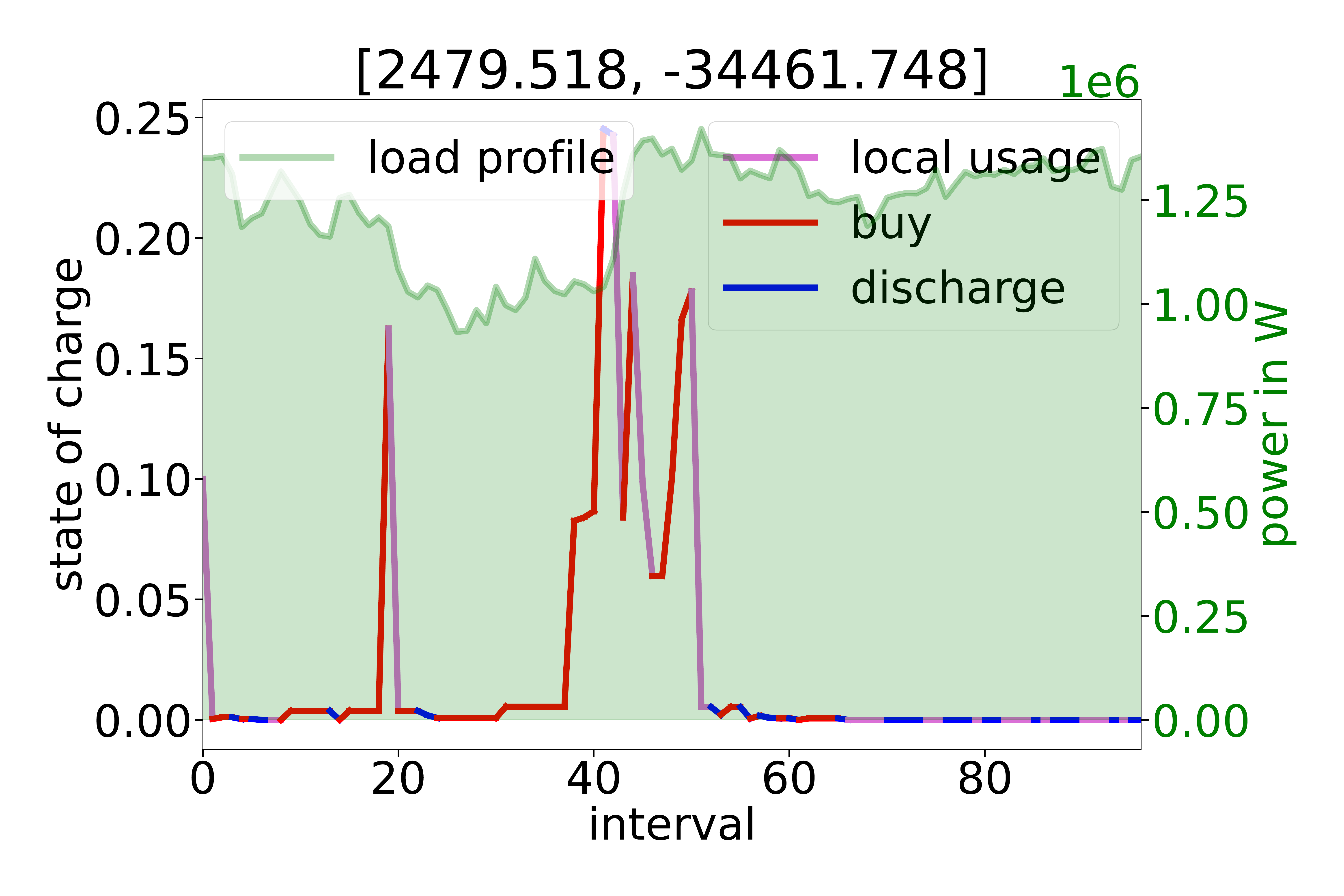}
    \caption{Industry three: chosen OS}
    \label{subfig:first_industry_three_pareto}
  \end{subfigure}
  \begin{subfigure}{0.24\textwidth} 
    \centering
    \includegraphics[width=1\textwidth]{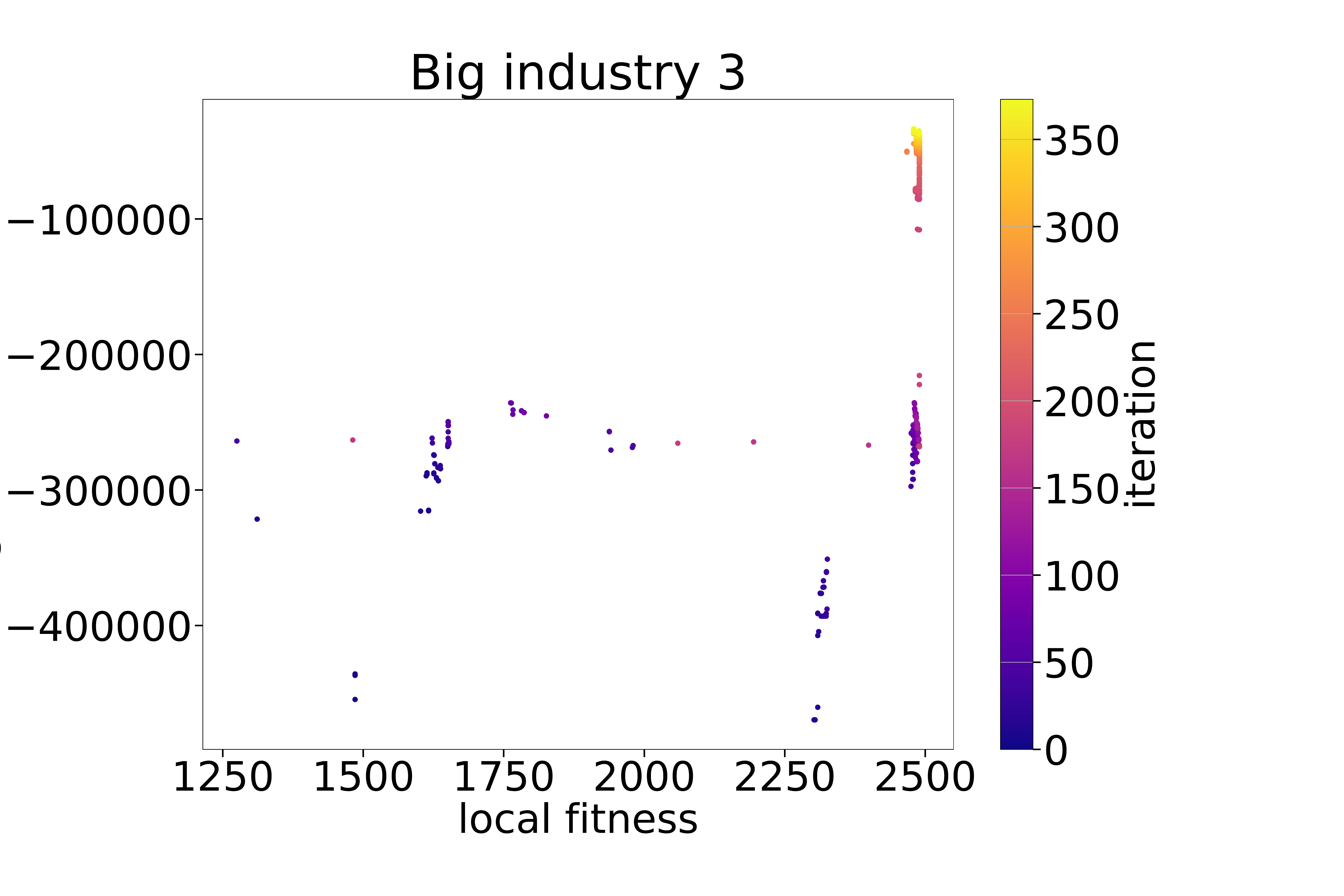}
    \caption{Industry three: Pareto front}
    \label{subfig:first_industry_three_pareto_fit}
  \end{subfigure}
  \begin{subfigure}{0.24\textwidth}
    \centering
    \includegraphics[width=1\textwidth]{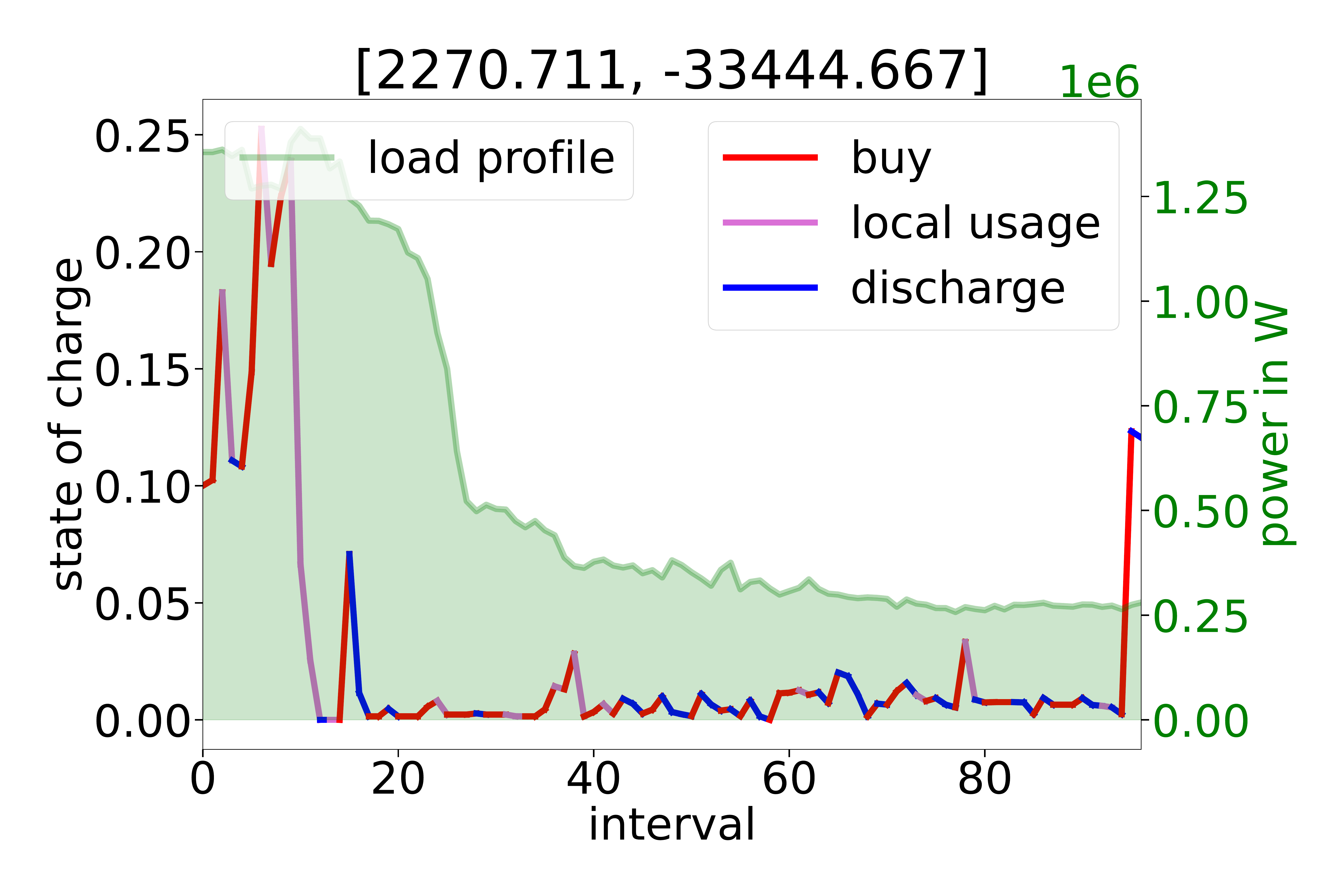}
    \caption{Industry four: chosen OS}
    \label{subfig:first_industry_four_pareto}
  \end{subfigure} 
  \begin{subfigure}{0.24\textwidth}
    \centering
    \includegraphics[width=1\textwidth]{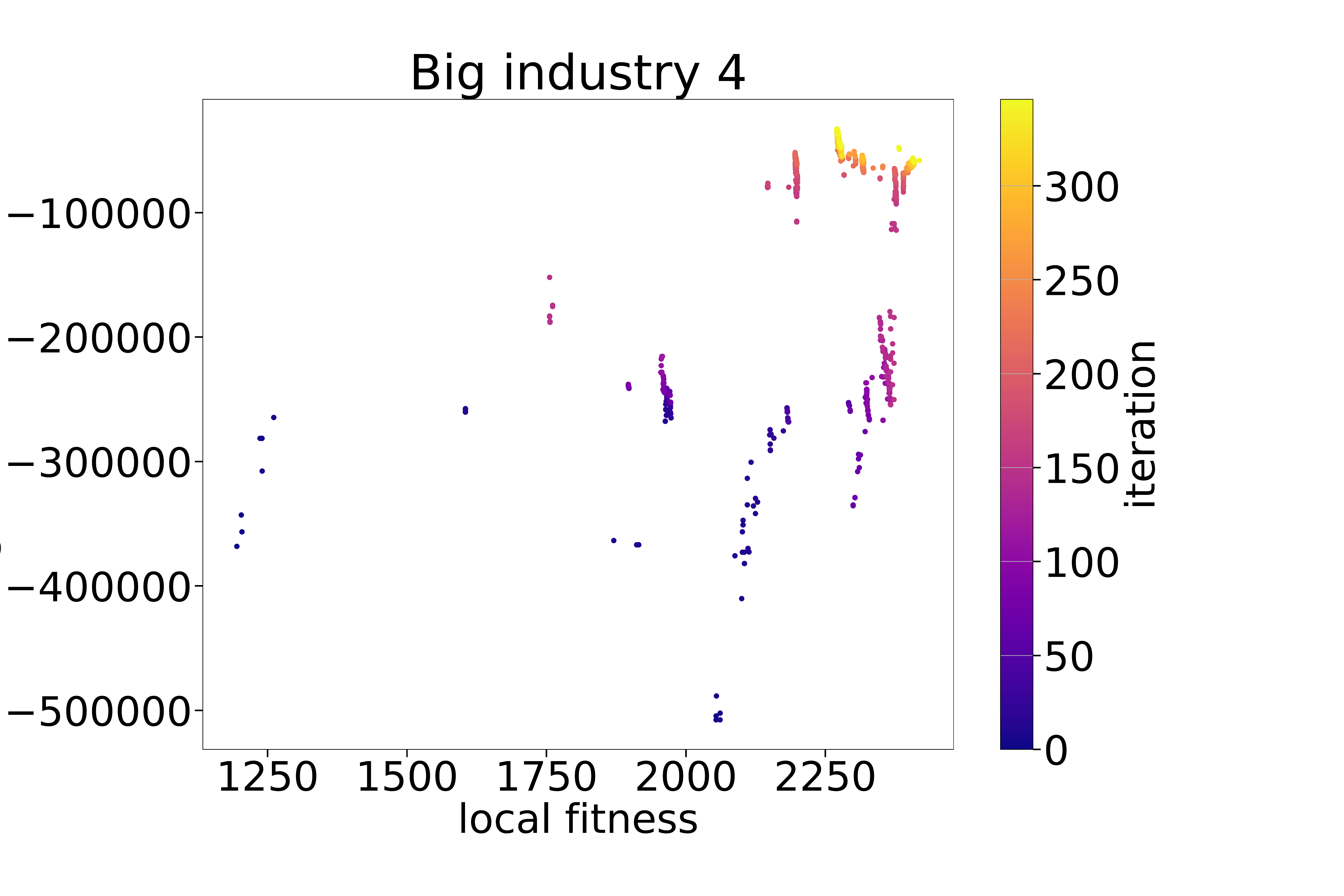}
    \caption{Industry four: Pareto front}
    \label{subfig:first_industry_four_pareto_fit}
  \end{subfigure} 
  \begin{subfigure}{0.24\textwidth}
    \centering
    \includegraphics[width=1\textwidth]{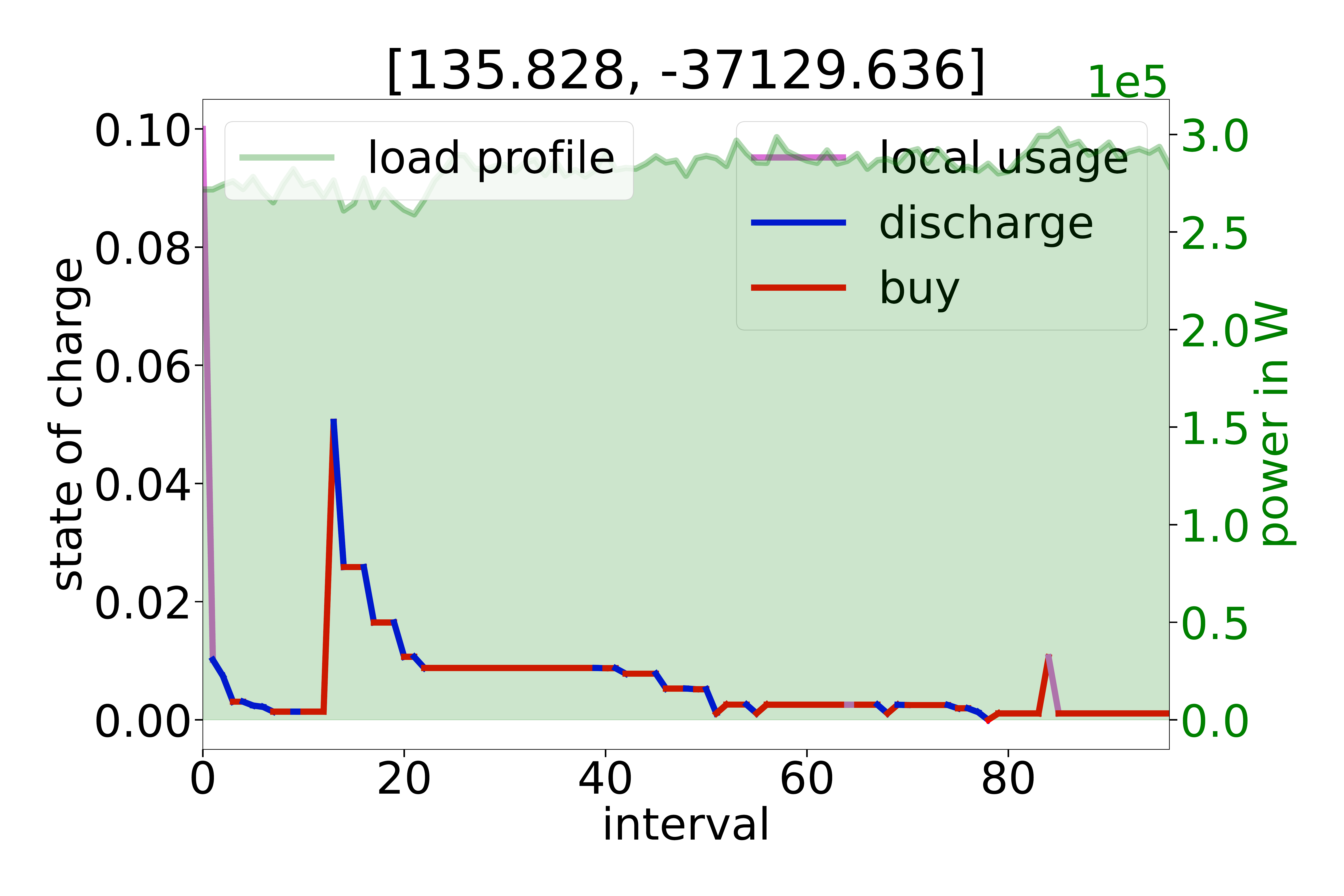}
    \caption{Industry five: chosen OS}
    \label{subfig:first_industry_five_pareto}
  \end{subfigure}
  \begin{subfigure}{0.24\textwidth}
    \centering
    \includegraphics[width=1\textwidth]{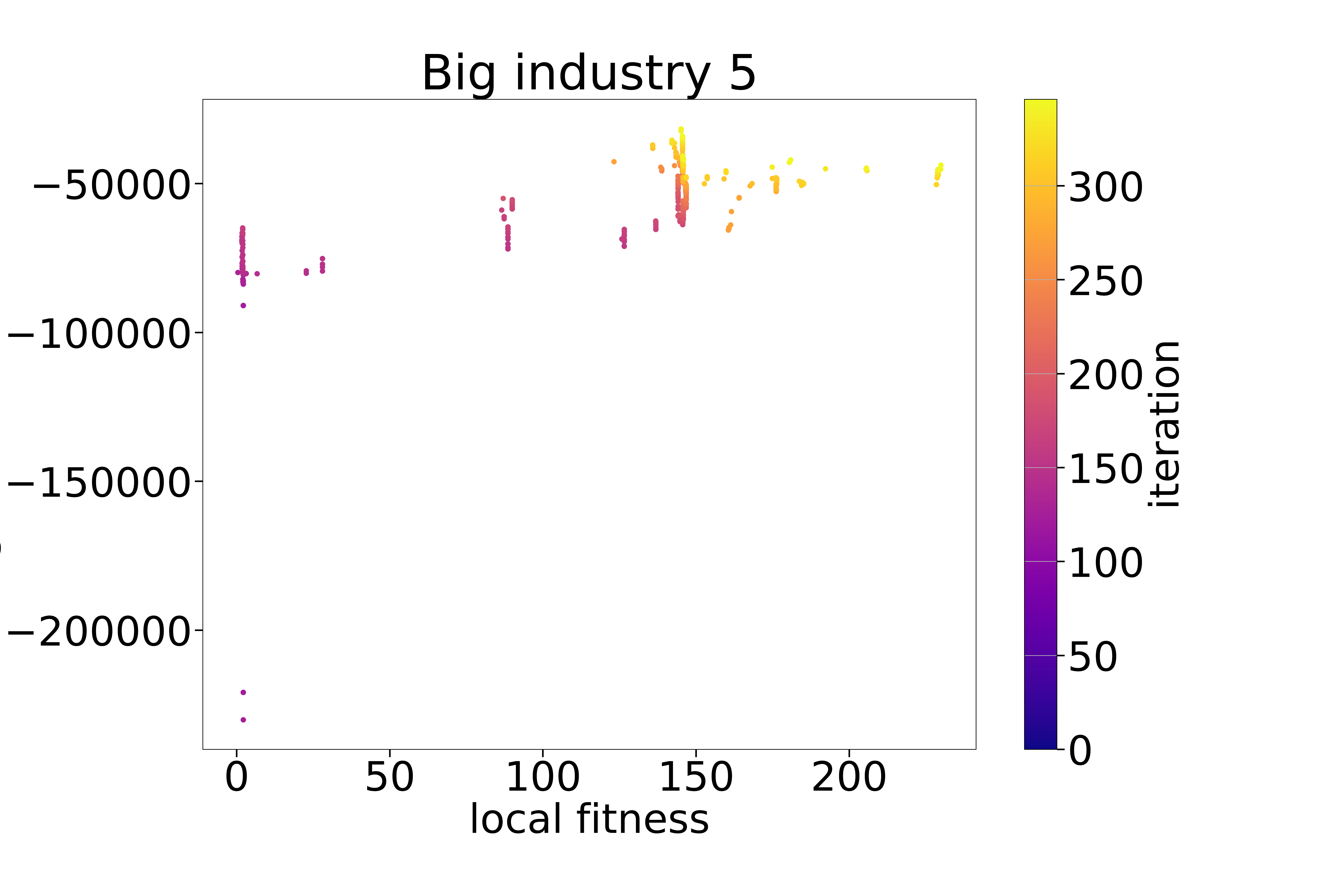}
    \caption{Industry five: Pareto front}
    \label{subfig:first_industry_five_pareto_fit}
  \end{subfigure}
  \begin{subfigure}{0.24\textwidth}
    \centering
    \includegraphics[width=1\textwidth]{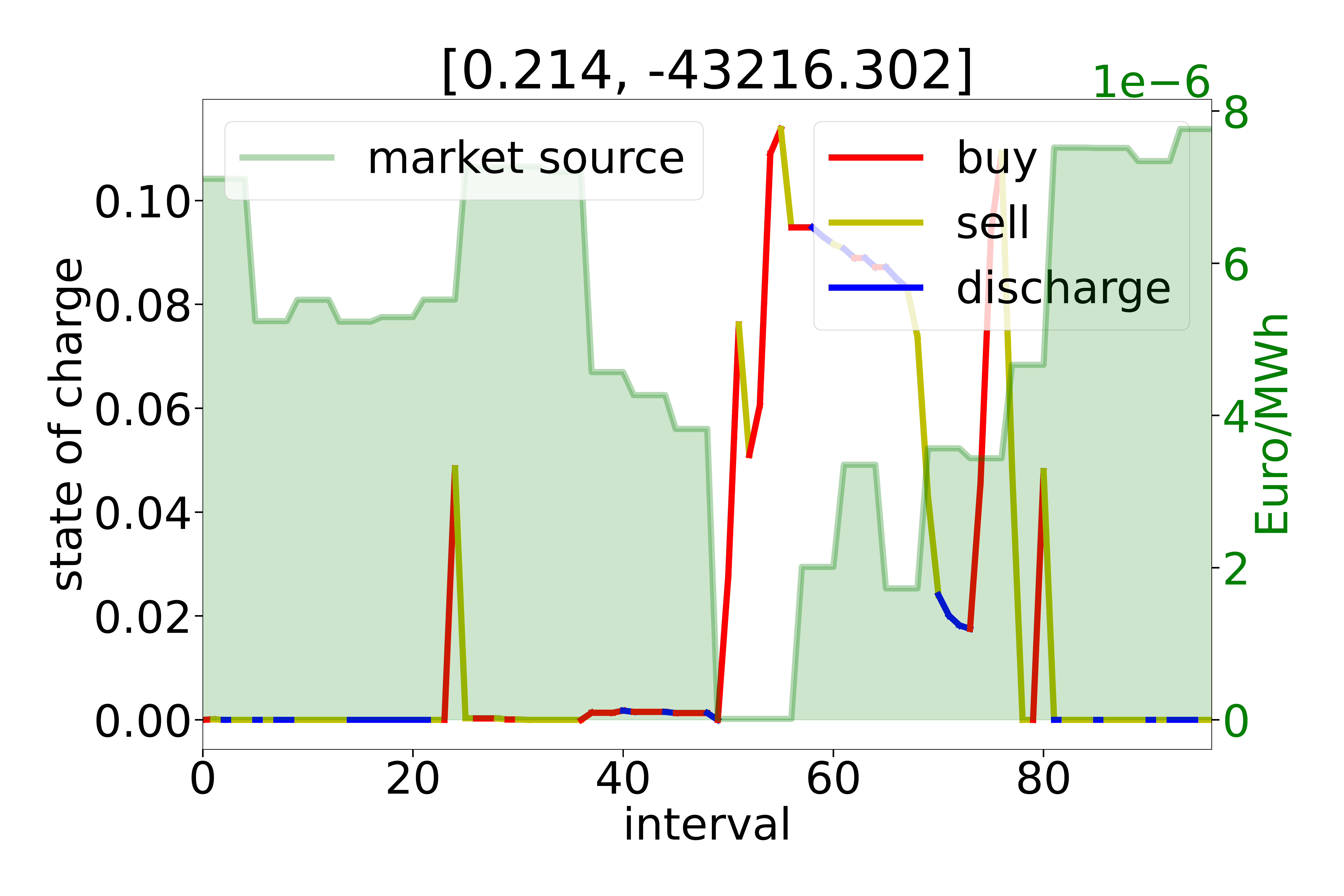}
    \caption{PSP: chosen OS}
    \label{subfig:first_psp_pareto}
  \end{subfigure}
  \begin{subfigure}{0.24\textwidth}
    \centering
    \includegraphics[width=1\textwidth]{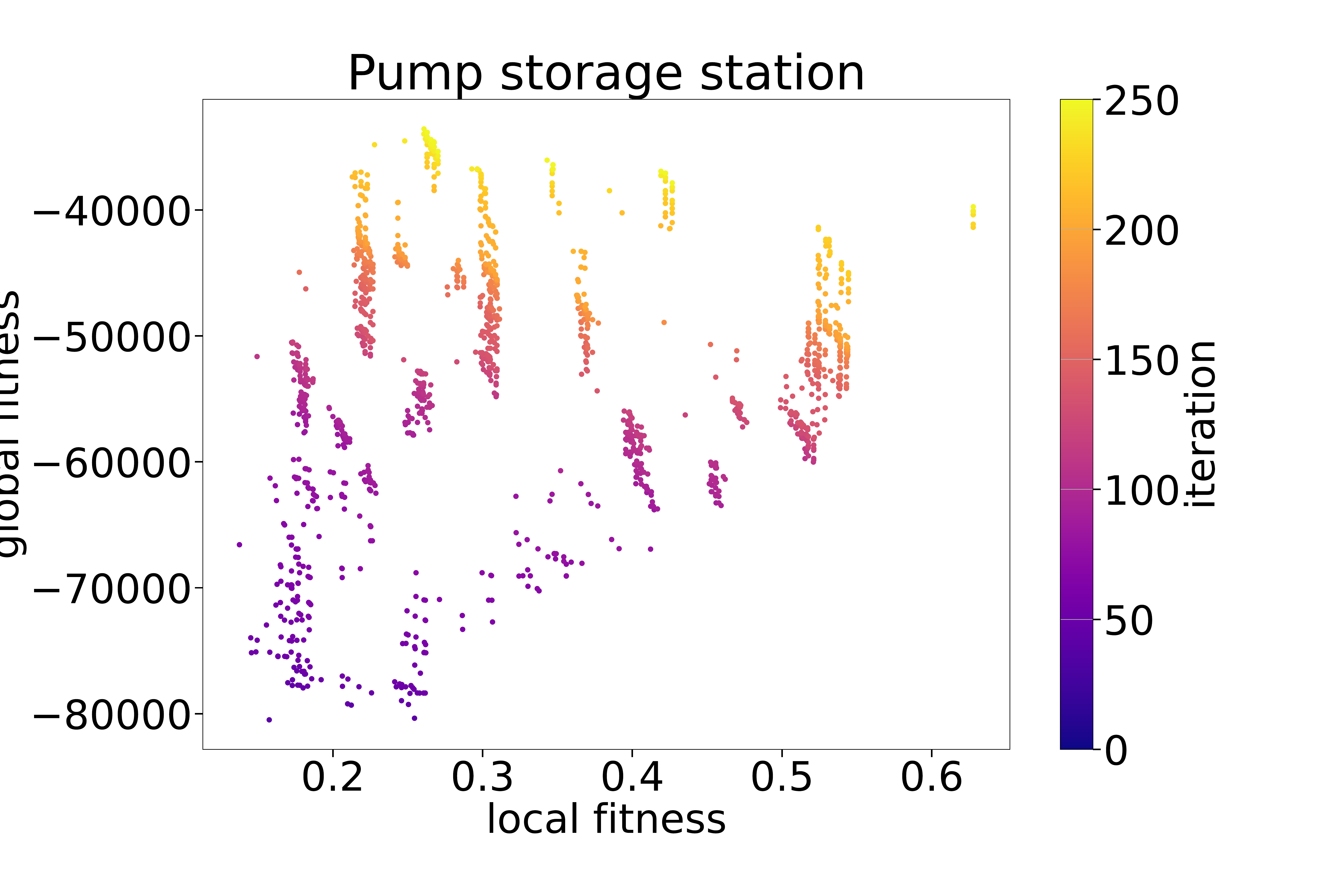}
    \caption{PSP: Pareto front}
    \label{subfig:first_psp_pareto_fit}
  \end{subfigure}
  \caption{Locally chosen OS -- 1. Scenario with Pareto optimization}
  \label{fig:first_industry_one_to_four_pareto}
\end{figure*}
\begin{figure*}[htbp]
  \centering
  \begin{subfigure}{0.24\textwidth}
    \centering
    \includegraphics[width=1\textwidth]{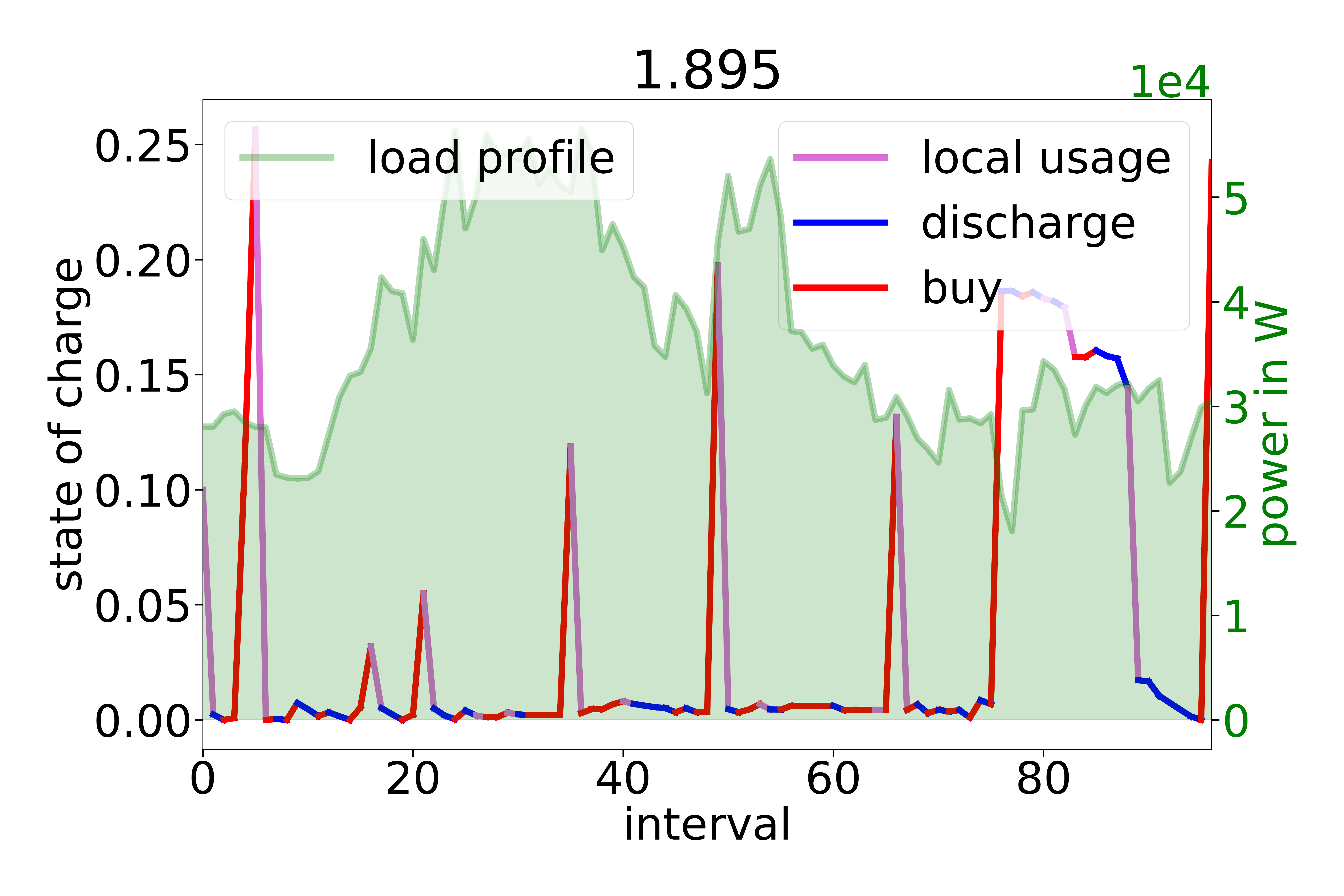}
    \caption{Industry one}
    \label{subfig:second_industry_one_norm}
  \end{subfigure}
  \begin{subfigure}{0.24\textwidth}
    \centering
    \includegraphics[width=1\textwidth]{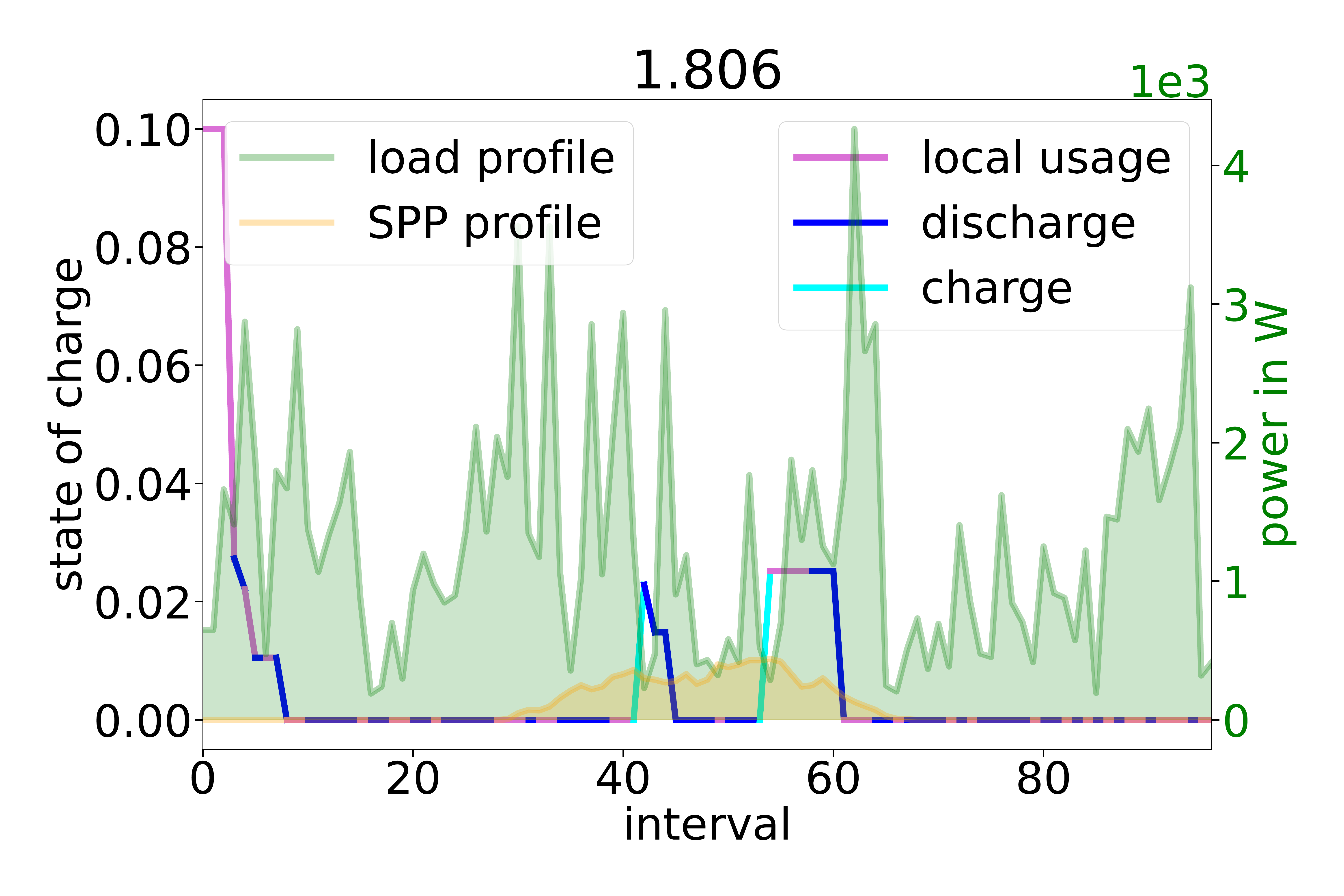}
    \caption{Household one}
    \label{subfig:second_hh_one_norm}
  \end{subfigure}
  \begin{subfigure}{0.24\textwidth} 
    \centering
    \includegraphics[width=1\textwidth]{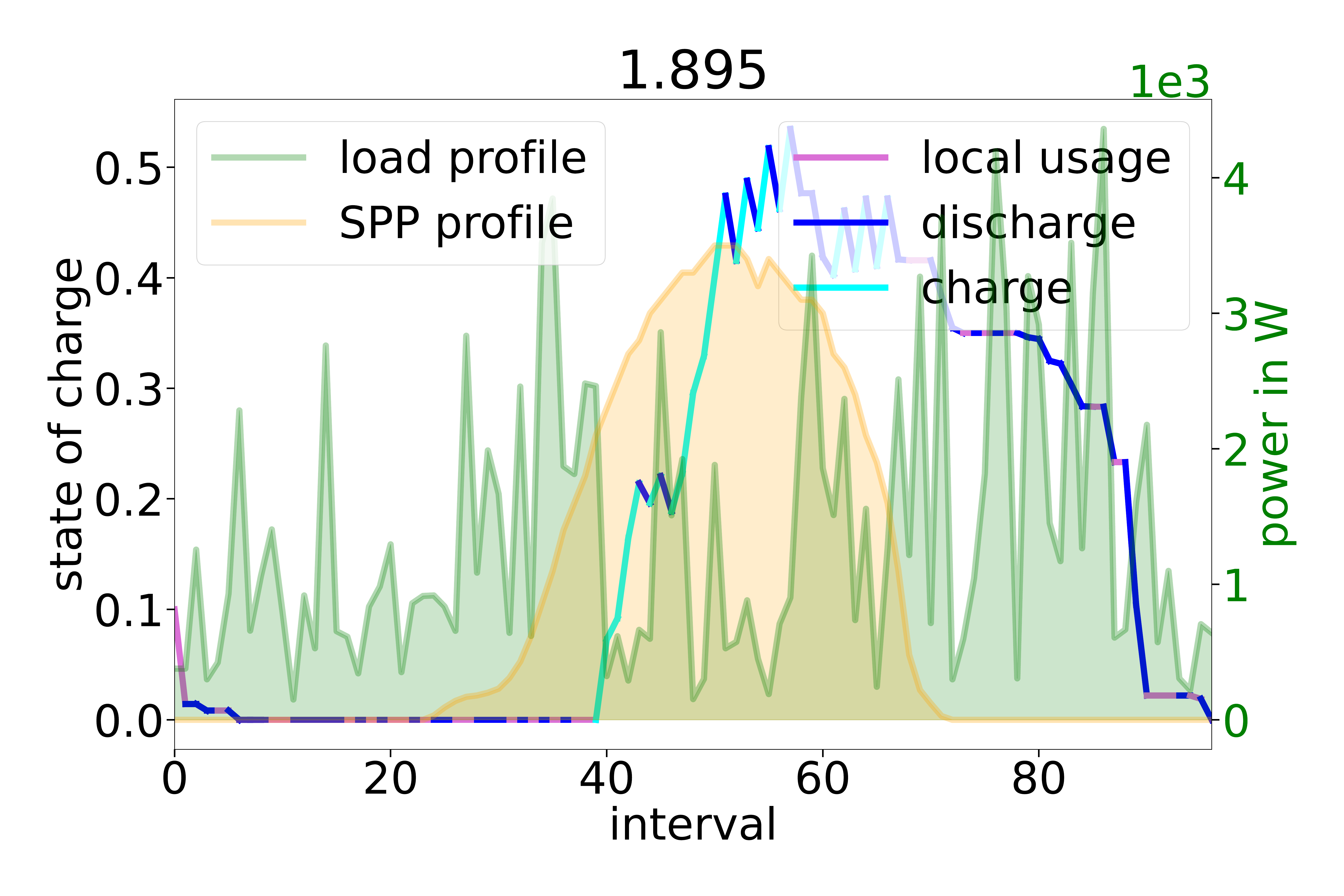}
    \caption{Household two}
    \label{subfig:second_hh_two_norm}
  \end{subfigure}
  \begin{subfigure}{0.24\textwidth}
    \centering
    \includegraphics[width=1\textwidth]{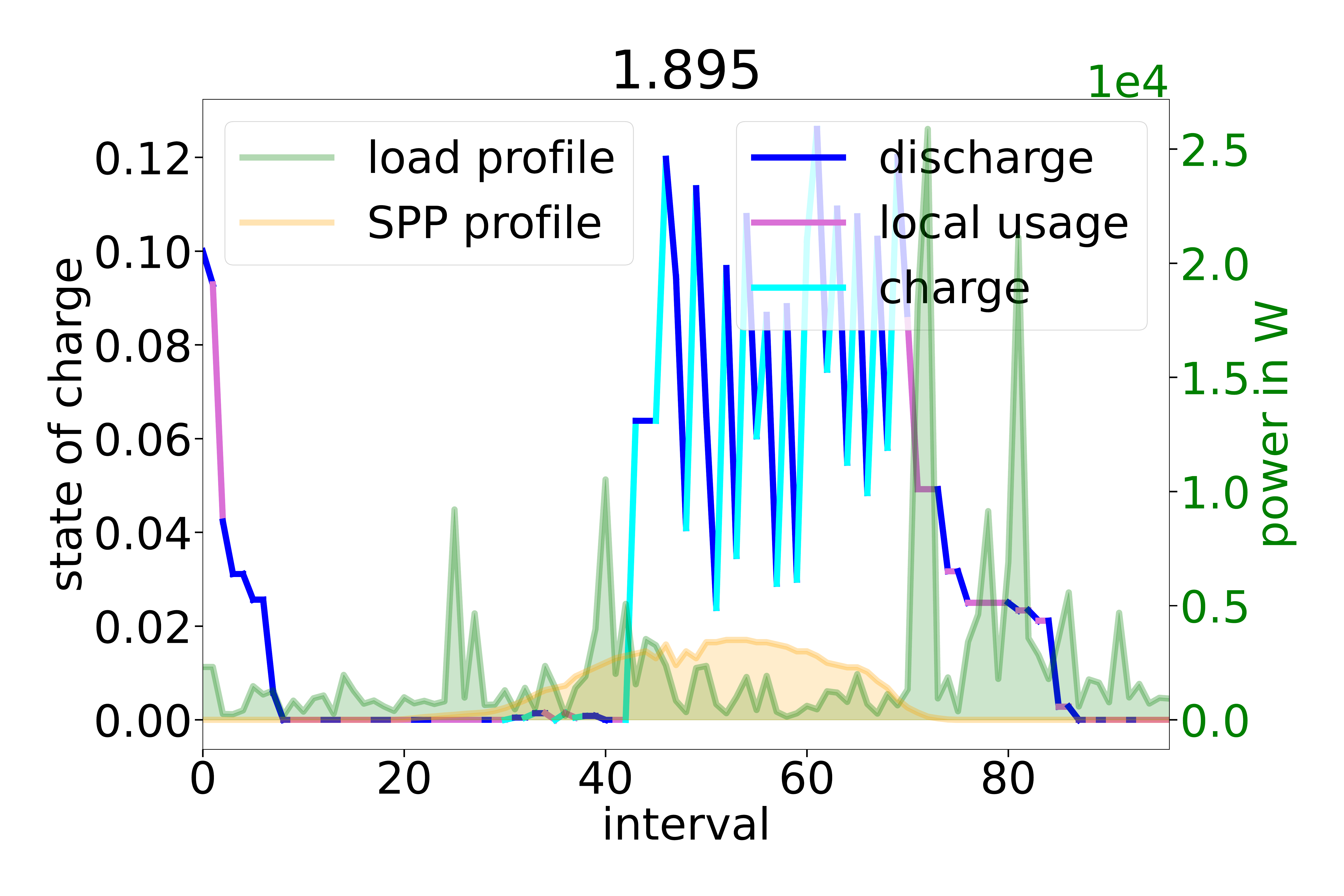}
    \caption{Household three}
    \label{subfig:second_hh_three_norm}
  \end{subfigure} 
  \caption{Locally chosen OS -- 2. Scenario with normalization}
  \label{fig:second_solutions_norm}
\end{figure*}
\begin{figure*}[htbp]
  \centering
  \begin{subfigure}{0.24\textwidth}
    \centering
    \includegraphics[width=1\textwidth]{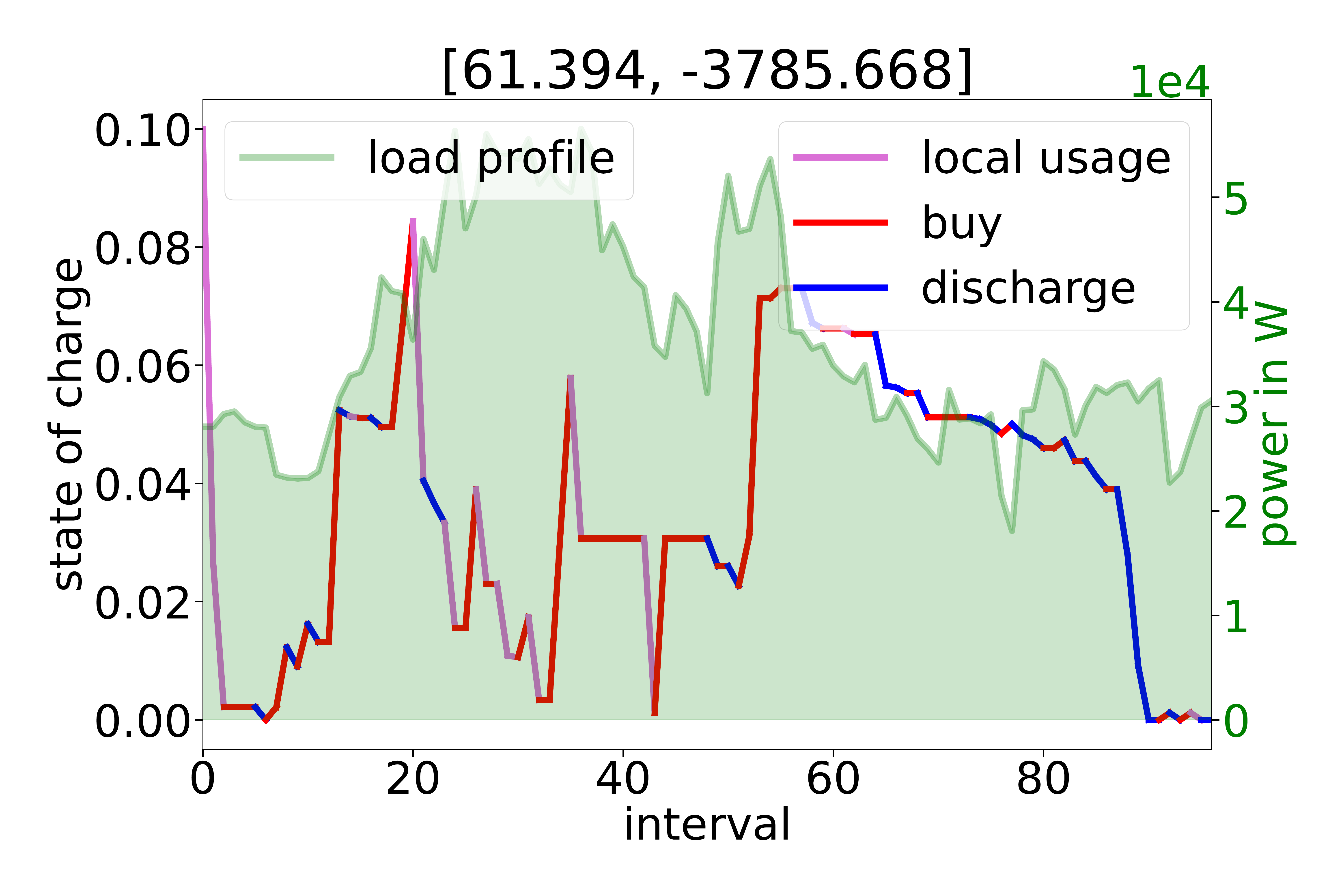}
    \caption{Industry one: chosen OS}
    \label{subfig:second_industry_pareto} 
  \end{subfigure}
  \begin{subfigure}{0.24\textwidth}
    \centering
    \includegraphics[width=1\textwidth]{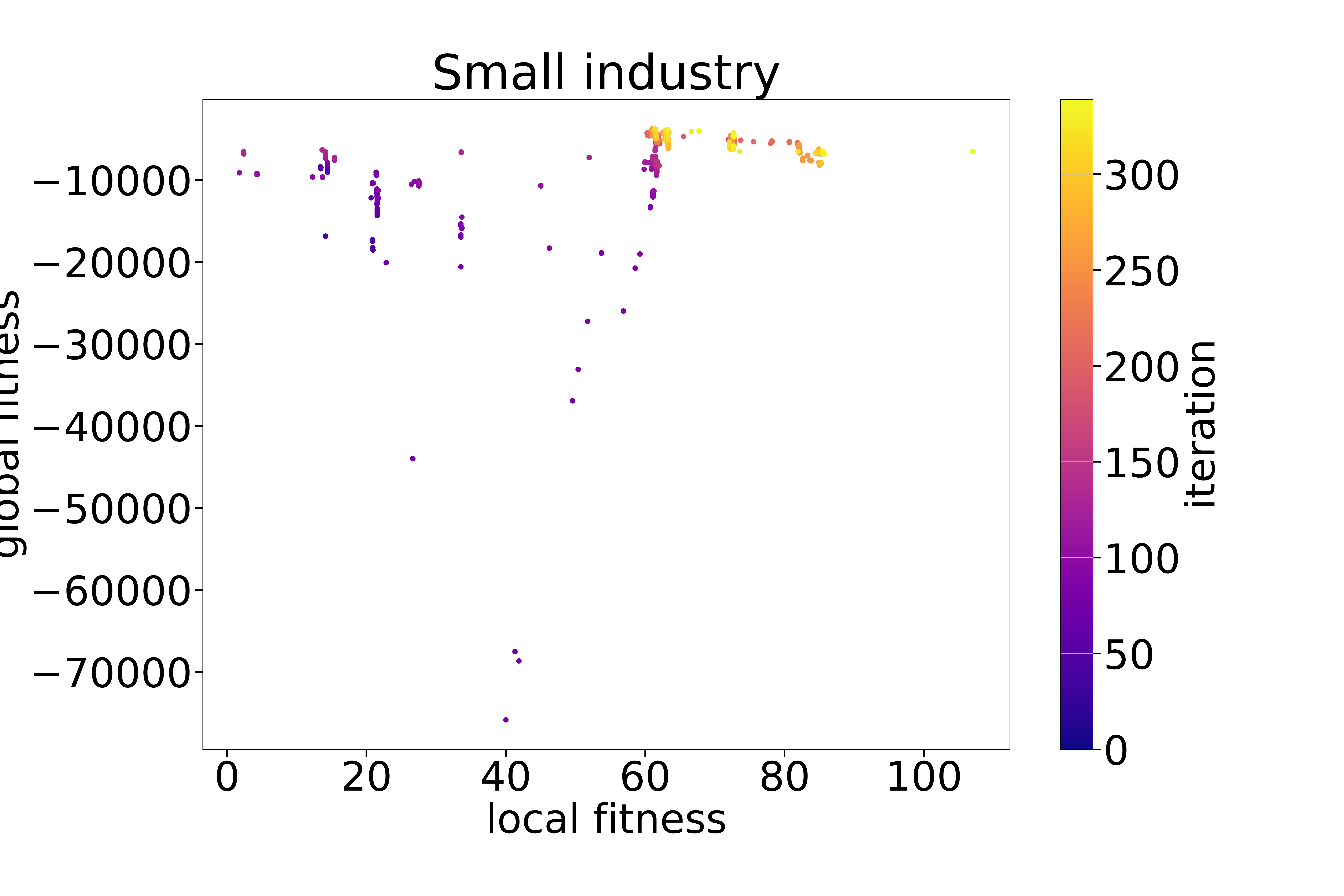}
    \caption{Industry one: Pareto front}
    \label{subfig:second_industry_pareto_fit} 
  \end{subfigure}
  \begin{subfigure}{0.24\textwidth}
    \centering
    \includegraphics[width=1\textwidth]{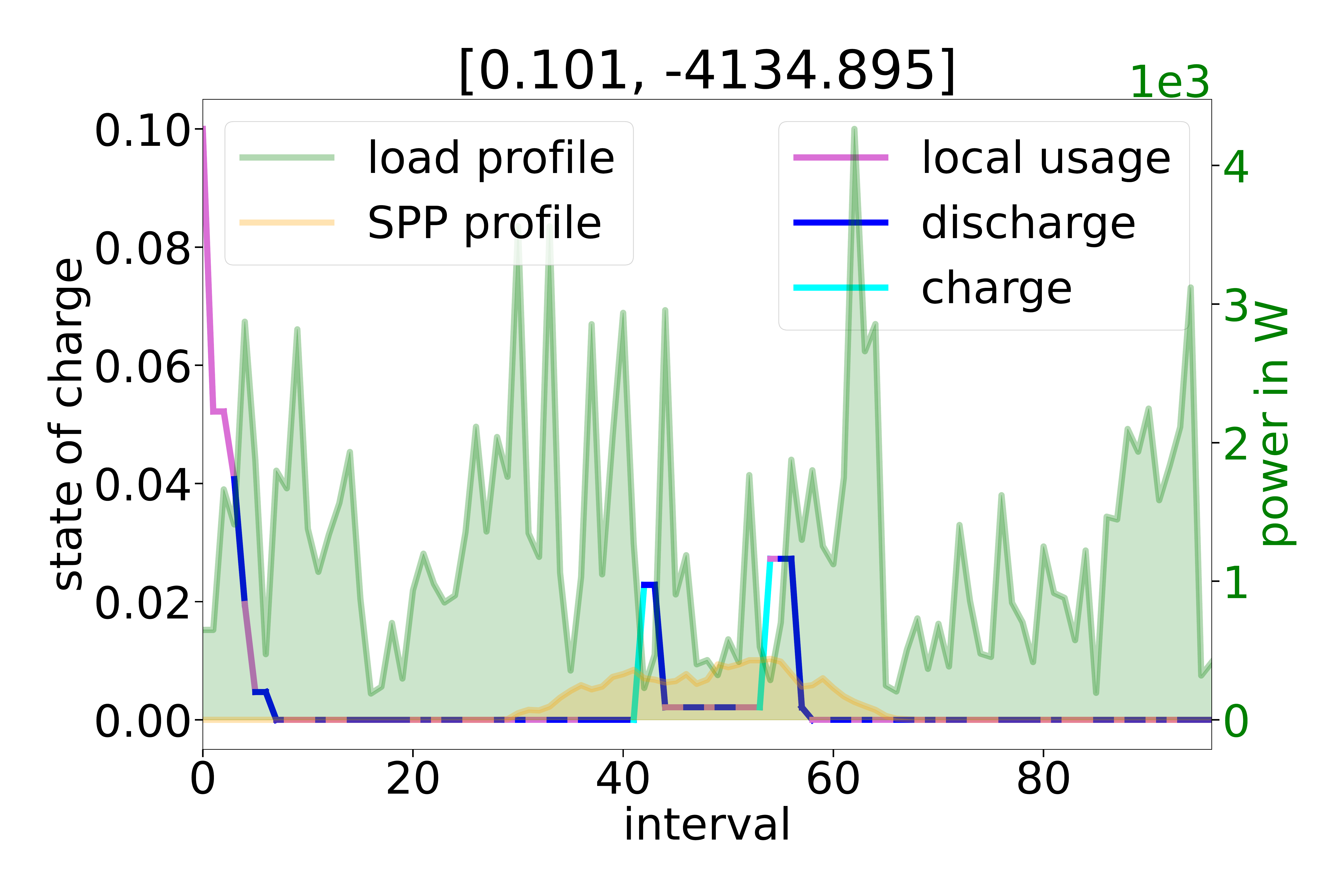}
    \caption{Household one: chosen OS}
    \label{subfig:second_hh_one_pareto}
  \end{subfigure}
  \begin{subfigure}{0.24\textwidth}
    \centering
    \includegraphics[width=1\textwidth]{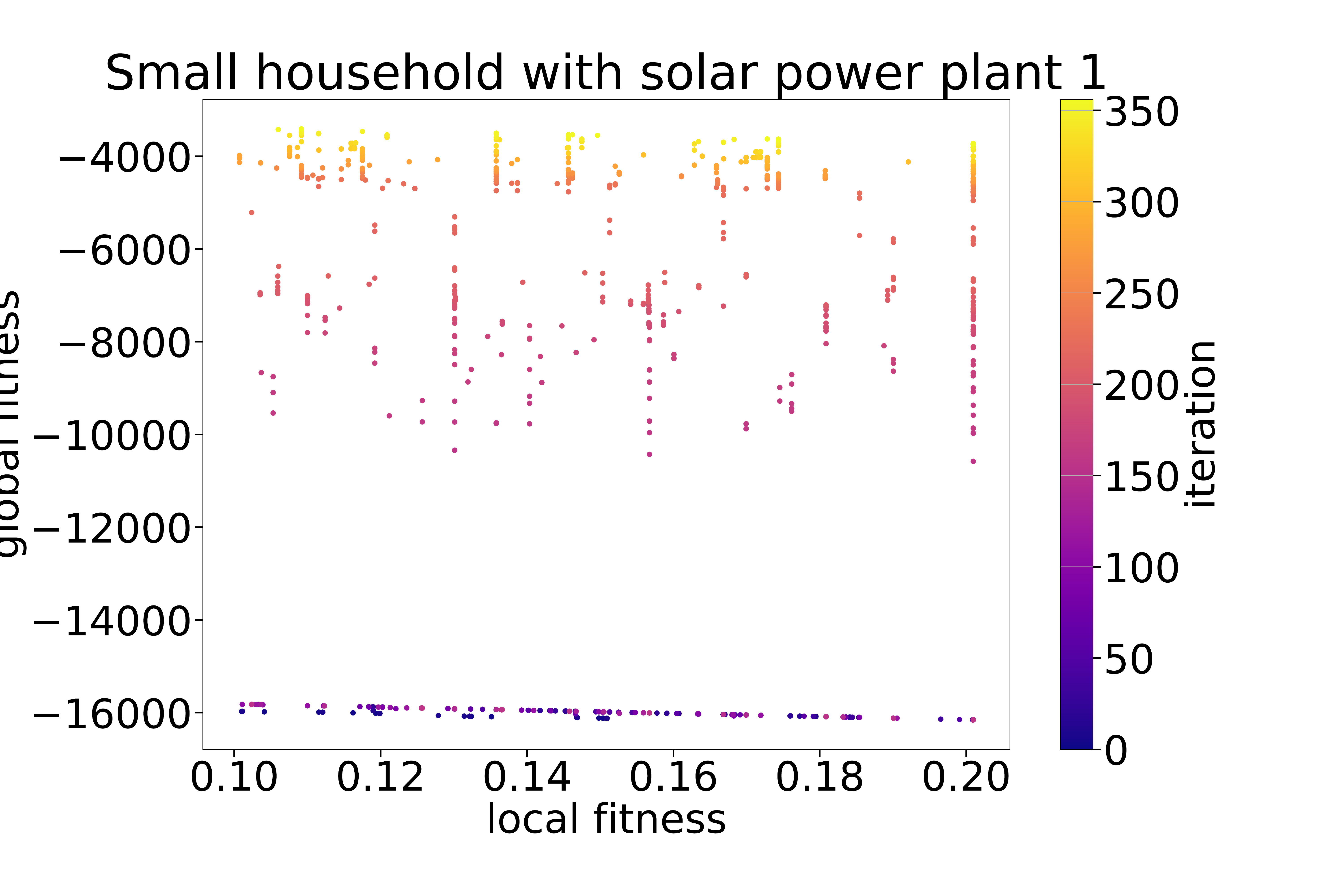}
    \caption{Household one: Pareto front}
    \label{subfig:second_hh_one_pareto_fit}
  \end{subfigure}
  \begin{subfigure}{0.24\textwidth} 
    \centering
    \includegraphics[width=1\textwidth]{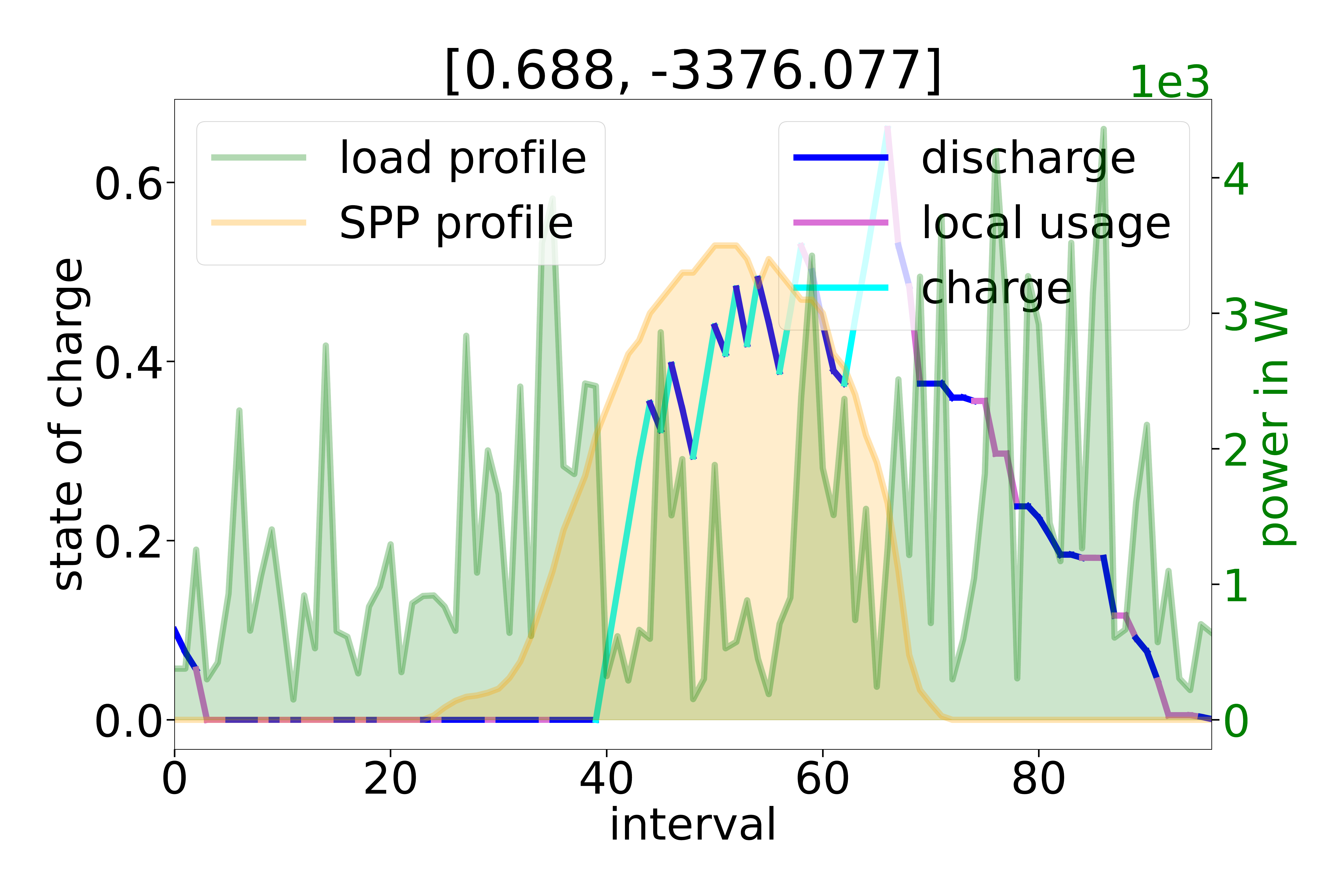}
    \caption{Household two: chosen OS}
    \label{subfig:second_hh_two_pareto}
  \end{subfigure}
  \begin{subfigure}{0.24\textwidth} 
    \centering
    \includegraphics[width=1\textwidth]{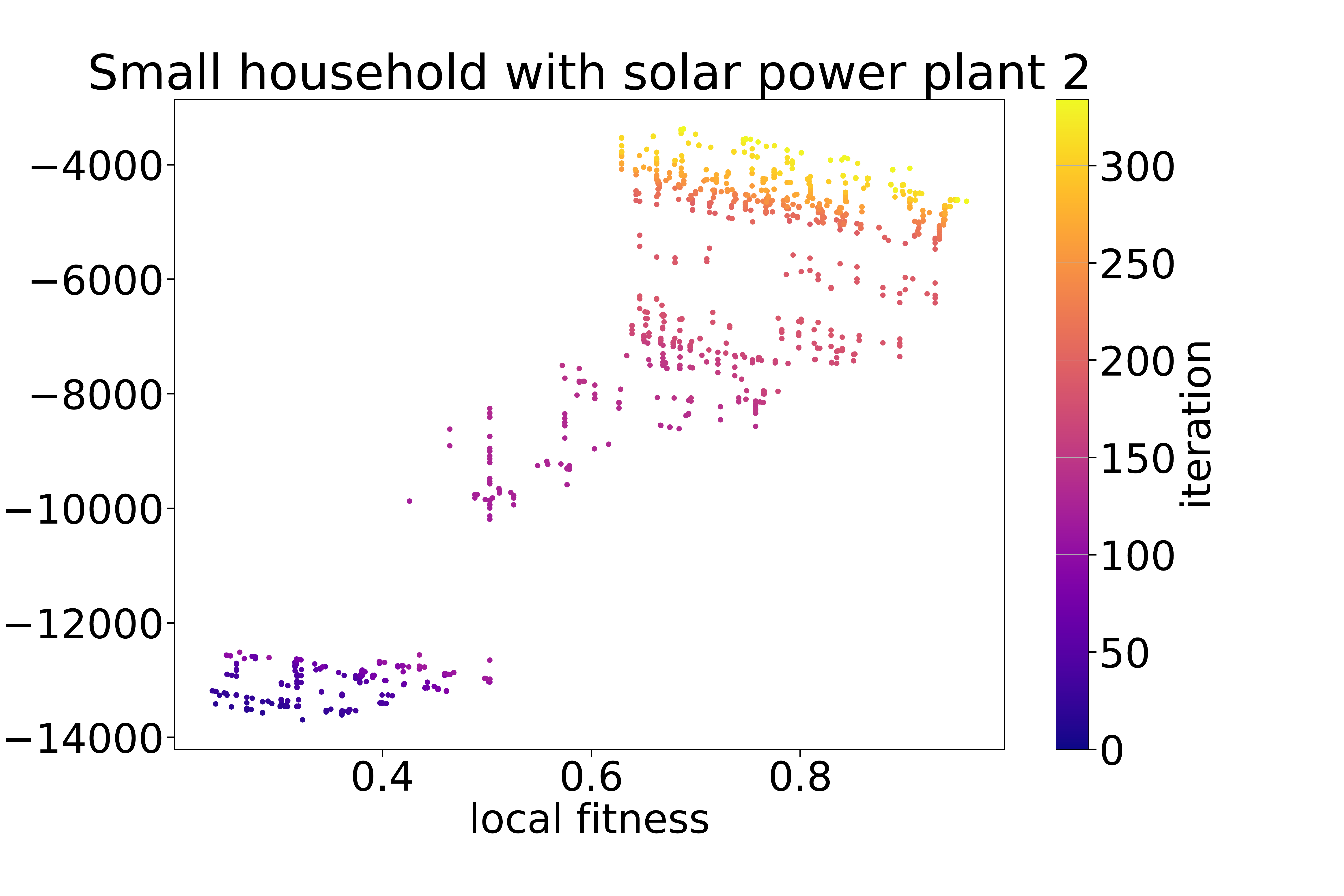}
    \caption{Household two: Pareto front}
    \label{subfig:second_hh_two_pareto_fit}
  \end{subfigure}
  \begin{subfigure}{0.24\textwidth}
    \centering
    \includegraphics[width=1\textwidth]{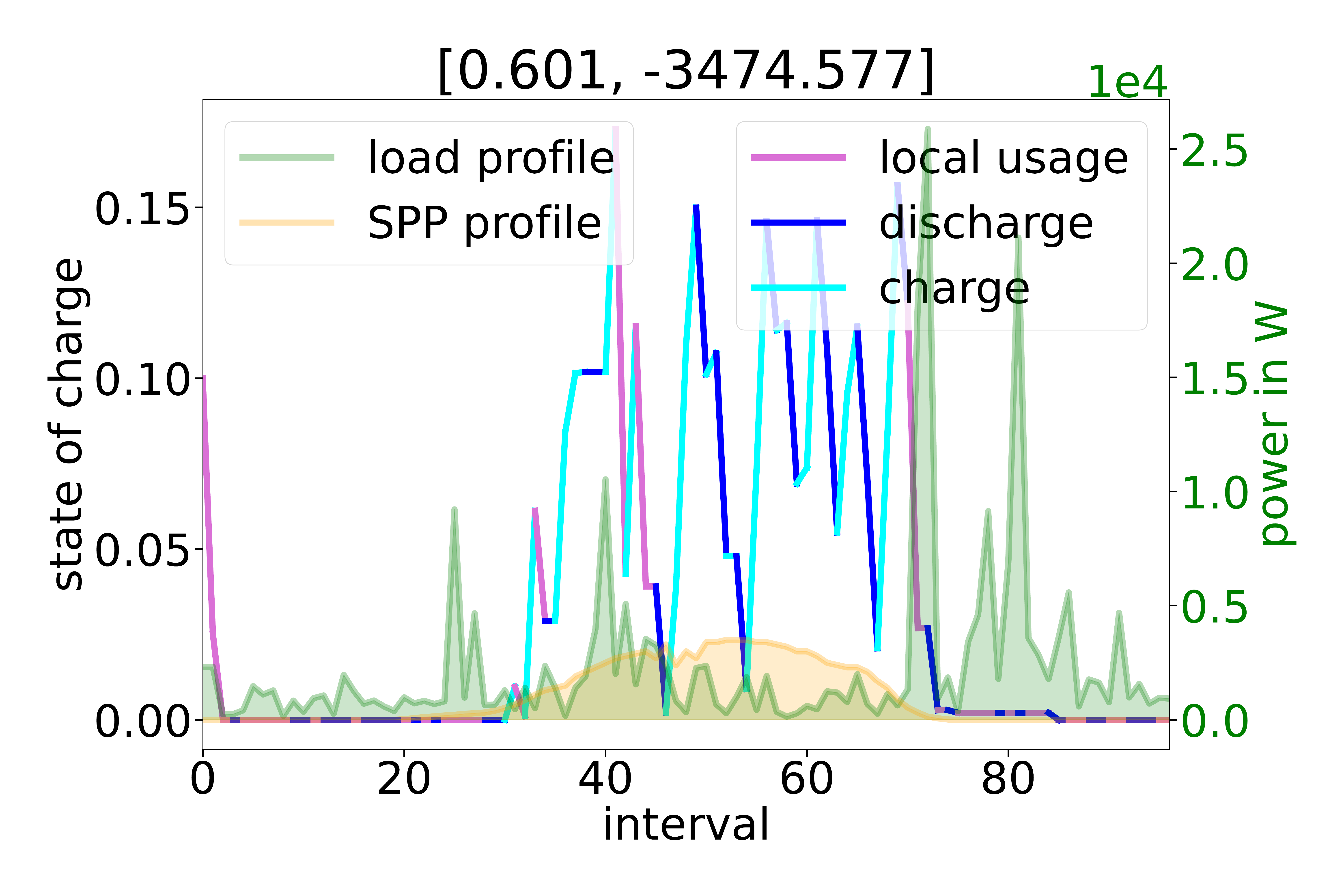}
    \caption{Household three: chosen OS}
    \label{subfig:second_hh_three_pareto}
  \end{subfigure} 
  \begin{subfigure}{0.24\textwidth}
    \centering
    \includegraphics[width=1\columnwidth]{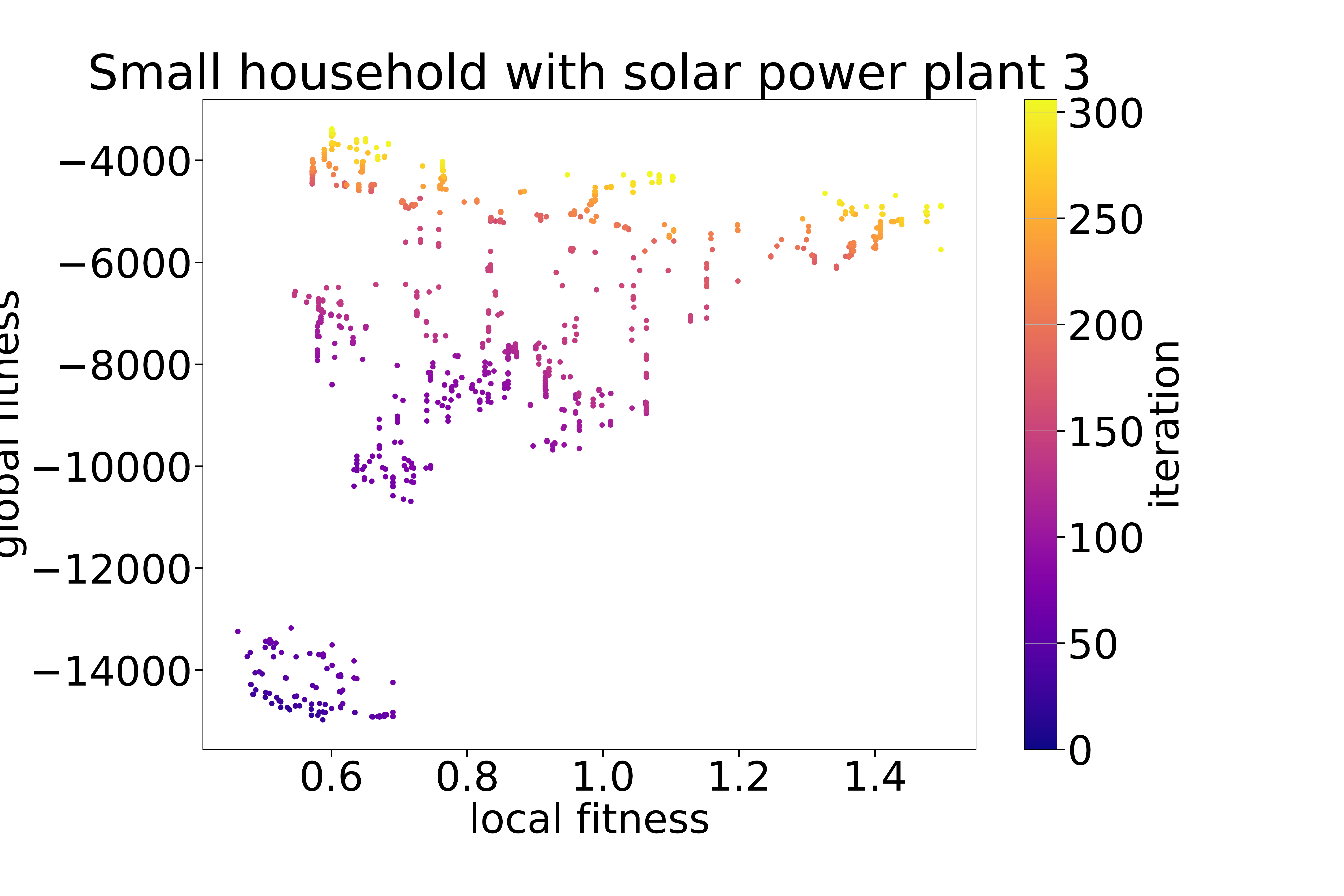}
    \caption{Household three: Pareto front}
    \label{subfig:second_hh_three_pareto_fit}
  \end{subfigure} 
  \caption{Locally chosen OS -- 2. Scenario with Pareto optimization}
  \label{fig:second_solutions_pareto}
\end{figure*}
\subsection{Parameter choice}\label{subsec:param_optimization}
In this section the three local test cases (\textit{arbitrage}, \textit{peak shaving}, \textit{local SDM}) presented in the evaluation are used to determine the optimal parameters for GABHYME. The relevant parameters are the number of generations $\numberofgenerations$, the population size $\populationsize$, and the number of parents $\numberofparents$. All test cases were executed $500$ times, and the resulting fitness plot is the average of those executions. In figure \ref{fig:evo_param_gensize} it is shown, how the fitness changes depending on different numbers of generations, figure \ref{fig:evo_param_popsize} shows fitness- and boxplots for different populations sizes $\populationsize$. The plots for the number of parents $\numberofparents$ are included in figure \ref{fig:evo_param_parents}. 
The evaluation of the impact of $\numberofgenerations$ shows that the fitness increased with increasing generation size. Looking at the box plots for the population size, it is clear that the most significant impact is the increasing robustness with increasing population size. Lastly, the evaluation of the number of parents is ambiguous. There are differences in the impact depending on the objective. For \textit{peak shaving}, it seems to be advantageous not to use the recombination at all, while for \textit{arbitrage}, the use of four parents seems to be optimal. As a result, the parameters listed in table \ref{tab:alg_param} are used in the evaluation.

\end{document}